# Space Plasma Physics: A Review

Bruce T. Tsurutani[1], Gary P. Zank[2], Veerle J. Sterken[3], Kazunari Shibata[4,5], Tsugunobu Nagai[6], Anthony J. Mannucci[1], David M. Malaspina[7,8], Gurbax S. Lakhina[9], Shrikanth G. Kanekal[10], Keisuke Hosokawa[11], Richard B. Horne[12], Rajkumar Hajra[13], Karl-Heinz Glassmeier[14], C. Trevor Gaunt[15], Peng-Fei Chen[16], Syun-Ichi Akasofu[17]

[1]Jet Propulsion Laboratory, California Institute of Technology, Pasadena, CA, USA; [2] Center for Space Plasma and Aeronomic Research (CSPAR), University of Alabama in Huntsville, Huntsville, AL 35805, USA; [3]Department of Physics, ETH Zürich, Zürich, Switzerland; [4]School of Science and Engineering, Doshisha University, Kyotanabe City, Kyoto, 610-0394, Japan; [5]Kwasan Observatory, Kyoto University, Yamashina, Kyoto, 607-8471, Japan; [6]retired, Tokyo, Japan; [7]Astrophysical and Planetary Sciences Department, University of Colorado, Boulder, CO, USA; [8]Laboratory for Atmospheric and Space Physics, University of Colorado Boulder, CO, USA; [9]Indian Institute of Geomagnetism, Navi Mumbai, 410218, India; [10]Goddard Space Flight Center, Greenbelt, Maryland, USA; [11]University of Electro-Communications, Chofu, Tokyo, Japan; [12]British Antarctic Survey, Madingley Road, Cambridge, CB3 0ET, UK; [13]Indian Institute of Technology Indore, Simrol, Indore 453552 India; [14]Technische Universität Braunschweig, Germany; [15]C. Trevor Gaunt, University of Cape Town, Cape Town, South Africa; [16]School of Astronomy and Space Science, Nanjing University, Nanjing 210023, China; [17]International Arctic Research Center, Fairbanks, Alaska, USA

## Abstract

Owing to the ever-present solar wind, our vast solar system is full of plasmas. The turbulent solar wind, together with sporadic solar eruptions, introduces various space plasma processes and phenomena in the solar atmosphere all the way to the Earth's ionosphere and atmosphere and outward to interact with the interstellar media to form the heliopause and termination shock. Remarkable progress has been made in space plasma physics in the last 65 years, mainly due to sophisticated in-situ measurements of plasmas, plasma waves, neutral particles, energetic particles, and dust via space-borne satellite instrumentation. Additionally high technology ground-based instrumentation has led to new and greater knowledge of solar and auroral features. As a result, a new branch of space physics, i.e., space weather, has emerged since many of the space physics processes have a direct or indirect influence on humankind. After briefly reviewing the major space physics discoveries before rockets and satellites (Chapter 1), we aim to review all our updated understanding on coronal holes, solar flares and coronal mass ejections, which are central to space weather events at Earth (Chapter 2), solar wind (Chapter 3), storms and substorms (Chapter 4), magnetotail and substorms, emphasizing the role of the magnetotail in substorm dynamics (Chapter 5), radiation belts/energetic magnetospheric particles (Chapter 6), structures and space weather dynamics in the ionosphere (Chapter 7), plasma waves, instabilities, and wave-particle interactions (Chapter 8), long-period geomagnetic pulsations (Chapter 9), auroras (Chapter 10), geomagnetically induced currents (GICs, Chapter 11), planetary magnetospheres and solar/stellar wind interactions with comets, moons and asteroids (Chapter 12), interplanetary discontinuities, shocks and waves (Chapter 13), interplanetary dust (Chapter 14), space dusty plasmas (Chapter 15) and solar energetic particles and shocks, including the heliospheric termination shock (Chapter 16). This paper is aimed to provide a panoramic view of space physics and space weather.

## 1. Space Physics Before Rockets and Satellites

William Gilbert, who was Queen Elizabeth I of England's personal physician, established that the Earth had a large, global magnetic field [1]. Magnetic surveys and changes of the Earth's magnetic field at that time and before were keen subjects during this period of history. When sailing was the main means of exploration of the world, many countries established magnetic observatories. In 1675, the British established the Greenwich Observatory. In 1817 a separate building was erected for magnetic observations as part of the Observatory. As part of the British empire, a Colaba Observatory was built in Mumbai, India in 1827 (this will be discussed further in Chapter 4: Storms and Substorms). The calibration for the Grubb magnetometer that was installed there [2] did not contain a schematic of the instrument, indicating that England wished to keep its technology secret from its European competitors.

A correlation between ground magnetic field variations and auroras were noticed in 1743 by Olav Hiorter [3]. Later, the famous naturalist Alexander von Humboldt performed an experiment in 1806 that found that magnetic needles oscillated when auroras were overhead. When the auroras disappeared, so did the magnetic needle oscillations. Von Humboldt named this phenomenon "magnetische ungewitter" or "magnetic storms" [4], a term we use today. In 1847, William Barlow performed an experiment that showed that railroad telegraph magnetic needles were deflected when auroras were overhead [5]. Barlow gave us evidence for geomagnetically induced currents (GICs) in conductors at Earth. GICs are of great concern today in this era of high technology and will be discussed in Chapter 11, Space Weather.

On 1 September 1859, both Richard Carrington and Richard Hodgson saw a brief solar flare on the Sun [6, 7]. Carrington noted that a magnetic storm occurred at Earth some ~17 hrs and 40 min later, but he reported "one swallow does not make a summer". Lord Kelvin who did not believe there was connection between the Sun and Earth said the magnetic storm on Earth must have been "mere coincidence" [8].

Edwin Maunder reported [9] that geomagnetic activity sometimes reappeared with ~27-day intervals, the same period as the rotation of the Sun. This is the same Maunder who reported that auroras were exceedingly rare from ~1645 to 1715 (recently named the "Maunder Minimum"). It was not until Charles Chree, the director of the Kew Observatory proved that Maunder's ~27-day quasiperiodic results were statistically significant, giving us the "Chree superposed epoch" statistical analyses [10], a method widely in use today. Although these periodic geomagnetic activities at Earth were substantiated by Chree, no identifiable optical features causing them were apparent on the visible Sun. Julius Bartels called these solar regions "M-regions" for magnetically active regions [11]. It was not until soft X-ray images of the Sun were available from the Skylab satellite that it was realized that high speed solar winds came from dark or low temperature regions not visible in optical wavelengths [12]. The M-regions were renamed "coronal holes". Coronal holes and related geomagnetic activity will be discussed in Chapter 2 (Coronal Holes, Solar Flares and Coronal Mass Ejections) and Chapter 4 (Storms and Substorms).

Following the idea of Kristian Birkeland [13], Sydney Chapman and Vincent Ferraro [14] considered the solar wind coming from the Sun. This plasma impact would form the magnetosphere. Ludwig Biermann and Thomas Cowling studied the ion tails of comets to deduce from ground observations that there was a solar wind emanating radially from the Sun [15]. The densities and speed estimates were high and those numbers were later revised downward with in situ information.

The ionosphere was well known to exist before the space age. Carl Friedrich Gauss [16] suggested that a conducting region of the atmosphere could explain some observed variations of the Earth's magnetic field (see [17] for an English translation of the Gauss paper [16]). Edward Appleton and Miles Barnett [18] proved that an ionosphere exists by experiments launching waves from the ground at an angle relative to the zenith. From the ground interference patterns the height of the reflecting layer (the ionosphere) was deduced.

Cosmic rays consist of high energy protons and other atomic nuclei which originate from outside our galaxy. When they impact the Earth's atmosphere, they create a "shower" of secondary and tertiary particles. The existence of cosmic rays was proven by Victor Hess [19] who used balloon experiments to show that radiation increased with increasing balloon altitude. Scott Forbush discovered rapid decreases in the observed galactic cosmic ray intensities at Earth [20]. These decreases are now known to be caused by solar/interplanetary high magnetic field events which mirror cosmic rays coming towards the Earth. Forbush made this discovery using ionization chambers. Today these cosmic ray decreases are known as Forbush decreases. John Simpson [21] invented neutron monitors which are standardly used today to monitor cosmic ray intensities on the ground.

The concept of electromagnetic hydromagnetic "Alfvén" waves was introduced before the space age. A short Nature article was published in 1942 [22]. The existence of electromagnetic waves in an electrically highly conducting medium was a ground breaking theoretical idea, experimentally proven by Lundqvist [23]. The Alfvén wave, with its transverse magnetic $\vec{b}$ and electric $\vec{E}$ perturbations, is driven by transverse polarization or inertial $\vec{J}_P$ and field-aligned $j_{||}$ currents in a plasma (Figure 1). The electric and magnetic field components are in-phase and perpendicular to each other. Thus, the electromagnetic Poynting vector is aligned with the local background magnetic field. This Alfvén mode is of paramount importance for the electro-magnetic momentum and energy coupling between different space plasma regimes.

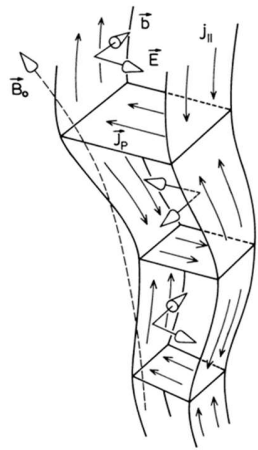

**Figure 1**. Schematic of the basic features of an Alfvén wave with its transverse magnetic $\vec{b}$ and electric $\vec{E}$ perturbations, is driven by transverse polarization or inertial $\vec{J}_P$ and field-aligned currents $j_{||}$. The figure is taken from [24].

We have mentioned in the abstract that new miniaturized instrumentation developed during the space age has enormously increased our knowledge of space plasma physics since 1958. However, it is not possible nor the intent of this paper to review space plasma instrumentation here. But at the request of a referee of this paper we mention a NASA book entitled "Small Instruments for Space Physics" (1993) that was the culmination of a weeklong workshop that was held in Pasadena, California for the purpose of developing miniaturized instrumentation for the Parker Solar Probe mission. It should also be mentioned that our laboratory plasma colleagues were also developing instrumentation and unique plasma devices to help study space plasma phenomena in parallel with space mission studies. Two excellent reviews are [25] and [26].

## 2. Coronal Holes, Solar Flares, and Coronal Mass Ejections (CMEs)

Our Sun is a main sequence star, with its interior composed of three layers, i.e., the core (0-0.25 $R_S$) where the hydrogen fusion powers the whole star, the radiation zone (0.25-0.7 $R_S$) where energy is transferred outward via radiation, and the convection zone (0.7-1 $R_S$) where energy is transferred out via convection. The turbulent convection in the rotating volume has many crucial consequences: under the Coriolis force, it produces differential rotation, i.e., the equator rotates faster than the poles [27]. It also produces the global circulation, i.e., the meridional flows. More importantly, the turbulent convection works as an effective dynamo, generating magnetic fields. The amplified magnetic fields in the convection zone emerge up through the solar surface into the solar atmosphere. With the everlasting mixing of the solar surface motions and the continual pumping of magnetic fields [28], the corona is heated to 1-2 MK (in contrast to the 6,000 K solar surface), forming the solar wind. Moreover, sporadic eruptions are driven in the background solar wind, such as solar flares and coronal mass ejections (CMEs). The resulting coronal holes, solar flares, and CMEs are the three main sources of geomagnetic activities at Earth.

**Coronal Holes**

The Sun has an atmosphere with four layers, which are, from inside to outside, the photosphere with a thickness of ~500 km and a temperature around 6,000 K, the chromosphere with a thickness of ~1,500 km or more and a temperature up to 20,000 K, the transition region with a thickness of ~100 km, and the corona with a temperature of 1-2 MK [27]. As demonstrated by Parker [29], the coronal pressure is so high that the solar gravity cannot hold the corona in a hydrostatic equilibrium state, leading to continuous solar wind in the corona and in interplanetary space. It has recently been argued that solar wind is also inevitable since both gravity and magnetic tension force cannot provide sufficient centripetal force for a hydrostatic equilibrium [30]. Still, some parts of the corona are trapped by relatively stronger magnetic fields which exit the surface and go back into the surface, forming coronal loops [31]. Therefore, the low corona is naturally divided into (1) the portions trapped by closed magnetic fields, and (2) the portions threaded by open magnetic field with fast solar wind. Since the trapped portions correspond to stronger magnetic fields, implying stronger heating and more chromospheric evaporation, they are brighter not only in white-light images, but also in EUV and X-ray images. In contrast, the open field portions are less heated and the plasma keeps flowing away from the Sun, therefore they are much fainter in white-light images, as well as in EUV or X-ray images, manifested relatively dark regions known as coronal holes.

When observed in EUV or soft X-rays on the solar disc, coronal holes are seen as intensity depleted areas surrounded by either quiet Sun or active regions. Above the solar limb, they are seen as intensity depletion regions bounded by ~10 times brighter streamers. The above-limb coronal holes should have already been observed during total solar eclipses for centuries prior to the space age. Maunder [9] noted a strong ~27-day recurrence in geomagnetic activity at Earth (see also [10]) and speculated that some "invisible" feature on the Sun was causing this. Though not seeing any feature on the Sun, Bartels [32] suggested that the periodic storms are produced by unseen "M-regions" on the Sun. Coronal holes above the limb were first quantitatively measured by Waldmeier [33] with the imaging observations in Fe XIV 5303 Å. In the space age starting from the 1960s, coronal holes were observed as discrete dark patches on the solar disc in UV, EUV or X-ray [34-35]. As expected, coronal holes are brighter than the quiet Sun in He I 10830 Å [36].

The ratio of total coronal hole area to the total solar disc varies from near zero at solar maxima to ~0.25 at solar minima [37]. During solar minima, coronal holes are mainly confined near the poles, with latitudes above 60°. These polar coronal holes may exist for ~7 years around solar minima, and are absent for ~1-2 years around solar maxima when the polar magnetic fields reverses their signs. After solar maxima, small polar holes appear, and merge together to grow into a large one [38-39]. When there are active regions on the solar disc, coronal holes might appear in low latitudes, with a lifetime of 1-2 years.

Coronal holes are nearly indistinguishable in the photosphere and low chromosphere from the quiet solar images (although there are subtle differences in the spectral profiles of some chromospheric lines, such as the Ca II H and K lines [40]. Coronal holes become eminent in the emissions above $10^5$ K. Interestingly, the coronal holes in UV or X-rays rotate almost like a rigid body, with the rotation rate slower than other features [38, 41]. It might be due to that their magnetic field is rooted in the deep interior [42] or due to interchange reconnection between the open field in the coronal holes and the closed field nearby [43]. Another remarkable difference between coronal holes and other areas is the so-called first ionization potential (FIP) effect, i.e., in quiet Sun and active regions, the elements with a low FIP are significantly enhanced in the corona compared to the solar surface. In contrast, the element abundance in coronal holes is the same between the corona and the solar surface [44-45].

Large polar coronal holes may contain many ray-like structures called polar plumes [46] and sporadic jets. The plumes were proposed to be generated via the interchange reconnection between the open field and ephemeral bipolar magnetic field, the same mechanism as that for coronal jets [47], although the jets are more impulsive and more energetic [48-49]. Compared to the interplume regions, plumes are 2-3 times brighter, their kinetic temperature is lower [50], and their outflows are slower [51]. It was shown that at heights around $r \approx 2\ R_S$, polar coronal holes are comprised of about 25% plume and 75% interplume plasma [52]. A plume decays after the minority-polarity flux is cancelled on a ~1 day timescale of the supergranular convection [46]. The sporadic jets, due to impulsive interchange reconnection especially near the edge of coronal holes, may become narrow CMEs [53-54].

Photospheric magnetograms indicate that coronal holes are more unipolar than other regions [46, 55], and the Potential Field Source Surface (PFSS) model revealed that they correspond to open magnetic fields, hence are the source of the fast solar wind [12]. It was found that when the source surface is assumed to be located at $r \approx 2.5\ R_S$, the open field in the PFSS model can best match the coronal holes [56].

The *Ulysses* mission indicated that there exist two types of solar wind, i.e., the steadier fast wind (~750-800 km s$^{-1}$) originating from coronal holes and the slow wind (~350-400 km s$^{-1}$) originating from a region around the Sun's equatorial belt that is known as the "streamer belt". Now there is accumulating evidence to show that fast solar wind comes from the core region of large coronal holes, whereas slow solar wind comes from the edge of large coronal holes and small coronal holes around active regions [57-61]. Note that the edges of coronal holes correspond to the boundaries of helmet streamers or pseudo-streamers, where interchange reconnection happens frequently. There are solar wind speeds between the above two values. These are from the outskirt of the large coronal holes and small coronal holes around active regions. In the low corona, the fast solar wind has a lower temperature, whereas the slow wind has a higher temperature. Beyond the heliocentric radius $r \approx 1.5\ R_\odot$, the fast wind temperature becomes higher than that of slow wind, and the wind speed is positively correlated with the in-situ temperature. While some authors have claimed that fast solar wind and slow solar wind are accelerated by different mechanisms, e.g., Alfvén waves for the former and reconnection for the latter, it seems that the expansion factor of the flux tube plays a key role in determining the terminal velocity, i.e., radially or slightly super-radially expanding flux tubes result in fast solar wind, the hyper-radially expanding flux tubes result in slow solar wind [46]. Empirical correlation technique in this framework has become ready for real-time prediction of solar wind speed [62-65]. We argue that the acceleration of solar wind involves nonlinear mode conversion of Alfvén waves [66-67], Alfvénic turbulence [68] and magnetic reconnection. Here the latter two processes are nearly inseparable in the corona, i.e., turbulent magnetic field induces reconnection and reconnection generates turbulence. On the other hand, it should be pointed out that kinetic processes are responsible for some properties of the solar wind [69], since minor ions, e.g., $O^{5+}$, have a strong temperature anisotropy ($T_\perp \sim 10 T_{//}$) with $T_\perp$ up to 200 MK and their outflows are faster than the proton wind by the local Alfvén speed [70]. Besides, $^3$He/$^4$He, $O^{7+}$/$O^{6+}$ and Fe/O ratios are often much larger in slow wind than in fast wind (it is not surprising that the kinetic properties of some slow wind streams are the same as those of fast wind in the case that the slow wind is due to hyper-radial expansion of the flux tubes). Tsurutani et al. [71] have recently reviewed kinetic processes in the solar wind with discontinuities, magnetic reconnection and possible intermediate shocks playing important roles.

When fast wind streams catch up with slow wind, their collision forms corotating interaction regions (CIRs: [72]). As the coronal holes rotate along with the Sun, fast streams and occasional shock waves at the CIRs hit the Earth magnetosphere, producing recurrent storms with $D_{st} \geq$ -100 nT in most cases [73-74].

It is mentioned in passing that when a CME happens, sometimes we can observe twin dimmings around the source region, which are called transient coronal holes [75]. They were believed to correspond to the footpoints of an erupting flux rope [76].

**Solar Flares**

Solar flares are transient brightening phenomena in the solar atmosphere, observed in all electromagnetic spectrum ranging from radio to gamma rays [77]. Their typical energies are $10^{29}$-$10^{32}$ erg, and time scales are a few minutes to a few hours, although there are no actual characteristic energies and time durations for flares. The flare frequency statistics show that the number of flares $N$ increases with decreasing flare energy $E$ with a power-law distribution: d$N$/d$E$ $\propto E^{-\alpha}$, where $\alpha$ = 1.6-2.0 [78].

Solar flares are often associated with mass ejections, such as jets, filament (prominence) eruptions, and coronal mass ejections (CMEs). The largest mass ejections, CMEs [54], play a fundamental role in generating geomagnetic storms, and will be described in detail later in this paper. Solar flares also emit solar energetic particles (SEPs), and especially fast-mode shocks ahead of CMEs are known to be the important source of SEPs.

The recent progress of space-based solar observations in last few decades has revolutionized solar flare research, and it has been established at least phenomenologically that solar flares are caused by magnetic reconnection (the release of magnetic energy stored in the solar atmosphere: [79]). Historically, it has sometimes been discussed that the origin of CMEs is different from that of flares [80], but there is increasing evidence that at least major CMEs and flares are simply different aspects of the same magnetohydrodynamic phenomena ([54], see discussions on CME mechanisms later).

The first solar flare that human beings observed was a white light flare observed by Carrington [6] and Hodgson [7]. This flare induced the largest geomagnetic storm in the recent 200 years, and caused several problems in human civilization even in the infancy of electromagnetic technology [81]. Telegraph systems, the high technology of the day, went down and fires were started in telegraph stations [82]. It has been considered that the Carrington flare was one of the most energetic flares (with energy of order of $10^{32}$ erg) observed so far [83].

Recently, Maehara et al. [84] discovered that many solar type stars have created superflares with energies of $10^{33}$-$10^{35}$ ergs (10-1000 times more than the Carrington-class flare). This suggested that our Sun has the possibility of generating such superflares. Although the frequency is low, once in a few 100 to a few 1000 years for a $10^{33}$-$4\times10^{34}$ erg superflare [85]. The consequences for humankind from some extreme events could be disastrous. Evidence of SEPs from solar extreme superflares 1000 years to 9,000 years ago have been found from $^{14}$C in terrestrial tree rings and ice core data [86-89].

In 2003, there were flares that were so intense that their X-ray emission saturated the NOAA 1–8 Å GOES detectors. As a reference point an X10-class solar flare is estimated to be approximately a $10^{32}$ erg flare. To approximate the saturated flare intensities, NOAA extrapolated the flare light curves to obtain a value of X28 for the November 4, 2003 flare and X17 for the October 28, 2003 flare. On the other hand, Thomson et al. [90] estimated the November 4 flare as having an X45 $\pm$ 5 intensity (using an ionospheric radio technique). Tsurutani et al. [91] noted that the October 28, 2003 flare was double the intensity of the November 4 flare in EUV wavelengths (the most intense EUV flare in recorded history), so it is possible that the total energy of the October 28 flare was even greater than the November 4 event if one integrates over the whole flare spectrum. It therefore is possible that some solar flares have already approached the lower end of superflare intensities (see also [92]). However, the effects on humankind have not been particularly bad so far.

**Coronal Mass Ejections (CMEs)**

CMEs were discovered in early 1970s [93-94], which were initially called coronal transients. The white-light emission is due to the Thomson-scattering of the almost-constant solar photospheric light by the free electrons in the corona. In 1976, the phenomenon was coined CMEs [95]. According to the early definition [96], CMEs are observable as discrete changes in the coronal structures moving outward.

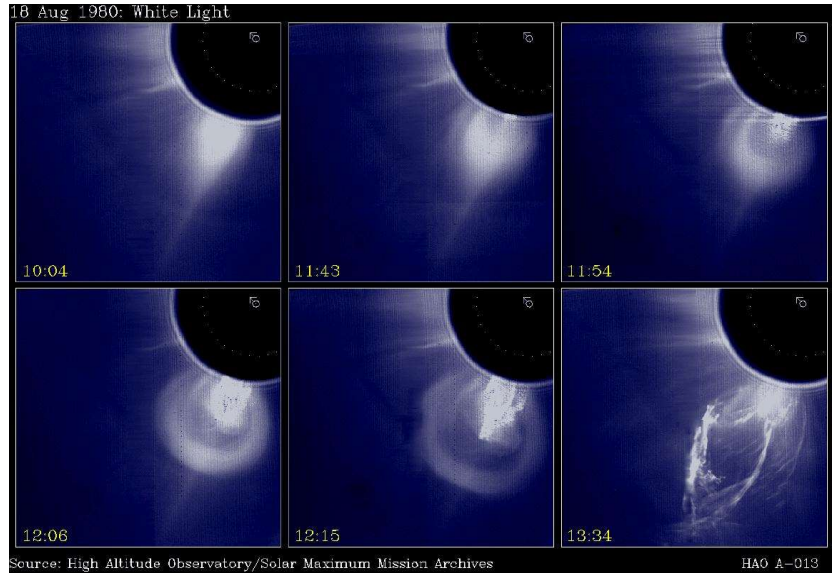

**Figure 2**. From top left to bottom right, a time sequence of images of a CME coming from the Sun. Note that the CME has three intrinsic parts. The figure is taken from Illing and Hundhausen [97].

A typical CME consists of three parts: a bright frontal loop, a cavity, and then a bright core at the center as shown in Figure 2 bottom left panel [97]. The "cavity" is filled with high intensity magnetic fields and low-density plasma [98] and thus the "magnetic cloud"/cavity [99] causes magnetic storms at Earth if the magnetic fields of the cavity are directed southward [100]. It is noted that not every CME clearly manifests all these three components in the white-light images, presumably due to the projection effects or the contamination from the surrounding corona. A piston-driven shock wave is believed and sometimes observed straddling over the CME frontal loop. However even if a shock is not detected along with the frontal loop, a shock will form as the CME propagates further from the Sun where the local magnetosonic wave speed decreases [101]. It is the cavity plasma/magnetic cloud which serves as the piston which drives the formation of the shock.

The CME occurrence is intimately related to magnetic activity at the Sun, hence their occurrence rate follows the sunspot cycle, with on average 10 events per day during solar maximum, and slightly less than 1 event per day during solar minimum. As a result, ~19,835 CMEs were observed in solar cycle 23, and ~15,685 CMEs in solar cycle 24 [102].

The central position angle is the angle between the central propagation direction of a CME and the solar north, measured counter-clockwise. It reflects the propagating direction of the CME projected onto the plane of the sky, or the latitude. It was found that CMEs are concentrated near the equator (mostly within ± 30° in latitude) during solar minimum, and can appear in any latitude (mainly within the latitude range of ± 60°) during solar maximum [103].

The angular widths of CMEs range from several degrees to 360°. Those events with the angular width < 20° are often called narrow CMEs [53], those events between 20° and 120° are called normal CMEs, and those wider than 120° but smaller than 360° are called partial halo CMEs [103-104]. CMEs encircling the solar disc are called halo CMEs. It was found the average angular width increases from 47° near solar minimum to 61° near the early phase of solar maximum [103]. As illustrated in Figure 3, narrow CMEs and other normal CMEs might have significantly different magnetic environments.

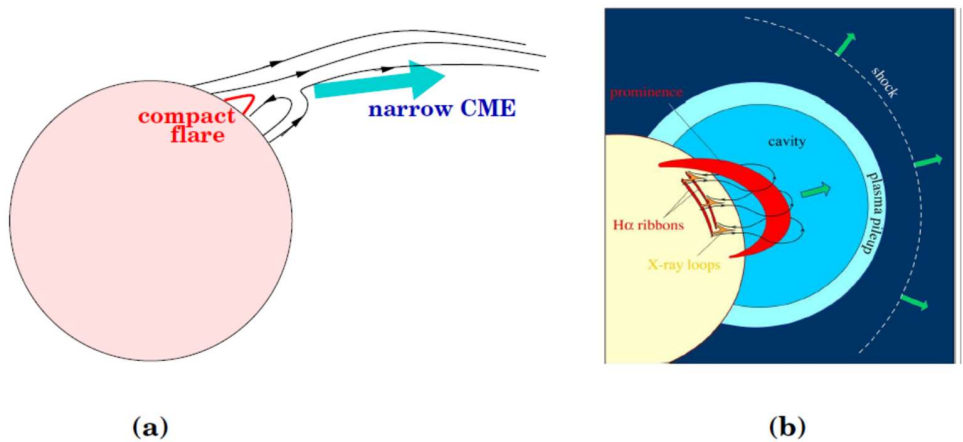

**Figure 3.** Schematic models to describe the eruption of narrow CMEs (left) and normal CMEs (right). The left panel is from Chen [54], and the right panel is taken from Forbes [105].

The CME mass distribution covers a wide range, down to $10^{12}$ g or even $10^{10}$ g depending on the threshold of identification [102]. However, most CMEs fall in the mass range of $3\times10^{13}$-$10^{16}$ g.

A CME generally starts with a slow rising motion, which is accelerated in the impulsive phase, reaching a relatively constant eruption speed [106]. The impulsive acceleration generally happens within an altitude of 1 $R_S$. The acceleration is about several hundred m s$^{-2}$, sometimes reaching 7,300 m s$^{-2}$ [106]. The final speed, which is also the apparent one, ranges from tens of km s$^{-1}$ up to 4,300 km s$^{-1}$. The CME final speed has a log-normal distribution [107], which peaks at 250-300 km s$^{-1}$. During propagation into interplanetary space, faster CMEs tend to decelerate and slower CMEs tend to accelerate. With the measurements of CME mass and centroid speed, the CME kinetic energy can be estimated as well, which has a log-normal distribution ranging from $10^{26}$ to $10^{32}$ erg, averaged at $2.3 \times 10^{29}$ erg [108].

Filaments on the solar disc are called prominences when appearing above the solar limb. Their eruptions are intimately related to CMEs. Roughly over 70% of CMEs are accompanied by filament eruptions, and also over 70% of filament eruptions are accompanied by CMEs [109-110]. There is even stronger relationship between CMEs and solar flares. Most CMEs are accompanied by solar flares, albeit some flaring brightness is too faint to be recorded by the *GOES* SXR light curves. However, the vice versa is not true. For example, from 1975 to 2011, more than 338,000 flares of all different levels were detected in the *GOES* SXR light curves [78], implying more than 100,000 flares per solar cycle. The CME association of solar flares increases with the flare brightness [102, 111].

The energy density of a CME, i.e., the kinetic energy divided by the volume, ranges from $10^{-2}$ to 10 J m$^{-3}$, whereas the density of both coronal thermal energy and potential energy is smaller than $5 \times 10^{-2}$ J m$^{-3}$ [54, 105]. This renders it the only possibility that all CMEs, except the extremely weak ones, are driven by the coronal magnetic field, whose energy density is ~40 J m$^{-3}$. Based on the strong correlation among CMEs, filament eruptions and solar flares, the classical standard flare model, i.e., the CSHKP model [112-115], was later extended to explain the eruption of CMEs. According to this model (see schematic in [116]), a filament supported by the core magnetic field is initially held in equilibrium with overlying closed magnetic field lines (i.e., the envelope magnetic field). Somehow the filament loses its equilibrium and starts to rise slowly. As a result, the field lines overlying the filament are stretched upward, and a current sheet is formed below the filament between the upward and downward field lines. The magnetic reconnection inside the current sheet leads to the flaring loops below the reconnection region and the fast eruption of the filament and the overlying closed field lines, forming the three-components of a CME.

While the standard model catches the essence of the eruptions, it should be mentioned that it is much simplified compared to real observations. It is a descriptive framework, and many details need to be supplemented [117], and care should be taken. More importantly, it should be kept in mind that the standard CME/flare model does not account for the pre-eruption structure and how such a structure is triggered to rise.

A missing piece in the standard model is: What is the pre-eruption structure (or progenitor for short)? Considering that the magnetic field at the solar surface changes little after a CME, it is generally believed that the required magnetic energy has already been stored in the corona. A magnetic structure with sufficient free energy must be either strongly sheared or even twisted as a flux rope. Whereas it is widely accepted that a flux rope exists in most normal CMEs during eruption, a long-standing debate is whether the CME progenitor is always a flux rope or it can also be sheared arcades.

An indirect method was proposed to distinguish between sheared arcades and flux ropes in the CME progenitor [118]. While a lot of papers were devoted to confirming the existence of a flux rope before eruption, a statistical study using the indirect method of Chen et al. [118] indicated that flux ropes exist in 89% of the CME progenitors, and sheared arcades exist in 11% of the CME progenitors [119].

With both emerging magnetic field and surface motions, magnetic free energy is accumulated in the corona along with the quasi-static evolution of the coronal structures [54]. At a certain stage, the core magnetic field of the CME progenitor might be triggered to rise significantly, paving the way for the current sheet formation and reconnection described in the standard CME/flare model. The triggering mechanisms fall into two types, reconnection type and ideal magnetohydrodynamic (MHD) type.

The reconnection-related mechanisms include (1) the tether-cutting model, where strongly-sheared arcades in the core field reconnect, and the upper post-reconnection field pushes the core field to rise up [120]. A similar mechanism is the magnetic cancellation model [121], where positive and negative magnetic polarities cancel each other at the solar surface below the core field. One effect of such magnetic reconnection is the increase of the poloidal magnetic flux, which was demonstrated to be able to trigger the flux rope to rise [122-123]; (2) the breakout model, where a magnetic null point exits above the core field [124]; (3) the emerging flux trigger mechanism, where newly emerging bipolar magnetic field reconnects with the envelope magnetic field, causing the envelope field along with the core field to rise [125]. The ideal MHD mechanisms include (1) kink instability [126]; (2) torus instability or catastrophe model [127-128].

## 3. Solar Wind

The existence of a solar wind was first inferred by Biermann from comet tail analyses [15]. Observational hints to the solar wind existence were provided by Gringauz [129] and it was first directly measured from spacecraft in-situ measurements by Neugebauer and Snyder [130]. For a more detailed description of the history of the solar wind concepts see Obridko and Vaisberg [131]. The solar wind is found to be a more general concept of stellar dynamics with stellar winds being a very common feature, e.g., Wood [132].

It is now known that there are two basic types: a slow solar wind and a fast solar wind [e.g., 130, 133]. The slow solar wind is believed to originate near or at helmet streamers at the Sun and has a speed of ~300 to 400 km s$^{-1}$, density of ~5 electrons cm$^{-3}$, and proton and electron temperatures of ~0.5×10$^5$ K and ~1.0×10$^5$ K and a magnetic field intensity of ~5 nT at 1 au. The fast solar wind originates from coronal holes [12] and has a speed of ~750 to 800 km s$^{-1}$ at 1 au. The proton density is ~3 cm$^{-3}$, and the proton and electron temperatures are ~2.8×10$^5$ K and ~1.3×10$^5$ K, respectively. The embedded magnetic fields have an intensity of ~5 nT at 1 AU.

High speed streams overtake the slow solar wind and form corotating interaction regions or CIRs [72]. CIRs have density and magnetic fields ranging from ~15 to 30 cm$^{-3}$ and ~20 to 30 nT, respectively at 1 AU.

Interplanetary Coronal mass ejections (ICMEs) have speeds ranging from ~300 to 3000 km s$^{-1}$ at 1 au. Their proton temperatures are ~0.1×10$^5$-1.2×10$^5$ K, densities are ~1 to 10 cm$^{-3}$ and embedded magnetic fields are ~5 to 50 nT. Fast CMEs form an upstream shock. The sheaths behind the shocks have temperatures of ~0.5×10⁵ to 3.0×10⁵ K and densities of ~5 to 25 cm⁻³. The magnetic field strengths inside sheaths are ~10 to 30 nT.

## 4. Storms and Substorms

Knowledge of magnetic storms was present well before the space age. The great naturalist Alexander von Humboldt noticed that magnetic needles oscillated as long as auroras were overhead from his home in Berlin, Germany [4]. In his paper, he called this a "Magnetische Ungewitter" or a magnetic storm.

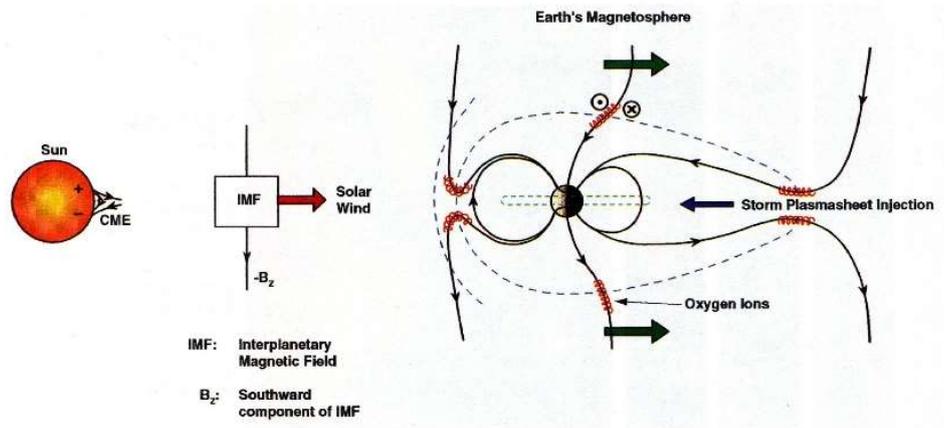

**Figure 4**. Magnetic reconnection causes solar wind energy input into the Earth's nightside magnetosphere. The schematic shows how ionospheric ions become part of the ring current. The figure is adapted from Dungey [134].

Magnetic storms are caused by reconnection of interplanetary southward magnetic fields with the Earth's dayside magnetopause magnetic fields (see Figure 4) as first proposed by James Dungey [134]. Echer et al. [135] have shown this to be the case for all 90 magnetic storms that occurred in solar cycle 23. The magnetotail magnetic reconnection causes convection of ~100 eV to 1 keV plasmasheet plasma into the nightside magnetosphere. The electrons, protons and Helium ions get energized to ~10 to 100 keV ring current energies by adiabatic compression [136]. Depending on the strength of the convection electric fields (and storm intensity) the ring current particles can be injection in as far as L = ~2.

At first it was believed that the ions in the ring currents were all protons. But in 1972 Shelley et al. [137] showed that singly charged oxygen ions were present as well. Daglis [138] showed that during intense magnetic storms, $O^+$ became the dominant ion in the ring current. Although such low charge state ions clearly must come from the ionosphere, Fok et al. [139] have argued from a computer simulation experiment that the oxygen ions come from the plasmasheet. However local oxygen ion acceleration through auroral zone double layers and then magnetospheric compression by the storm electric field cannot be ruled out in contributing to the ring current.

Today we know that magnetic storms have to do with energetic ~30 to 300 keV ion and electron injection into the magnetosphere, causing an enhanced particle ring current [100]. The electrons gradient and curvature drift from midnight through dawn and then continuously around the Earth. Because the ions drift in the opposite direction, the energetic particles form a ring of current extending from L ~2 to 7 shells (the McIlwain L parameter is the radial distance in $R_E$ at the magnetic equator for a dipole approximation of the Earth's magnetic field: [140]). This diamagnetic ring current causes decreases in the Earth's surface, near-equatorial magnetic fields. The decreases in the horizontal component of the field measured by ground magnetometers are recorded as a Dst (hourly) or SYM-H (minute resolution) index. It has been shown theoretically that the decrease in the magnetic field is related to the total energy of the injected energetic electrons and ions [141-143].

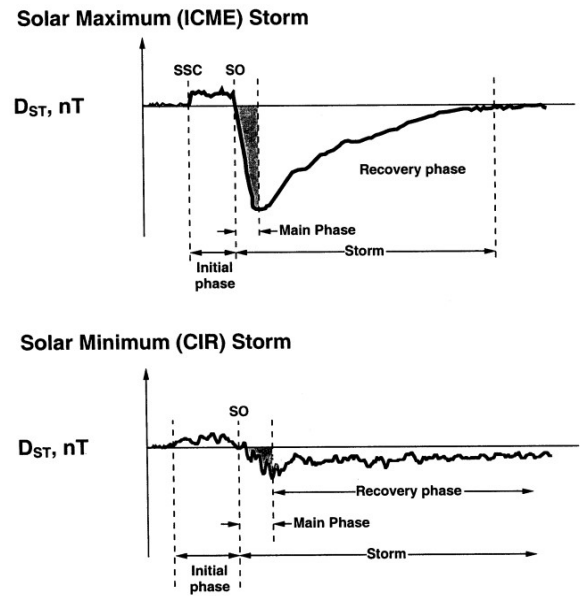

**Figure 5**. Two types of magnetic storms identified in the ground based Dst(/SYM-H) indices. The top panel shows a schematic of a magnetic storm caused by an ICME. This type of storm is most prevalent during the maximum sunspot phase of the solar cycle. The bottom panel shows a magnetic storm caused by a CIR and the following high-speed stream. The latter geomagnetic activity is called a HILDCAA. The figure is taken from Tsurutani [144].

Figure 5 shows two types of magnetic storms, one due to an ICME (top panel) and the other due to a CIR and following HSS. In the top panel the shock ahead of a fast ICME creates a sudden increase in the Dst index. This is due to the compression of the magnetosphere and a sudden impulse (SI$^+$) on the ground. The storm main phase is caused by the southward magnetic fields and magnetic reconnection, leading to an increase in the radiation belt flux (ring current). This southward IMF can occur in either the sheath or the following magnetic cloud (MC). The bottom panel shows a magnetic storm caused by a CIR. The magnetic storm main phase is caused by southward IMF components within the CIR. The following HSS contains large amplitude Alfvén waves whose southward components pump more energy into the outer magnetosphere through sporadic magnetic reconnection. This latter region, called a HILDCAA, is not really a storm recovery phase proper since more energy is being injected into the magnetosphere during this interval. The HILDCAA interval can last days to weeks.

There has been a recent focus on an optical solar flare occurring on September 1, 1859 that was observed by two London scientists, Richard Carrington [6] and Richard Hodgson [7]. In particular there was a large magnetic storm at Earth that occurred some ~17 hrs 40 min after the flare event. It is now recognized as one of the largest magnetic storms that have occurred in recorded history [11, 81, 145]. Part of the reason for this renewed interest in this ancient magnetic storm is that there were strong GIC effects observed on the ground [82]. It is realized that if such an intense magnetic storm occurred today, the GIC effects would be far more damaging in our highly technological society [146]. Substorms were originally defined by a sequence of auroral forms where their evolution occurred over ~30 min to an hr in the midnight sector of the auroral zone (~60°-70° geomagnetic latitude). This morphology was first identified by Syun-Ichi Akasofu using All-Sky camera data [147]. Substorms are now known to also involve plasma and magnetic field dynamics within the magnetosphere and magnetotail [136, 148]. Substorms were so named because they were thought to constitute a fundamental subelement of a magnetic storm. Substorms can occur as isolated events as well [149]. They can also occur in a series of events called "High Intensity Long-Duration Continuous AE Events" or HILDCAAs that can last for days to weeks [150-151]. It has also been shown that under certain circumstances, magnetic storms can occur without substorms [152]. Tsurutani and Gonzalez [153] have suggested that the convection systems of storms and substorms are separate and can at times be superimposed.

## 5. The Magnetotail and Substorms

The Earth's magnetotail exists as a coherent structure until at least 240 $R_E$ in the antisunward direction (see review by Tsurutani et al. [154]). Southward interplanetary magnetic fields connect to the Earth's magnetopause magnetic fields to form "open" magnetotail magnetic fields [134]. When these magnetic fields are convected antisunward by the solar wind, they form the magnetotail lobe fields. The magnetotail consists of the northern and southern low β tail lobes and a high β plasma sheet sandwiched between the two lobes. At ~240 $R_E$ downtail, the north-south magnetotail dimension is ~55 $R_E$ and the east-west dimension ~45 $R_E$. The plasma sheet in the center has a north-south size of ~12 $R_E$ [155-156]. The near-Earth plasma sheet is approximately the same size [157-158] whereas the near-Earth lobes are slightly smaller.

The near-Earth plasma sheet contains hot (typically > 100 eV) electrons and protons whose enhanced energy density stretch across the magnetotail from the dusk to the dawn borders with flaring near the borders [159]. The crosstail current through the plasmasheet (and closing currents over the lobes) sustains the antiparallel magnetic fields of the tail. The plasmasheet also contains the neutral sheet.

The most dynamical phenomenon in the magnetotail is produced by sporadic magnetic reconnection in the near-Earth magnetotail leading to a substorm [149, 160]. The magnetosphere and the ionosphere are strongly coupled with field-aligned currents (electric currents along the magnetic field lines). Visible and dynamic auroras are the result of intense 1-10 keV electron precipitation into the auroral zone atmosphere. These aurora-producing electrons are accelerated in the electric field parallel to the magnetic field lines called "double layers" [161-162].

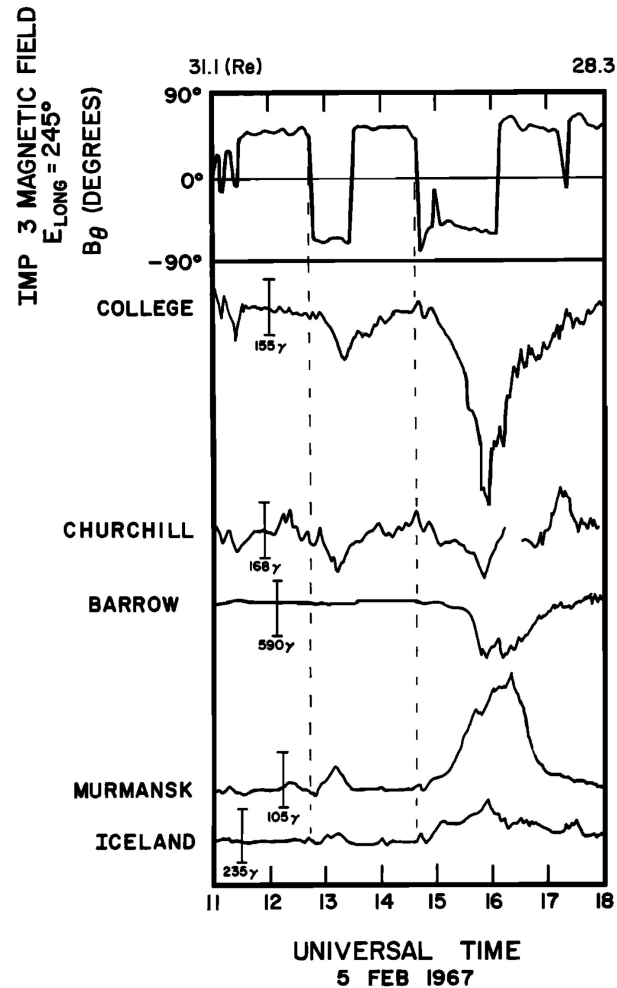

**Figure 6**. Southward interplanetary magnetic field impingement onto the magnetosphere and the short delay times before the start of intense auroral (substorm) activity. The figure is taken from Tsurutani and Meng [149].

There are believed to be many scenarios for substorms. We will focus on only one here. We will discuss magnetic reconnection that occurs in the near-Earth tail within 30 Re of the Earth [163-165]. The interplanetary magnetic field turns southward and enhanced magnetic reconnection occurs at the dayside magnetopause. Although it appears that magnetic reconnection on the nightside can start at any timing during the "loading process", substorms have been detected in time as short as 20 to 40 min from the dayside reconnection process [149, 166]. Examples of the immediate response of substorms to interplanetary southward magnetic fields impinging onto the magnetosphere are shown in Figure 6. The accumulated field lines and the plasmas in the northern and southern tail lobes are transported to the equatorial plane forming inflows for tail magnetic reconnection. Magnetic reconnection proceeds explosively with the X-line geometry of the magnetic field and produces high-speed plasma outflows near the equatorial plane [167], later called plasmoids [168-170]. After the expansion phase of the substorm, the magnetosphere and ionosphere return to the pre-substorm condition during a period referred to the recovery phase.

A new era emerged for detailed kinetic observations of substorms using the Geotail spacecraft [171]. Various aspects of Hall physics in the ion-electron decoupling region of magnetic reconnection were revealed [172-173]. At the center of the magnetic reconnection site, an intense electron current layer is formed with dawnward-moving electrons (the electric current is duskward), which sustain the antiparallel magnetic fields. In the current layer, earthward and tailward electron jets [173] are produced as outflows from the magnetic reconnection. Since the electron current layer is thin relative to the ion Larmor radius, the ions exhibit meandering motion [174]. In the thin antiparallel magnetic field configuration, an ion cannot perform a full Larmor (gyro) motion. A macroscopic feature of Hall physics is the formation of the Hall current system [175]. Figure 7 shows a schematic of the Hall current system and the electron dynamics in the vicinity of the X-line in the magnetic reconnection site. In the separatrix layer, electrons flow into the X-line creating outward flowing electric currents. In the equatorial plane, the difference in the electron jet speed and the ion flow speed creates inward flowing electric currents. Therefore, one current loop forms in each quadrant of the magnetic reconnection site meridian, and the four loops can produce the significant dawn-dusk (By) component of the magnetic field, the Hall magnetic field. The latter has a quadrupole structure (By < 0 in the northern hemisphere earthward of the magnetic reconnection site), and it is easily detected in space.

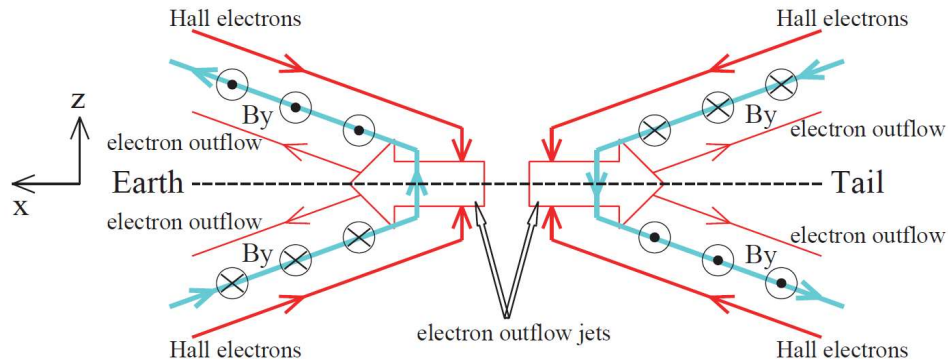

**Figure 7**. Schematic of electron motion in the vicinity of the X-line for the Hall current system illustrated in the 2-D x–z plane. Inflowing Hall electrons produce almost field-aligned currents flowing out of the diffusion region, while outflowing high-speed electrons produce currents flowing into the diffusion region. These electron motions make the four-loop Hall current system, which results in a quadrupole magnetic field structure. The magnetic field configuration inevitably becomes 3-D, and the central region should be located inside the plane. This figure is adopted from Nagai [176].

The electron diffusion region, which is situated at the center of the above-mentioned electron current layer, has been explored by the MMS mission [177-178]. The MMS spacecraft approached the X-line as close as one electron inertial length of approximately 30 km during the July 11, 2017 substorm [179].

There are other scenarios for substorms, possibly hundreds of them. We mention just a few: [180-187]. There may be many different types of substorms fitting these different scenarios [Akasofu, personal comm., 2018]. Interplanetary shocks are known to trigger substorms within a few minutes after initially reaching the magnetopause [188-189]: the mechanism for this is still not well understood. It has recently been shown that supersubstorms (SML < -2500 nT) triggered by interplanetary shocks may not have the dominant energy deposited in the midnight sector [190]. In one case such a supersubstorm occurred primarily on the dayside at ~10 am local time [191]. Thus, substorm onset mechanisms are becoming even more complex with time.

## 6. Radiation Belts/Energetic Magnetospheric Particles

The Earth's magnetosphere is filled with charged particles with energies in the range of a few keV to hundreds of MeV [192]. They are comprised predominantly of protons and electrons, and smaller amounts of alpha particles and oxygen ions. Most of the particles are stably trapped in the magnetosphere between ~1.1 and 9.5 $R_E$, and gradient and curvature drift around the closed magnetic field lines, ~8.0 $R_E$, however transient populations such as solar energetic particles may be present for periods ranging from days to months (see for example, [193, 194]). The main sources of these particles are galactic cosmic rays, magnetic storm and substorm energetic particle injections, acceleration of terrestrial ionospheric (thermal) plasma, and solar wind particles. The magnetospheric energetic particles are lost by several physical processes, including charge-exchange, Coulomb collisions, wave-particle interactions, and convection out the front side of the magnetopause [195].

Discoveries of the Earth's radiation belts [192, 196] by *Explorer* 1 and *Sputnik* 2, respectively, were the most important discoveries at the beginning of the space age. The present understanding of the Earth's radiation environment is a "two-belt" structure popularly known as the "Van Allen radiation belts" where energetic particles are organized by the McIlwain L parameter. The two radiation belts are located approximately at 1.1 < L < 2, called the "inner belt", and at 3 < L < 8, called the "outer belt". The two belts are separated by a "slot region" where there is a minimum of energetic particles. This is generally located around L of ~2.5. Tsurutani et al. [197] have suggested that the electron slot is created by the interaction of relativistic (~MeV) electrons with coherent electromagnetic plasmaspheric hiss. The two belts are significantly different in particle population and variability.

The inner belt has a stable population of ~10-100 MeV protons. The higher energy tail of the protons is produced by cosmic ray albedo neutron decay (CRAND: [198]). In this production process, the incoming galactic cosmic rays collide with the nuclei of atmospheric atoms and molecules creating albedo neutrons. The neutrons undergo β-decay, transferring most of their kinetic energy to the daughter protons (~100 MeV) and lesser energy to the electrons (< 1 MeV) (see [199]). A secondary source of the inner zone protons, particularly at the lower energy tail of the spectrum (< 50 MeV), is solar energetic protons from production at solar flares and interplanetary coronal mass ejection (ICME) shocks [200-203]. These protons become trapped as they enter the inner zone and get scattered owing to their gyroradii being larger than the local magnetic gradient scale-lengths. Although the inner belt comprises predominantly of protons, recent studies [e.g., 204-205] have shown that CRAND also produces energetic electrons of ~200-800 keV that are trapped near the inner edge of the inner belt.

The ring current which is co-located with the highly variable outer belt is comprised of ~10-100 keV electrons, ~10-300 keV protons, and ~10-300 keV singly-charged oxygen ions ($O^+$) during geomagnetically active periods. The enhanced storm-time ring current is created by the injection of ~100 eV to ~1 keV magnetotail plasmasheet particles into the midnight sector of the magnetosphere [136]. As these particles are convected into the magnetosphere to as low as L = 2, they are adiabatically compressed (conserving the first two adiabatic invariants; [206-207]), energizing the electrons and ions to their full ring current energies. Energetic electrons gradient and curvature drift in the opposite direction from the energetic ions to form the diamagnetic current system, or ring current. The ring current causes the magnetic field magnitude depressions at the equatorial surface of the Earth [11, 141]. This magnetic depression signature is one feature of a "geomagnetic storm" [4, 100].

The existence of ring current protons was experimentally confirmed by the *Orbiting Geophysical Observatory* (OGO) 3 satellite observations [208-210] for the first time. This was followed by several other missions like *Active Magnetospheric Particle Tracer Explorer* (AMPTE), *Combined Release and Radiation Effects Satellite* (CRRES), and IMAGE, contributing significantly to the current understanding of the ring current particles and their variability [211-214, and references therein]. While ~10-300 keV protons carry the majority portion of the ring current particle energy, the 10 to 300 keV $O^+$ ions, accelerated from the ionospheric thermal population plus adiabatic compression effects, are known to be an important part of the ring current particles [211], particularly during intense geomagnetic storms [212, 215].

Highly variable relativistic (~MeV) electrons constitute the outer zone or the outer radiation belt [216-220]. The MeV electrons are currently believed to be predominantly locally accelerated in the outer radiation belt through wave-particle interaction with electromagnetic chorus waves [221]. Radial transport may play a role as well [222-225]. Measurements over 7 years of Van Allen Probes have established the complex nature of electron energization in the outer radiation belt [226].

Wave-particle interactions may be stochastic, resonant and nonlinear [for reviews see 227-231]. The mechanism may be broadly described as follows: temperature anisotropy of ~10-100 keV electrons leads to plasma instabilities generating whistler-mode chorus waves (frequency range of ~100 Hz to ~10 kHz) in the magnetosphere [232-234]. Cyclotron resonant interaction of the ~100 keV electrons with chorus waves leads to ~MeV electron acceleration, in a "bootstrap" mechanism from lower to higher energy [235-237]. It is to be noted that wave-particle interactions also result in loss or removal of energetic electrons by pitch angle scattering into the loss cone [231]. Direct observations of pitch angle scattering have been made both by the Van Allen Probes [238] and the Arase missions [239].

**Energetic Particle Space Weather and Extreme Events**

As discussed above, a significant part of the (outer) magnetospheric energetic particle (protons) enhancements are associated with geomagnetic storms. Magnetic reconnection [134] between the southward interplanetary magnetic fields (IMFs) and northward (dayside) geomagnetic fields is the primary mechanism for magnetic storms [100, 135]. Magnetic cloud portions of the ICMEs and upstream sheaths are sometimes characterized by intense southward IMFs with hours duration [240-241]. During such events, the plasma injection is deeper into the magnetosphere and the particles get energized to ~50-500 keV that constitute the enhanced ring current particle population.

High-speed (~550-800 km s$^{-1}$) solar wind streams (HSSs) emanate from solar coronal holes [12, 242]. Embedded in these streams are nonlinear Alfvén waves [71, 133]. The waves (and plasma) are strongly compressed as the HSSs interact with upstream slow solar wind, forming corotating interaction regions [CIRs; 72, 243-244], leading to both intense southward and northward IMFs. The southward IMF components within CIRs cause magnetic storms of limited intensities [245].

The HSS Alfvén waves also cause geomagnetic activity, but of a different nature. Magnetic reconnection due to the southward component of the Alfvén waves cause substorms [147] and DP2 events [246] for several days to weeks with only moderate ring current particle injections. These are classified as high-intensity long-duration continuous auroral electrojet (AE) activities (HILDCAAs; [150]). While HILDCAAs do not contribute much to the ring current protons, long-intervals of ~10-100 keV electron injections during the events lead to acceleration of MeV electrons through the wave-particle acceleration process. As a result, a one-to-one association of HILDCAAs with radiation belt relativistic electron fluxes has been reported [237, 247]. The schematic in Figure 8 relates interplanetary HPSs, CIRs and HSSs to variations in relativistic electron fluxes in the Earth's outer radiation belt. The impingement of a high-density HPS onto the magnetosphere compresses it, depleting the relativistic electrons. The mechanism is as follows: compression of the pre-existing, anisotropic 10-100 keV protons in the magnetosphere generates coherent EMIC waves. The waves cause rapid loss of relativistic electrons. The magnetospheric compression also causes the escape of relativistic electrons out the dayside magnetopause due to magnetopause shadowing [248]. The impingement of the CIR and HSS cause sporadic substorms and DP2 events and the generation of chorus waves. The chorus accelerates ~100 keV substorm electrons to relativistic energies, repopulating the magnetosphere with equal or even higher fluxes of relativistic electrons from values prior to the depletion event.

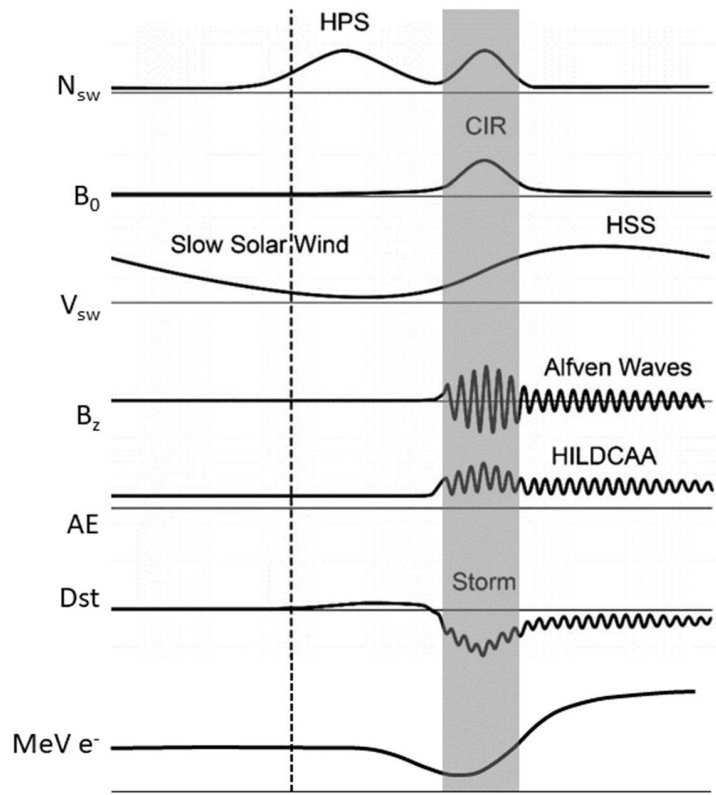

**Figure 8**. A schematic of CIR and HSS solar wind impacts on the magnetosphere and their associated geomagnetic and radiation belt effects. From top to bottom, the panels show the solar wind plasma density $N_{sw}$, the IMF magnitude $B_0$, the solar wind plasma speed $V_{sw}$, the IMF Bz component, the auroral electrojet index AE, the ring current index Dst, and the relativistic (MeV) electron flux at geosynchronous orbit. The dashed vertical line is the heliospheric current sheet (HCS), and the density associated with it is the heliospheric plasma sheet (HPS). The corotating interaction region (CIR) is shown by gray shaded region. The schematic is adopted from Tsurutani et al. [74].

Interplanetary shocks can inject energetic electrons deep into the magnetosphere; these electrons may be extremely energetic as in the case of the famous March 1991 event [200] which resulted in electrons of up to 50 MeV at L shells as low as 2. More recently Kanekal et al. [249] and Foster et al. [250] have reported on multi-MeV electron injected into the outer zone magnetosphere within a few minutes. While these recent events have not been of the same magnitude as the March 1991 event, the phenomenon of shock injection is clearly important especially during current times when human reliance on space-based technologies has increased vastly (e.g., GPS navigation). Furthermore, these events can result in GICs which can cause power system failures.

## 7. Ionosphere

The ionosphere is a thin layer of lightly ionized plasma that extends from ~80 km above the surface of the Earth to ~1,000 km. This electrically conducting layer in the atmosphere was conjectured by Gauss [16] (see English translation by Glassmeier and Tsurutani [17]). Arthur Edwin Kennelly introduced the idea of an ionosphere in more detail. The ionosphere was finally experimentally shown to exist by Appleton [18]. The primary cause of the ionosphere is photoionization of atmospheric atoms and molecules by solar UV to X-radiation that occurs when that portion of the ionosphere is exposed on the dayside [251]. As the Earth rotates and the ionosphere is no longer exposed to the solar radiation, much of the plasma recombines back into neutral atoms and molecules, although significant ionization remains at night.

An important driver of the ionosphere and its variability is the chemical composition and thermodynamics of the neutral component of the upper atmosphere known as the thermosphere. Seasonal variations of the ionosphere are generally attributed to seasonal changes in the composition of the thermosphere [252]. The fraction of thermospheric atomic (e.g., O) to molecular species (e.g., $N_2$) is a significant determining factor for the ionosphere. The molecular species generally have higher ion recombination rates than atomic species, and thus a larger fraction of $N_2$ relative to O will tend to reduce the plasma density. Since the thermosphere is influenced by the lower atmosphere through upward propagating waves, the ionosphere can be strongly influenced by lower atmospheric conditions such as those associated with sudden stratospheric warmings [253].

Energetic ~1-100 keV electron precipitation also creates ionospheric plasma. This precipitation which forms auroras occurs primarily on the night side at high auroral (65° -70°) magnetic latitudes in both the northern hemisphere (aurora borealis) and southern hemisphere (aurora australis). The most energetic electrons deposit their energy at ~80 to 85 km whereas the ~1 keV electrons deposit their energy at higher altitudes, ~110 -130 km [254].

The intensity of solar radiation varies with activity on the Sun [255] and influences ionospheric densities. In particular, during solar maximum (the interval around the maximum number of sunspots during the ~11-year solar cycle), active regions (clusters of sunspots) cause many flares per day. On October 28, 2003, the most intense EUV flare in recorded history occurred [91]. Using a relatively newly developed scientific tool, GPS dual-frequency wave transmissions detected with ground-based receivers [256], it was ascertained that the EUV portion of the solar flare spectrum caused the near equatorial ionosphere to increase in total vertical electron column density by ~30% in a matter of minutes! The enhanced ionization lasted for ~3 to 4 hours, far longer than the duration of the flare. The long duration was caused by the slow recombination rates at high altitudes.

Ionospheric coupling to the magnetosphere strongly influences the ionosphere from the poles to sub-auroral latitudes. Magnetospheric anti-sunward flow, driven by the solar wind, causes large scale noon-to-midnight motion of ionospheric plasma over the geomagnetic polar cap via electrodynamic coupling [257]. At subauroral latitudes, sunward "return" flow of the ionosphere associated with Earthward convection of the inner magnetosphere leads to plasma structuring at the interface regions between anti-sunward and sunward flows [258].

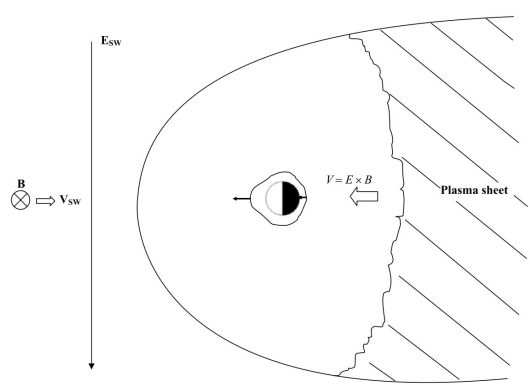

**Figure 9**. The Earth's magnetosphere as viewed from above the north pole. The Sun is on the left out of view. The solar wind is flowing from the left to the right. Associated with the magnetic reconnection process are dawn-to-dusk directed electric fields. The magnetic reconnection in the Earth's magnetotail leads to Earthward convection of the plasmasheet plasma into the nightside magnetosphere. During magnetic storms, this magnetotail electric field also penetrates into the nightside and dayside equatorial and midlatitude ionospheres. On the dayside, the **E×B** convection lifts up the F-region ionosphere to higher altitudes and latitudes. The figure is taken from Tsurutani et al. [259].

Although it was discovered in 1968 that near polar ionospheric current systems were connected to equatorial current systems [246, 260-261], it was not until fairly recently that it was found that this connection could have profound effects for the near-equatorial ionosphere. During magnetic storm main phases, prompt penetration electric fields (PPEFs) cause an uplifting of the dayside ionosphere to heights well above 400 km [259, 262-263], the height where many polar orbiting spacecraft fly. A schematic of this mechanism is shown in Figure 9. For one storm case, significant oxygen ion densities were found at over ~850 km altitude [264]. The current worry is that if an extreme magnetic storm occurred [265], satellite drag effects could cause many polar orbiting satellites to be lost to tracking radar for days to weeks, and possibly lose significant altitude and orbiting lifetime.

Figure 10 shows the impact of a prompt penetrating electric field (PPEF) during the October 30, 2003 magnetic storm. The blue trace shows TEC above the CHAMP satellite (orbiting at ~400 km altitude) before storm onset. The two peaks at ± 10° geomagnetic latitude are persistent daytime ionospheric structures known as the equatorial ionization anomalies (EIAs). Just after the storm onset the red trace shows the TEC above CHAMP for the dayside pass. The EIA peaks are located at ~ ± 20° with a peak value of ~200 TECU. On the next dayside pass (the black trace), the EIAs are at ~ ± 30° with the peak TEC values near 300 TECU.

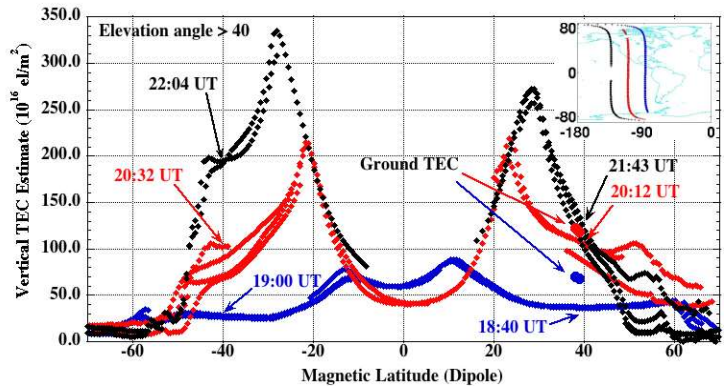

**Figure 10**. Integrated electron content (TEC) measured above the CHAMP satellite (400 km altitude) during the geomagnetic storm of October 30, 2003. Ground tracks of the CHAMP satellite for the three dayside tracks are shown in the upper right inset. The blue curve is TEC above CHAMP prior to the storm. The red and black curves are the TEC values during the first and second passes after storm onset. The figure is taken from Mannucci et al. [262].

## 8. Plasma Instabilities, Waves and Wave-Particle Interactions

In the 1950s the field of plasma physics grew considerably. It was driven by the goal to achieve controlled nuclear fusion with the aim of providing almost limitless energy for human society. Attempts to achieve nuclear fusion were plagued by the problem of plasma instabilities where the plasmas were lost to the walls of the confinement devices. In space, plasma instabilities excite plasma waves that react back on the particle distribution leading to acceleration and loss. One very positive effect of space plasma instabilities and wave-particle interactions is they create the diffuse aurora including 5-15 s optical and X-ray pulsations [266]. More on this topic can be found in Chapter 10, Auroras.

### Adiabatic invariants

To understand how wave-particle interactions operate in space it is important to understand particle motion in the geomagnetic field and the three adiabatic invariants [206]. Electrons have three types of cyclic motion, cyclotron motion around the magnetic field, bounce motion along the field between two mirror points in the northern and southern hemispheres of the Earth and drift motion around the Earth caused by the gradient and curvature of the magnetic field. The period of each motion depends on location and energy, but for a 1 MeV electron at L = 4 the cyclotron, bounce and drift periods are few milliseconds, a few seconds, and several minutes, respectively. Provided changes in the system are slow compared with each period there is an invariant (conservation law) associated with each periodic motion. Wave-particle interactions break the first and hence all three adiabatic invariants allowing the efficient exchange of energy and momentum.

**Doppler shifted cyclotron resonance**

Doppler shifted cyclotron resonance is of central importance in wave-particle interactions. It arises out of kinetic theory and weak turbulence theory [227, 230, 267].

The classic example is for a circularly polarised whistler mode wave travelling along the geomagnetic field where the wave frequency $\omega$ lies below the electron cyclotron frequency $\Omega = 2\pi f_{ce}$. Resonance is possible when the frequency is shifted by the phase velocity $\omega/k_z$ of the wave relative to the electron velocity $v_z$ along the magnetic field so that the electron and wave electric field rotate around the background field in unison. Equation 1 gives the relation where $\gamma$ is the Lorentz factor and $n$ is an integer

$$\omega - \frac{n\Omega}{\gamma} - k_z v_z = 0 \qquad (1)$$

For cyclotron resonance with circularly polarized whistler mode waves travelling exactly along the magnetic field the waves and the resonant electrons must travel in opposite directions. Cyclotron resonance is also possible between whistler mode waves and ions, but the ions must travel in the same direction and overtake the waves (anomalous cyclotron resonance). For oblique waves there are multiple harmonic resonances. Cyclotron resonance breaks the first invariant and leads to particle scattering in pitch angle (the instantaneous direction of the particle velocity relative to the ambient magnetic field) and energy [228, 268]. A discussion of particle pitch angle and energy diffusion with a broad band of incoherent whistler mode waves is given by Gendrin [269]. Lakhina et al. [270] discuss wave-particle interactions with a narrow band of coherent waves.

**Plasmaspheric Hiss**

Plasmaspheric hiss is a seemingly structureless electromagnetic whistler mode wave observed primarily inside the high density plasmasphere (L < 6) as a broad band set of wave emissions with frequencies below the electron cyclotron frequency [232, 271-272]. Weak turbulence theory shows that hiss waves are responsible for electron precipitation into the atmosphere and the formation of the so-called slot region between the inner and outer electron radiation belt [273]. As a result, it was suggested that the quiet-time two-zone structure of the electron radiation belts was formed by two processes: inward radial diffusion across the geomagnetic field from a source in the outer magnetosphere; and electron loss by plasmaspheric hiss closer to the Earth [274]. This basic idea remained in place until the late 1990s until it was recognised that other waves play an important role during active times.

The origin of plasmasphere hiss is a topic of considerable debate. One idea is that plasmaspheric hiss is generated by a plasma instability near the geomagnetic equator [271]. The waves would grow as they propagate along the geomagnetic field and reflect to fill the entire plasmasphere. Data analysis suggested that wave growth by this mechanism was sufficient to explain the observations [275].

Santolik et al. [276] and Bortnik et al. [277-278] suggested that hiss originated from chorus waves generated outside the plasmasphere (6 < L < 10). In this scenario [277], chorus first propagates to high latitudes and then gains access to the plasmasphere by penetration through the high latitude plasmapause. Furthermore, although the magnetic local time distribution (MLT) of hiss extends further into the afternoon sector than chorus, ray tracing showed that the MLT distribution can be explained by wave refraction from azimuthal density gradients [279]. Tsurutani et al. [197, 280] have provided additional support for the idea of chorus being the origin of plasmaspheric hiss. They have shown that plasmaspheric hiss is composed of ~3 to 5 cycles of coherent emissions, similar to that of chorus sub-elements [281]. Furthermore Tsurutani et al. [197] predicted that with plasmaspheric hiss being coherent, the electron slot will be formed within days and not months. However observational confirmation has not been obtained yet.

Plasmaspheric hiss is a possible important loss process for the radiation belts inside the plasmasphere for electron energies up to a MeV or more [282]. The wave properties are used to calculate the parasitic pitch angle and energy diffusion rates and electron loss timescales for energies between a few hundred to several MeV assuming incoherent waves [283]. Electron pitch angle and energy diffusion by hiss is now an essential component of global radiation belt models.

Energetic ~10 to 100 keV anisotropic electrons injected during magnetic substorms [136] increase wave instability which then reacts back on the particle distribution causing more precipitation [227]. By balancing wave growth against wave energy transported out of the unstable region a limit on the trapped flux can be obtained. This idea has been used to assess the electron spectrum and maximum electron flux during active times [284]. The method has also been used to assess the electron energy spectrum at the magnetised planets [285]. The method has an important modern application, for example, to help assess the maximum flux and the risk of satellite charging for a severe space weather event [e.g., 286].

Plasmaspheric hiss has been observed at the magnetised planets. At Saturn, studies show that there are regions where the plasma density is sufficiently small that hiss can actually accelerate electrons and contribute to the Saturnian radiation belt [287].

**Chorus waves**

Chorus waves are electromagnetic whistler mode waves which consist of a repeated series of intense, short-lived (~0.1 to 0.5 s) rising frequency tones [232-233, 288]. A repeated train of chorus emissions can last for hours, essentially as long as the dispersive ~10 to 100 keV substorm anisotropic pitch angle electron cloud lasts [289]. Very often satellite observations show that chorus is split into two bands, one below and one above half the electron cyclotron frequency [288]. The signals became known as "dawn chorus" as the signals, when converted into sound, resemble the chirping of birds at dawn [290]. They were first detected by ground-based radio equipment with an occurrence rate that tends to peak near dawn. However, the waves are generated in space by natural plasma instabilities which are highly nonlinear in nature and require special conditions to propagate to the ground [291]. An example of rising tone chorus waves recorded at Halley, Antarctica is shown in Figure 11. The chorus comes out of a weak background of hiss waves.

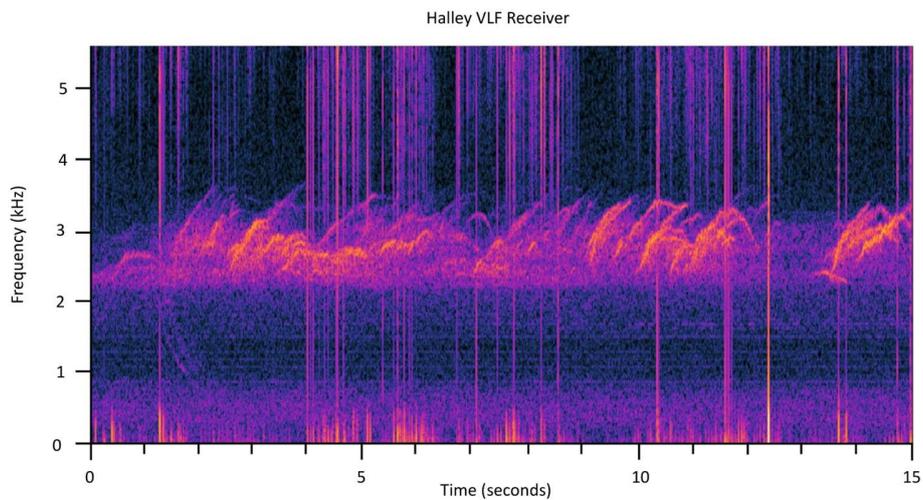

**Figure 11**. Chorus waves detected at Halley, Antarctica, in 2012. The strong rising frequency elements are chorus whereas hiss waves are a weaker background of broad band waves near 2-3 kHz. The vertical lines are impulsive signals from lightning.

Chorus is one of the strongest set of plasma waves observed in the magnetosphere and perhaps the most important one. The characteristic rising frequency elements indicate that these waves are generated by nonlinear wave-particle interactions. The basic interaction is via Doppler shifted cyclotron resonance between waves and electrons propagating in opposite directions, but propagation effects and other the nonlinear aspects such as energetic electron phase bunching are required to explain the change in frequency and the limited duration of each chorus element. The main aspects of the interaction were shown in an early series of plasma simulations [292] and numerous other studies, notably by Kyoto University scientists [e.g., 236, 293-294] have addressed nonlinear effects. One possible scenario can be summarised as follows. Waves are excited below the electron cyclotron frequency, typically around ~0.2 to 0.3 $f_{ce}$, by a temperature anisotropy in the electron distribution. The waves grow initially with a linear and then nonlinear growth rate, such that a narrow band of frequencies dominates the spectrum. The waves react back on the electrons and cause the nonlinear phenomena of phase bunching whereby electrons become grouped with a particular phase with respect to the wave electric field. The initial bandwidth is important as phase bunching can only take place if a narrow band of waves dominates (the chorus subelements), otherwise the phase becomes randomised. The phase bunched electrons now form a resonant current that rotates around the background magnetic field. That part of the resonant current in phase with the wave electric field enables wave growth while the component in phase with the magnetic field modifies the wave dispersion and causes a change in the frequency [293]. As the wave propagates along the background field the non-homogeneity of the plasma enables a second order resonance that leads to rising (or falling) tones characteristic of chorus [294].

An alternative idea is the backwards wave oscillator mechanism [295]. In this idea a step function in the velocity distribution is assumed to form because of interactions with noise-like waves, e.g., hiss. A monochromatic wave is excited near the equator and serves as a trigger for higher frequency waves as it propagates along the geomagnetic field. Measuring such a steep shoulder to test this idea is beyond present capability.

Chorus wave-particle interactions involve phase bunching and trapping which is omitted in the quasi-linear approach of Kennel and Petschek [227]. Currently a full modification of the Kennel-Petschek [227] theory for coherent chorus with monochromatic subelement frequencies needs to be developed [296].

Several studies have surveyed the occurrence and distribution of chorus. Early work showed that chorus occurred outside the plasmapause after local midnight in association with substorms [234, 288]. Observations also indicate two main source regions, one near the magnetic equator and two at higher latitudes on the dayside of the Earth, in the northern and southern hemisphere, where the geomagnetic field has a minimum due to solar wind compression. In general chorus is strongest just outside the plamaspause but extend to L=10 where it becomes much weaker. It is observed from midnight through dawn to just after noon in magnetic local time [297-298]. The waves are most intense at latitudes a few degrees above (below) the geomagnetic equator and have a very small wave normal angle [299-300] indicating propagation along the geomagnetic field. Wave power extends to at least 40º in latitude but becomes weaker with higher latitude. Wave properties such as power, frequency banding, polarization, direction of the wave normal angle, latitude and MLT distribution as well as dependences on thermal plasma density and background magnetic field strength are essential for determining the effectiveness of the waves for electron precipitation and acceleration. Several models assuming a variety of these properties have been developed [e.g., 301].

High altitude balloon observations of bremsstrahlung X-rays [302] and an auroral rocket flight with energetic electron detectors [303] have detected short duration ~0.1 to 0.5 s energetic electron precipitation events. These short burst events have been named "microbursts". More recently Lorentzen et al. [304] have observed microburst events in the magnetosphere measured by low altitude satellite particle detectors. Hosokawa et al. [305] have detected optical pulsations with the same time scale (see Chapter 10: Auroras). The time scale of the microbursts and chorus elements are approximately the same. Tsurutani et al. [306] and Lakhina et al. [270] have shown that microburst precipitation can only occur by wave-particle interactions involving coherent waves and not by incoherent waves. By comparing the trapped content of the radiation belts with the precipitation flux it is estimated that chorus induced microbursts could deplete MeV electrons in the outer radiation belt over a period of 1-2 days [307]. However, Tsurutani et al. [308] has shown that MeV trapped electrons at L = 6.6 "disappear" within 1 hr after the magnetosphere is compressed by a high plasma density phenomena such as a heliospheric plasmasheet or an interplanetary shock [309]. At this time, it is not known if this rapid loss rate is due to precipitation associated with interaction with coherent EMIC waves generated by the magnetospheric compression, magnetopause shadowing (the relativistic electrons exiting out the dayside magnetopause: [248]), or both.

Chorus has grown in importance due to its role in radiation belt dynamics. By the late 1990s satellites had shown that the electron flux in the outer belt can vary by orders of magnitude on timescales of hours to days [309]. The theory at the time applied to the equilibrium structure of the radiation belts but could not explain these rapid variations. As a result, two leading ideas emerged, one on chorus wave acceleration and the other on acceleration by enhanced radial diffusion driven by ULF waves. Only the former is discussed here.

The suggestion that chorus could be responsible for electron acceleration in the radiation belts was put forward in 1998 [221, 310]. Using a satellite survey of chorus wave power [297] and quasilinear theory to calculate electron diffusion coefficients, it was shown that electron energy diffusion by chorus was very effective at large pitch angles and could extend up to several MeV [311]. Furthermore, the process was more efficient in regions of low density (more accurately low $f_{pe}/f_{ce}$), such as just outside the plasmapause [312-313]. Using chorus observations in different MLT sectors it was shown that the MeV electron flux could increase by an order of magnitude or more over a timescale of a day or so [311].

Horne and Thorne [314] showed that in the scheme of wave-particle interactions with quasi-linear theory, electrons with large pitch angles are diffused in energy to even higher energies and remain trapped in space. The effect of pitch angle diffusion is a transfer of energy from many electrons at low energies to the waves which then accelerate a fraction of the trapped population to relativistic energies.

Global models of the radiation belts solve the Fokker Planck equation using diffusion rates that are based on a quasi-linear approach. They show that chorus can generate a peak in the electron phase space density inside geostationary orbit [315-321]. Evidence for a peak in the phase space density that grows with time has been found in RBSP data [322]. Such a growing peak cannot be formed by radial diffusion alone as radial diffusion acts to remove peaks. The growing peak is clear evidence to support local wave acceleration. Simulations also show that chorus wave acceleration should form a characteristic electron pitch angle distribution [323] which has been found experimentally in satellite data [324]. Chorus electron acceleration now plays a central role in the formation of the Earth's radiation belts and has transformed ideas which have been held for 40 years or more [325].

Chorus waves are now used in global models to forecast the Earth's electron radiation belts for space weather applications. They have moved from a challenging theoretical problem to an important practical application, namely, to help protect satellites on orbit from harmful electron radiation [326-327].

Chorus waves have also detected at Jupiter [328-329] and Saturn [330-331]. At Jupiter chorus has been suggested as providing the missing acceleration needed to produce 50 MeV electrons that emit synchrotron radiation from the planet [332]. At Saturn, similar wave acceleration processes involving chorus [333], Z mode waves [334] and hiss [287] are now suggested to play a key role in the origin of the Saturnian radiation belt.

Chorus waves also play an important role in certain types of aurorae, such as pulsating aurora (see Chapter 10, Auroras). Tsurutani and Smith [288] showed that lower band chorus often occurred in ~5-15 second "trains" of bursts. These bunches of chorus elements with gaps of 5 to 15 seconds could cause wave-particle interactions with precipitation giving rise to the ~5 to 15 second optical pulsations. Recent observations of optical pulsations [335] are in agreement with this assessment. However, at this time, there is no explanation for the bunching of chorus elements noted above. Tsurutani and Smith [288] did not find 5-15 second micropulsations in the equatorial plane which could modulate the electron pitch angle distributions as suggested by Coroniti and Kennel [336]. A mechanism suggested by Davidson [337] remains a possible explanation.

**Magnetosonic waves**

Magnetosonic waves were first called equatorial noise [338-339] but now with greater wave diagnostics, they are known for what they are, magnetosonic waves. ELF/VLF magnetosonic waves are linearly polarized waves having magnetic oscillations parallel to the ambient magnetic field and electric components orthogonal to the ambient magnetic field [265, 340-344]. Magnetosonic waves have become very important as they can accelerate electrons to relativistic energies both inside and outside the plasmapause and contribute to the electron radiation belts [345]. These waves propagate across the geomagnetic field at frequencies between the harmonics of the proton cyclotron frequency below the lower hybrid resonance frequency. The waves have peak amplitudes within about ±5º of the magnetic equator but have been detected as far away as ±60° MLAT [346].

Electron diffusion is mainly by Landau resonance and as a result the waves resonate with relatively large pitch angles. They do not diffuse electrons into the loss cone on their own, but when combined with other wave modes, such as plasmaspheric hiss, they can contribute to electron loss over a wider range of pitch angles [347].

Magnetosonic waves are generated by a proton ring distribution where the ring speed exceeds the Alfvén speed [341, 348]. The waves propagate across the geomagnetic field [341]. Near the plasmapause ray tracing shows that these waves can be refracted to change their MLTs [349]. Since the waves are generated by protons in the ring current and parasitically interact with the electrons in the radiation belt, they couple the ring current and electron radiation belts together [348].

**VLF transmitters**

After the first world war electromagnetic VLF signals were used to communicate with submarines. The advantage of this frequency range is that it can penetrate seawater to a greater depth than waves at higher frequencies. Signals from VLF transmitters can travel halfway around the Earth or more due to propagation in a wave-guide mode existing between the conducting Earth (and sea) and the ionosphere (for more information see the introduction in [350]). While it was realized that these signals could also leak out into space and cause electron precipitation [351] it was only in the late 1990s that their importance for the radiation belts was established.

Abel and Thorne [352-353] showed that VLF transmitter waves caused a major reduction in electron lifetimes in the radiation belts between L = 1.5 and 2.5. Gamble et al. [354] measured the electron flux when the VLF station in Western Australia was transmitting compared to when it was not (down for maintenance). Gamble et al. [354] found definitive evidence for electron pitch angle diffusion into the drift loss cone when the transmitter was on. Observations by the Demeter satellite also showed a characteristic energy time signature of particle precipitation due to the transmitter. Research confirmed that as electrons drift around the Earth they are lost primarily over the south Atlantic anomaly region where the bounce loss cone is larger due to the weaker magnetic field in the southern hemisphere [355].

VLF transmitters have been invoked to explain the so-called "impenetrable barrier", that is, that during the Van Allen Probes satellite era (2012–2019) radiation belt electrons at relativistic energies have not been observed inside approximately L = 2.5. Since the inner edge of the outer radiation belt overlaps with the outer edge where VLF transmitter signals are observed it has been suggested that transmitters are responsible for electron precipitation and the so-called impenetrable barrier [356]. However, careful modelling of the transmitted frequencies show that VLF signals do not resonate with MeV electrons and therefore cannot be responsible for electron loss at MeV energies [357]. It should also be noted that the idea of an impenetrable barrier is somewhat misleading as relativistic electrons have been observed at lower L during very active periods prior to the Van Allen Probes mission [358-359].

The idea of using transmitters to remove radiation and protect space assets is also known as radiation belt remediation. Electron loss by ground-based transmitters prompted the idea of controlled precipitation by transmitters on satellites [360]. The DSX satellite was launched in 2019 to test this idea and will provide new opportunities to test theory against experiment.

**ECH waves**

Electrostatic electron cyclotron harmonic (ECH) waves were first observed in the magnetosphere in the early 1970s [361-362]. The waves are electrostatic in that the k vector of the wave is almost parallel to the ambient magnetic field and thus the induced magnetic field is negligeable. The waves are observed between the harmonics of the electron cyclotron frequency in multiple bands up to the upper hybrid frequency, and sometimes above it. For this reason they are known as (n + ½)$f_{ce}$ emissions [363]. The cold plasma density effectively controls how many bands can be excited [364]. They occur at almost all local times outside the plasmapause [365] and resonate with typically 1-10 keV electrons injected from the plasma sheet during substorms.

Johnstone et al. [366] suggested that chorus could be responsible for the diffuse aurora. Thorne et al. [367] used the remnant shape of the pitch angle distribution to further argue that chorus is the most important for creating the diffuse aurora. Where we presently stand is that both ECH waves and upper band chorus contribute to the diffuse aurora depending on the particular energy of the electrons being precipitated into the atmosphere [368].

Electrostatic ECH waves also play a role as a source of electromagnetic waves emitted from the Earth and planets [369]. However, the method by which energy is converted from ECH into electromagnetic O and X mode waves remains unresolved. In the linear mode conversion theory [370-371] ECH waves refract in the large density gradient at the edge of the plasmapause and mode convert into O mode waves at the so-called radio window. The radio window is where the refractive indices of two dispersion branches are the same. Ray tracing in a hot plasma shows that wave growth and refraction is possible without significant damping if the density gradient is large [372]. Other theories suggest that energy is converted via a nonlinear three wave interaction [373]. Mode conversion between ECH and free space O and X mode waves has also been suggested to explain radiation emitted from Jupiter [371] and Saturn [374].

**EMIC waves**

Electromagnetic ion cyclotron (EMIC) waves are propagating waves below the proton cyclotron frequency (as opposed to field line resonances). In an electron-hydrogen plasma the waves are left-hand circularly polarised, or more generally left-hand elliptically polarized for propagation at an angle to the background field. In a multi-ion plasma such as the magnetosphere there are stop bands where no left-hand polarised waves are possible, and bands where the waves are right-hand or right-hand elliptically polarised [375]. More details concerning EMIC waves can be found in Chapter 9: Electromagnetic Pulsations.

Several studies have shown that EMIC waves can heat heavy ions, at the second harmonic of the oxygen cyclotron frequency [376], at the bi-ion resonance frequency [377]. The waves are also very effective in heating up-flowing oxygen from low altitudes [378-379] and trapping the ions in the magnetosphere.

EMIC waves resonate with high energy (MeV) electrons and are important in radiation belt dynamics due to the relatively high amplitudes and effectiveness in pitch angle diffusion and precipitation [380-381]. Typically, the energy range is 1-10 MeV for typical cold plasma densities found in the magnetosphere [382]. One of the recent objectives has been to determine the lowest energy electrons these waves can precipitate which may be as low as ~400 keV [383]. Pitch angle diffusion is energy dependent such that the distribution that remains trapped in space becomes increasingly narrow and peaked near 90º with increasing energy [384-385].

EMIC waves are responsible for precipitation of ultra-relativistic electrons [386] and the separation of the outer radiation belt into two regions where the inner one is known as the storage ring and may persist for many days. These waves are now an essential part of any global radiation belt model.

## 9. Electromagnetic Pulsations

### Definition of Pulsations

Electromagnetic pulsations are defined as the short period fluctuations in the geomagnetic field in the range of 0.2 to 600 seconds. This unique type of perturbation of the terrestrial magnetic field has been interpreted as magnetohydrodynamic waves [387-388], a low-frequency wave type suggested by Alfvén [22]. Such short period variations in the geomagnetic field are driven either directly or indirectly by the solar output, and sometimes are called magnetic pulsations or micropulsations. The frequencies of magnetic pulsations fall in the range of ultra-low-frequency (ULF: ~0 to 30 Hz) electromagnetic waves. The amplitudes of the magnetic pulsations are found to increase as their time periods increase (or their frequencies decrease), varying over a large range, from several hundreds of nT at the longest time periods (~600 seconds) to fractions of a nT at the shortest time periods (~0.2 s). The electromagnetic pulsations are detected on the ground by D.C. magnetometers. In the magnetosphere of the Earth, they are recorded by magnetic and electric field sensors onboard spacecraft. Figure 12 shows a figure that combines both ground and satellite observations.

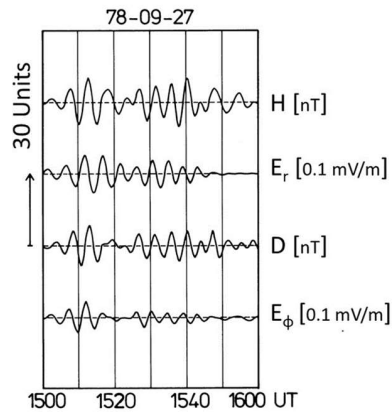

**Figure 12**. ULF wave recordings at a ground station in Northern Scandinavia and the GEOS 2 spacecraft. H and D denote the two horizontal components of the ground magnetic field variation, $E_r$ and $E_\phi$ the radial and azimuthal electric field vector components in the equatorial region at geostationary orbit (adapted from Glassmeier [24]).

It is not that ULF waves are observed only at the Earth, these waves are present in other planetary magnetospheres [389-392]. Khurana et al. [391] found that ULF wave power is maximum for the Earth and the amplitudes decrease with greater distance from the Sun: Jupiter, Saturn and then Uranus.

ULF wave activity is also observed in the foreshock regions of planetary magnetospheres [393-399] as well as in the induced magnetospheres of Mercury [400-401], Venus [402] and Mars [403]. Smith and Tsurutani [404] discovered heavy ion cyclotron waves inside the magnetosphere of Saturn. ULF waves have been observed near the magnetic pileup boundary of Comet P/Halley [405]. An excellent review of the ULF waves observed in the magnetospheres throughout the solar system is given in Glassmeier and Epsley [392]. See also Chapter 12: Planetary Magnetospheres and Solar/Stellar Wind Interactions with Comets, Moons and Asteroids and Chapter 8: Plasma Instability Waves and Wave-Particle Interactions.

ULF waves have also been detected upstream of interplanetary shocks [406-408]. Similar to the Earth's foreshock case, ULF waves have speeds that are too low to propagate upstream against the solar wind flow. Energetic particle beams propagate into the upstream region and through beam or ring-beam instabilities, generate waves there.

## Classification of Magnetic Pulsations

Magnetic pulsations observed on ground can be classified into two broad types depending on the waveforms and periods. Such a classification has formally been done by the International Association of Geomagnetism and Aeronomy (IAGA) [409]. Magnetic pulsations having quasi-sinusoidal waveforms are called pulsation continuous (Pc) whereas those with irregular waveforms are known as pulsation irregular (Pi). Pis may possibly be caused by nonlinear distortions of Pc waves. This has not been explored thoroughly, but discussion of interplanetary Alfvén waves have been examined (see Chapter 13: Interplanetary Discontinuities, Shocks and Waves). Each main class is further subdivided into period (or frequency) bands that roughly characterize a particular type of pulsation as shown in Table 1.

**Table 1: Classification of Magnetic Pulsations**

|  | Continuous pulsations | | | | | Irregular pulsations | |
|---|---|---|---|---|---|---|---|
|  | Pc1 | Pc2 | Pc3 | Pc4 | Pc5 | Pi1 | Pi2 |
| Period range (seconds) | 0.2-5 | 5-10 | 10-45 | 45-150 | 150-600 | 1-40 | 40-150 |
| Frequency range (mHz) | 200-5000 | 100-200 | 22-100 | 7-22 | 2-7 | 25-1000 | 7-25 |

During periods of disturbed geomagnetic activity, such as substorms and magnetic storms, giant magnetic pulsations, called Pc6 or Ps6, are observed with periods ranging from ~600-900 seconds.

Waves in the ULF frequency range can also be generated by plasma instabilities. These waves are identified by different names which will be discussed below.

## Generation Mechanisms of Magnetic Pulsations

A variety of plasma processes occurring in the solar wind and in the Earth's magnetosphere are responsible for the magnetic pulsations that are observed on the ground or in near-Earth space [410]. The fluctuations in the magnetospheric currents as well as the presence of nonthermal charged particle distributions existing in planetary magnetospheres can drive several plasma instabilities that can lead to the generation of magnetic pulsations [392, 411]. Therefore, changes in solar wind parameters can produce dramatic effects on the type of waves observed at a particular location at the Earth as well as in planetary magnetospheres.

It has been observed that Pcs having the same frequency range can have different characteristics, e.g., their harmonic structure, polarization, or spatial location may not be the same. These differences hint that there can be different generation mechanisms for Pc pulsations. For the long–period (Pc3 to Pc5) magnetic pulsations, two popular mechanisms have been discussed in the literature. The first mechanism is based on field-line resonance (FLR) theory. In the FLR mechanism, a monochromatic surface wave is excited by some plasma instability, such as the Kelvin-Helmholtz [412-413] and the drift mirror mode [414] instabilities, both possibly occurring at the magnetopause. These instabilities resonantly couple with a shear Alfvén wave associated with local field line oscillations [415-417]. Archer et al. [418] have provided an unambiguous direct observational proof that the magnetopause motion and magnetospheric ULF waves at well-defined frequencies, excited during a rare isolated fast plasma jet impinging on the magnetopause boundary, can only be explained in terms of the magnetopause surface eigenmode mechanism [417]. This resonant mode coupling or field line resonance (FLR) mechanism occurs quite commonly in planetary magnetospheres [419]. The second mechanism is based on cavity modes where sudden impulses in the solar wind excite global fast magnetosonic modes throughout the entire magnetospheric cavity [411, 420]. The cavity modes couple to the field line resonances that drive currents in the ionosphere producing magnetic pulsations. Shen et al. [421] analyzed the dayside ULF wave event observed by THEMIS-A on 11 January 2010, and concluded that the ULF waves were excited by the cavity mode mechanism [420]. Wang et al. [422] have shown from a multi-spacecraft study, that the quiet-time Pc 5 ULF waves are driven by the combined effects of Kelvin-Helmholtz instability and ion foreshock perturbations. In addition, some daytime Pc3 to Pc5 pulsations are believed to be driven directly by solar wind pressure fluctuations [423].

Short-period continuous pulsations (Pc1 and 2) have entirely different generation mechanisms, one being an electromagntic ion cyclotron (EMIC) instability. EMIC waves are generated by an anisotropic energetic proton distribution [227, 424-425], primarily as left hand elliptically polarized waves as these have the lowest resonant energies [375, 426]. However, the waves can become linear or even right-hand elliptically polarised because of propagation. The presence of heavy ions leads to magnetospheric wave reflection of EMIC waves [426-428]. Solar wind pressure pulses can enhance the proton perpendicular temperature anisotropy to generate EMIC waves in the dayside outer magnetosphere as well [308, 382]. Remya et al. [382] have shown that some EMIC waves are coherent and these can play an important role in ion precipitation, ion heating in the ring current and in relativistic electron loss from the radiation belts. Obliquely propagating EMIC waves acquire a large parallel electric field that can damp the waves and accelerate electrons leading to excitation of the red aurora [429], also known as stable auroral red (SAR) arcs. Further, EMIC waves can heat heavy ions, at the second harmonic of the oxygen cyclotron frequency [376] and at the bi-ion resonance frequency [377].

Although, the preferred magnetospheric location for EMIC wave growth is a region just inside the plasmapause which overlaps with the injection of protons during substorms, many surveys of EMIC waves show the occurrence of these waves in regions outside the plasmapause [430], near the magnetopause [431-432], and in the magnetosheath [433], all excited by the EMIC instability. Kasahara et al. [349] have observed EMIC waves well inside the plasmasphere which probably originate by mode conversion from magnetosonic waves [434] rather than from the EMIC instability.

Tsurutani and Smith [435] identified electromagnetic proton cyclotron waves associated with substorms in the distant tail ($X < -220$ $R_E$) plasmasheet boundary layer. Cowley et al. [436] identified energetic > 35 keV proton beams launched from an x-line tail reconnection site earthward of the spacecraft. Gary et al. [437] confirmed theoretically that the waves were generated by a proton beam instability. In a joint work, Tsurutani et al. [438] concluded that the Cowley et al. [436] picture of a distant tail signature of substorm magnetic reconnection earthward of the ISEE-3 satellite location is correct. The energetic proton beams are accelerated by the magnetic reconnection process and the beams generate the ion cyclotron waves.

The largest proton cyclotron waves on record to date is ~14 nT (peak-to-peak) detected in the Earth's polar cap/polar cusp boundary layer [379]. It was suggested that damping of a nonlinear Alfvén wave was the mechanism for generating a proton temperature anisotropy which then led to the generation of the waves.

Low-latitude irregular magnetic pulsations (Pi2) are believed to be generated mainly by the coupling of global cavity modes excited by plasma injections from the magnetotail with the low-latitude field-line resonances [439]. Most nighttime Pi1 and Pi2 pulsations are generated by earthward propagating fast mode waves launched at substorm onset by large-scale magnetic reconfiguration associated with cross-tail current disruptions [440-441].

Yumoto et al. [442-443] have argued that Pc3 and Pc4 waves generated in the Earth's foreshock region can propagate/ be convected into the magnetosphere proper.

**Mirror Modes and Planetary Magnetosheaths**

Mirror modes are magnetic and plasma pressure fluctuations that are driven by proton temperature anisotropy instabilities [414, 444]. The wave magnetic pressure and plasma pressure are 180° out of phase with each other so there is total pressure balance across the train of structures. Mirror modes have been detected in planetary magnetosheaths of the Earth, Jupiter, Saturn and Venus [445-453]. The mechanism for instability is quasi-perpendicular shock compression of the solar wind protons [454] and magnetic field line draping around the planetary body [455].

Although mirror modes are the dominant identifiable wave mode in planetary magnetosheaths, ion cyclotron waves can dominate when the solar wind/magnetosheath plasma beta is low [456].

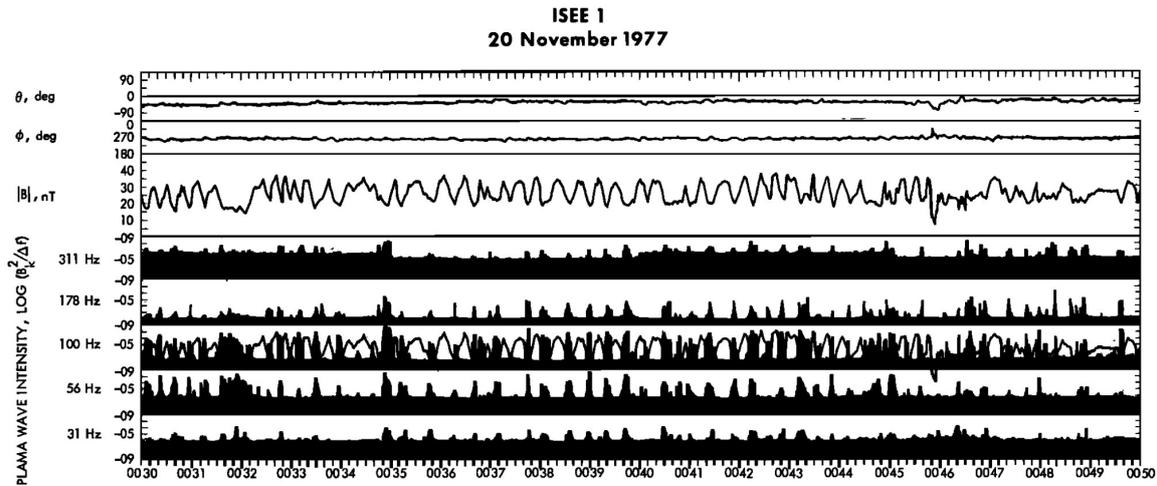

**Figure 13**. An example of mirror mode structures detected in the Earth's magnetosheath. The large magnetic field magnitude dips are complemented by enhanced plasma densities (not shown) so the entire mirror mode structures are in pressure balance. Electromagnetic plasma waves known as lion roars are detected in the magnetic dips. The figure is taken from Tsurutani et al. [445].

Mirror modes have also been detected at comets [405, 457-458], in interplanetary space [459] and in the distant downstream magnetosheath of the Earth [435]. Mirror mode waves have also been detected upstream of the heliospheric termination shock in the heliosheath [460-461].

Figure 13 is an example of mirror modes in the Earth's magnetosheath. The top three panels are the magnetic field angles and the magnetic field magnitude. The bottom five panels are wave spectrum channel data with 311 Hz to 31 Hz center frequencies. The mirror mode structures are identified as the quasiperiodic magnetic decreases with little or no changes in magnetic field direction. Mirror modes are pressure balance structures where plasma pressure supplants the decreased magnetic pressure in the magnetic dips (not shown for brevity). The plasma waves present in the magnetic dips are a parallel propagating whistler mode wave known as lion roars [462].

**Relationship between magnetic pulsations and geomagnetic activity**

Magnetospheric magnetic field configuration and plasma populations undergo drastic changes during geomagnetic disturbed times such as substorms and magnetic storms. Therefore, all magnetic pulsations that are generated internally are directly affected by geomagnetic activity. For example, characteristics of Pc1 and Pc2 waves excited by the EMIC instability, Pc3 and Pc4 driven by drift mirror instability, and Pi1 and Pi2 associated with the formation of substorm current wedges get modified during substorms and magnetic storms. Pc6 magnetic pulsations are usually generated during substorms and magnetic storms [463]. The IMF variations can sometime cause double substorm onsets and corresponding two consecutive Pi2-Ps6 band pulsations. Planetary distribution of Pc5 pulsations is also found to undergo changes during magnetic storms [464-465]. The frequency of long period-pulsations, especially Pc5s, tend to decrease by the injection of energetic oxygen ions into the magnetosphere during magnetic storms. Generally, the ULF wave power gets increased substantially during magnetic storms, and possibly during storm recovery phases as well [466]. The occurrence and characteristics of Pc3 pulsations at low latitudes are found to undergo seasonal and solar cycle modulation [467]. In another study based on the data from Arase spacecraft and groundbased stations, Takahashi et al. [468] proposed that ULF waves can play an important role in accelerating radiation belt electrons up to relativistic energies [469; see earlier [470]). From a 3-D MHD model, Degeling et al. [471] have found that the convection of plasma density controls the accessibility of dayside ULF wave power to the radiation belt region.

**Present status**

Observations by several spacecraft, such as the Time History of Events and Macroscale Interactions during Substorms (THEMIS), Van Allen Probes (VAP), Cluster, Magnetospheric MultiScale (MMS) mission and others, have advanced our knowledge about the properties of ULF waves, their excitation mechanisms, and their effects on the dynamics of energetic particles in the Earth's radiation belt [472-482]. Using the Cluster data, it is shown that poloidal mode ULF waves are more efficient for the acceleration of electrons and ions in the inner magnetosphere [478-479].

Zhang et al. [483] have done a statistical study of the Pc 5–6 ULF waves in the magnetotail using 8 years of THEMIS data, and found that the ULF waves are more frequently observed in the post-midnight region and their frequency decreases with increasing radial distance from Earth.

Claudepierre et al. [484] have done simulations of resonant ULF waves using the Lyon-Fedder-Mobarry (LFM) global magnetohydrodynamic (MHD) model. The inclusion of a plasmasphere leads to a deeper (more earthward) penetration of the compressional (azimuthal) electric field fluctuations, due to a shift in the location of the wave turning points. Furthermore, it is found that higher-frequency compressional (azimuthal) electric field oscillations penetrate deeper than lower frequency oscillations.

Li et al. [481] have analyzed the ULF waves and low energy ion fluxes observed by MMS on 20 January 2017. They report that long wavelength ULF waves could drive low-energy ions to drift in the direction normal to the plane defined by the electric and magnetic fields. The maximum measured low-energy ion energy flux peak agreed well with the theoretical calculation of $H^+$ ion $\mathbf{E} \times \mathbf{B}$ drift energy. Heynes et al. [485] identified that long period pulsations couple strongly with ground-based power systems even in subauroral regions.

To conclude, electromagnetic pulsations form an important component of space physics. They play an important role in the transport of energy from region of the magnetosphere to another region. Magnetic pulsations can significantly affect the dynamics of the inner magnetosphere and outer radiation belt during geomagnetic storms. Ground- studies of magnetic pulsations offer a unique and simple way of monitoring the conditions in the magnetosphere and solar wind. Measurements of magnetic pulsations can be utilized for geophysical surveys to probe the subsurface conductivity structure of the Earth.

## 10. Auroras

Auroras are divided into two broad categories, discrete and diffuse auroras [367, 486]. In this Chapter, we summarize the fundamental characteristics and generation processes of discrete and diffuse auroras. In addition, we briefly describe red auroras sometimes observed at high latitudes and at midlatitudes (during intense magnetic storms: SAR arcs).

**Discrete Auroras**

Discrete auroras are characterized by regions of emission showing sharp boundaries [486]. They are generally brighter than diffuse auroras (> 1 kR at 557.7 nm in most cases). The most common type of discrete aurora is a "quiet arc" [486], a thin structure mostly elongating in the east-west direction [487]. The statistical characteristics of quiet arcs (width, length and brightness) are summarized in a review by Karlsson et al. [488]. Quiet arcs are typically seen in the growth phase of auroral substorms [489]. The cause of the quiet discrete arcs has been attributed to electron acceleration by a quasi-static electric potential structure existing above the aurora [490-493]. An upward-directed parallel electric field in its central part accelerates electrons downward up to several keV leading to the formation of quiet discrete arcs [162, 494-495]. Measurements inside the electrostatic "double layers" were made by the Fast Auroral Snapshot (FAST) Explorer satellite [496]. More precise description of the acceleration mechanism is given in a review by Lysak et al. [497].

Discrete auroras sometimes show highly structured shapes whose scale size are ~1 km or less [498-500]. Such thin arcs often behave more dynamically (i.e., not quasi-static), some of which is closely related to time-varying magnetic field-aligned electric fields associated with dispersive Alfvén waves [501-504]. Details of the small-scale dynamic discrete aurora can be found in the review of Kataoka et al. [505]. The above-mentioned types of discrete aurora (quasi-static arc and small-scale dynamic arc) are mainly observed within the main auroral oval (see Figure 1 of [506]). However, discrete auroras are often observed at higher latitudes near the dayside cusp as a manifestation of direct energy input from the solar wind through magnetic reconnection at the dayside magnetopause [507, and references therein]. Discrete auroras are also observed within the polar cap (poleward of the auroral oval) during quiet intervals (northward IMF conditions) as reviewed by Hosokawa et al. [305].

**Diffuse and pulsating auroras**

Here, we discuss the optical characteristics and drivers of diffuse auroras, in particular "pulsating auroras" which is one of the outstanding features within the diffuse aurora category. Pulsating auroras are the end result of pitch angle scattering of ~5 to 30 keV energetic electrons through interaction with electromagnetic chorus waves in the outer magnetosphere. It has been known that from a global perspective, there is more energy deposition in the diffuse aurora than the discrete aurora [367].

**Optical characteristics of diffuse and pulsating auroras**

Diffuse auroras are characterized by regions of relatively dim (a few hundreds of R to ~10 kR at 557.7 nm emission) auroras that do not contain any shear or rotational motion of the features [508-509]. The occurrence of diffuse auroras is higher in the post-midnight sector [e.g., 510-511]. The majority of diffuse auroras have been detected in the morning sector, especially during the recovery phases of substorms. Quasi-periodic intensity variations of diffuse auroras are called pulsating auroras [512-516, and references therein). See Tsurutani et al. [266] for a review of X-ray pulsations, the higher energy portion of optical pulsations. Unlike static and less structured diffuse aurora [e.g., 508], pulsating auroras appear as patches having irregular shapes whose horizontal extent ranges from 10 to 200 km [509, 514, 517-518]. Pulsating auroras occur near the equatorward boundary of intense auroras [519].

It has been known that two characteristic periodicities are seen in the time-series of pulsating auroras, summarized in Hosokawa et al. [335]. The most prominent/outstanding one is called the "main pulsation" whose period ranges from a few to a few tens of seconds [509]. This is the long quasi-periodic component of diffuse auroral pulsations. The other quasi-periodicity has been called an "internal modulation" which are sub-second luminosity fluctuations embedded in the bright phase of the main pulsations. The periodicity of the internal modulation is ~3 Hz [518, 520]. In balloon X-ray observations these are called "microbursts" [302, 521]. These ~3 Hz fluctuations are detected in more than 50% of all pulsating auroras in the midnight and morning sectors, and the amplitude of modulation using panchromatic imaging is as large as 20% [518]. Recent studies have suggested the existence of even faster modulation at ~15 Hz [522] and ~54 Hz [523]. These latter features may be associated with pitch angle scattering of the energetic electrons associated with chorus subelements. Studies concerning this possibility are currently being undertaken.

The color of pulsating aurora is dominated by the green oxygen emission at 557.7 nm, but a blue/violet 427.8 nm color from molecular nitrogen ions is sometimes noted at the lowest altitude (the bottom) of pulsating patches. When this occurs, relatively higher energy precipitating electrons must be present [524]. The altitude of pulsating aurora has also been studied since the 1970s using stereoscopic observations [e.g., 525]. Recent stereoscopic observations by Kataoka et al. [526-527] have demonstrated that the altitude of pulsating auroras is 85-95 km, slightly lower than that of the other types of auroras. This tendency was confirmed by incoherent scatter radar observations of enhanced ionospheric ionization [528-531]. In some cases, ionization down to an altitude of 65 km have been noted.

Pulsating auroras are almost always observed in the post-midnight sector during the recovery phases of auroral substorms [532]. After a substorm expansion phase, dim patches of diffuse aurora occur. After a few tens of minutes, these diffuse patches start pulsating. These pulsations can be present for ~1-3 hours [511]. During magnetic storms pulsations can be present for as long as ~15 hours [533].

**Mechanisms causing optical pulsations**

The ultimate source of the precipitating electrons is the plasma sheet in the Earth's magnetotail. The ~100 eV to 1 keV plasmasheet electrons are convected into the midnight sector outer magnetosphere by substorm convection electric fields. As the electrons are convected into higher magnetic field strengths, they conserve their first two adiabatic invariants [206] and become energized to ~10 to 100 keV energies. The electrons are not only convected inward to lower L but also gradient and curvature drift towards local dawn [136]. The pitch angle anisotropy of the energetic electrons generated by the inward convection [534] lead to plasma instability [227] and the generation of electromagnetic chorus waves [288]. The chorus waves pitch angle scatter the electrons into the loss cone. The precipitating energetic electrons lose their energy by both excitation of atmospheric atoms, molecules and ions. The excited atoms, molecules and ions decay, giving off their characteristic auroral light. When the energetic electrons pass close to atmospheric nuclei, bremsstrahlung X-rays are created. The X-rays can be observed by high altitude balloons.

Besides auroral zone high altitude balloons flown in the 1960s and 1970s and more recently by the Balloon Array for Radiation Belt Relativistic Electron Losses (BARREL) program [521], several rocket observations showed that pulsating auroras are produced by temporal variations of precipitating electrons whose energies range from a few keV to ~100 keV (e.g., [303, 535-538]; see also recent review of Nanosat and balloon measurements by Sample et al. [539]). Wave-particle interactions through cyclotron resonance lead to losses of magnetospheric ~10-100 keV electrons into the atmosphere and the occurrence of diffuse aurora [367].

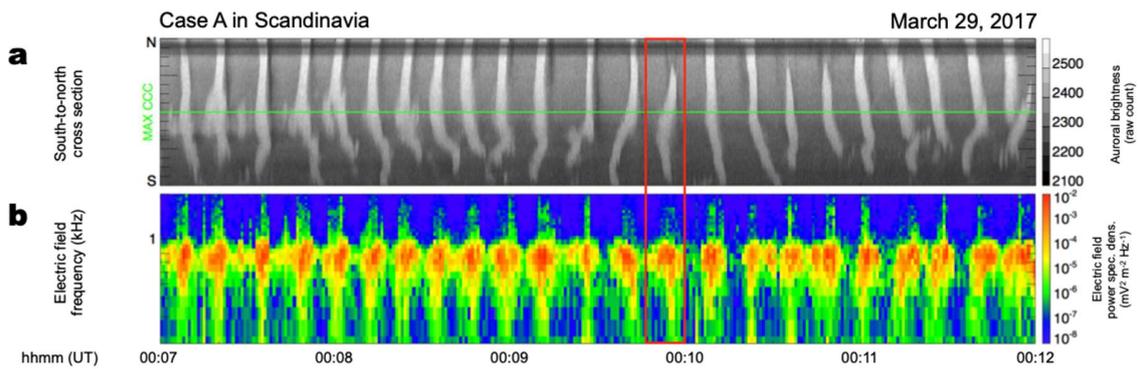

**Figure 14**. A case of pulsating aurora detected in Soldankyla, Finland on March 29, 2017. Panel a) shows a time-series of optical intensity along the south-to-north cross-section of the all-sky camera images. Panel b) is a frequency-time diagram of the chorus E-field power spectral density taken by the conjugate Arase satellite. The figure is taken from Hosokawa et al. [335].

The quasi-periodic on-off switching of pulsating aurora, i.e., the main pulsation, can be explained by periodic groupings of chorus wave bursts causing periodic scattering of energetic electrons into the atmospheric loss cone [288, 540]. Nishimura et al. [541] employed THEMIS spacecraft and ground-based all-sky camera data to confirm the one-to-one correspondence between the temporal variations of chorus intensity in the magnetosphere and optical pulsations at the magnetic (ionospheric) footprint of the satellite. More recently, Kasahara et al. [239] reported in-situ magnetospheric observations showing a one-to-one correspondence between the electron flux in the loss cone and the amplitude of chorus. During these in-situ observations, pulsating auroras were detected at the magnetic footprint of the satellite. The optical intensity of the ionospheric pulsations was well correlated with that of the magnetospheric loss cone energetic electron flux. This result further confirms that the electron scattering by chorus is the main driver of pulsating aurora. Figure 14 shows an example of the correlation between the "main pulsation" and successive burst of chorus [335] obtained from recent conjugate ground/satellite observations.

During this 5-minute interval, one-to-one correlations were noted between the brightness of auroras observed from the ground (panel a)) and the intensity of chorus waves at the magnetospheric counterpart (panel b)).

Pulsating auroras also exhibit ~3 Hz luminosity fluctuations embedded within the main pulsation. It has been suggested that such sub-second fast variations are caused by intermitted electron precipitation triggered by the fundamental chorus feature, known as chorus "elements" [233, 288, 291, 542]. Figure 15 shows recent auroral pulsation images and corresponding plasma wave data. Conjugate observations demonstrated a one-to-one correspondence between the successive appearance of chorus elements in the magnetosphere and ~3 Hz optical modulation at the ionospheric magnetic footprint [335, 543-544]. See Figure 15 panels c) and d) for an example of this correlation. Hosokawa et al. [335] demonstrated that when the chorus contained multiple chorus elements, ~3 Hz optical modulation were clearly detected at the footprint of the satellite. In contrast, when chorus is less discrete (i.e., unstructured), pulsating aurora did not exhibit obvious signatures of sub-second scintillations (see Figure 15 panels a) and b)). "Unstructured" chorus can either be hiss-like emissions or overlapping tones (see [288] for examples and [266] for discussion). From these recent studies employing conjugate satellite and ground-based observations it was concluded that the temporal variations of pulsating auroras are almost perfectly controlled by the intensity variations of chorus waves in the corresponding region in the magnetosphere.

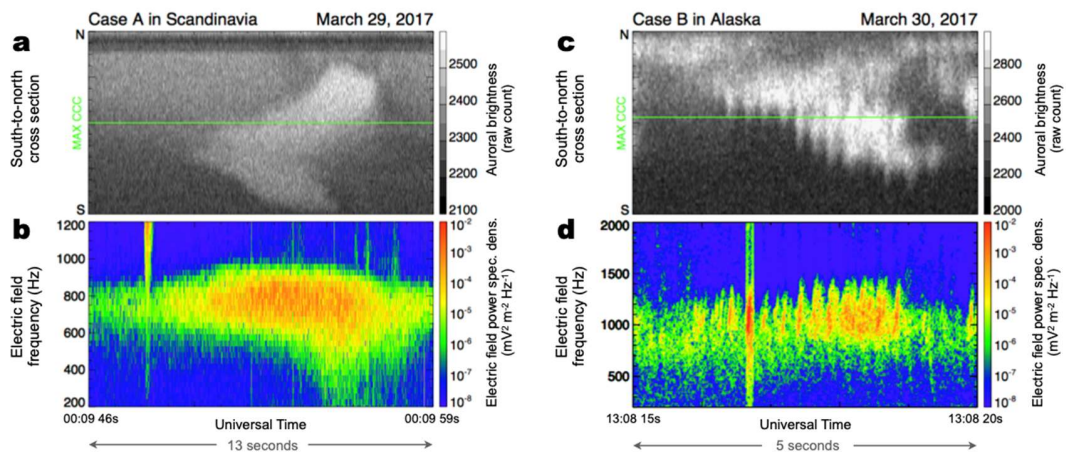

**Figure 15**. Two cases of a direct comparison between the ~3 Hz modulation of pulsating aurora and chorus elements. Panel a) shows the optical image of the pulsating aurora and panel b) is the corresponding plasma waves (the single chorus burst structure). Panels c) and d) are the same as panels a) and b), however there is discrete structures noticed in panels c) and d). This figure is taken from Hosokawa et al. [335].

It should be noted that in 2003 a new feature of chorus was discovered. Santolik et al. [281] noted that chorus elements contained substructures called subpackets or subelements. Tsurutani et al. [306] noted that such chorus subelements were coherent. Without this feature of chorus, the ~3 Hz modulation/microbursts could not be explained theoretically [270]. Also, such cyclotron resonant interactions of energetic electrons with the coherent chorus subelements might explain the super high frequency oscillations (higher than ~3 Hz) mentioned previously.

The altitude profiles of the optical emission and ionization during pulsating aurora has suggested that the energy of pulsating aurora electrons is higher than that causing other types of auroras [e.g., 531]. Recent observations reported that sub-relativistic electrons are sometimes observed within patches of pulsating aurora [545-546]. Turunen et al. [308] have implied that low altitude ionization caused by such precipitation can cause enhancement of $NO_x/HO_x$ and consequent decrease of $O_3$ at the mesospheric altitudes. Tsurutani et al. [308] has suggested that relativistic electron precipitation might modify atmospheric wind patterns. In this sense, diffuse/pulsating auroras are phenomena connecting the magnetospheric plasma environment with the middle and even lower atmosphere.

**Red Auroras**

Red-color auroral emission, the 630.0 nm emission from metastable atomic oxygen, is usually seen in the top part of discrete auroral features at altitudes from 200 to 300 km [547]. Such red auroral "fringes" are caused by the precipitation of the < 1 keV portion of the energetic electron spectrum. Red auroras are more prominent than the green aurora at 557.7 nm in the dayside cusp or within the polar cap because those regions are dominated by relatively soft electron precipitation (see reviews of [305, 507]).

Red auroras are sometimes seen from low-latitude regions during the main phase of magnetic storms which is called "low-latitude aurora" [548-552]. One of the causes of the low-latitude red aurora is the electron precipitation in a broad range of energy from ~30 eV to 30 keV seen at the subauroral latitudes [553].

Stable auroral red (SAR) arcs are often seen at subauroral latitudes during magnetic storms [554-555]. SAR arcs are composed of "purely" reddish emission at 630.0 nm from atomic oxygens excited by ambient heated electrons. They occur from altitudes of ~800 down to 200 km, the typical central altitude being ~400–600 km [556-558]. The temperature of ambient electrons can be enhanced by the heat conduction or low-energy (< 10 eV) electron precipitation in a region where the ions of storm time ring current interact with plasmaspheric electrons in their overlapping region [555, 559-560]. Because of the bloody color of SAR arcs, red auroras have been omens for war and bloodshed in ancient times.

## 11. Space Weather

"Space weather" is a relatively new name for a very old topic, the effect of dynamic features of our Sun affecting interplanetary space, planetary magnetospheres, ionospheres and atmospheres which are of concern to humankind and society. Space weather is therefore a subset of Space Plasma Physics. See Figure 16 for the many Space Plasma Physics topics that are concerns for space weather. Many of the topics covered in other Chapters of this paper are also space weather issues. For a nearly comprehensive review of space weather and future space weather problems, we refer the reader to Buzulukova and Tsurutani [561] and Tsurutani et al. [562]. We will only mention a few space weather topics below.

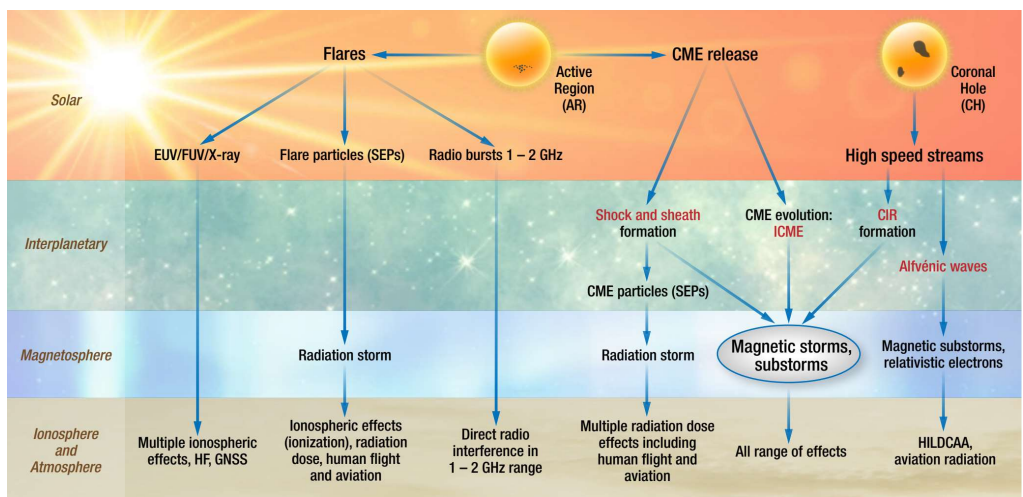

**Figure 16**. The space weather schematic shows solar phenomena that cause interplanetary, magnetospheric, ionospheric, atmospheric and ground level effects. Essentially all of these phenomena are discussed in this review, but from a Space Physics point of view.

## Geomagnetically Induced Currents (GICs)

The biggest focus of space weather features has been on those that can and have affected human society, e.g., humans on the ground or satellites and humans in space. Concerning the former, Loomis [82] reported that fires were set by arcing from currents induced in telegraph wires during the September 1-2, 1859 "Carrington" magnetic storm [6, 81]. The telegraph was the high technology of its day. It has been shown that magnetic storms of substantially greater intensities than the Carrington event can occur [265]. In a world increasingly reliant on electrical technology and space communication, extreme space weather endangers power grids, pipelines, railway systems and satellite communications, with enormous consequences for humankind [563].

During an intense magnetic storm in 1989, the Hydro Quebec power system had an outage for ~9 hours [564-565]. Another magnetic storm disrupted the power supply in southern Sweden (Malmo) in 2003 [566] and initiated damage to several large power station transformers in Africa [567]. Love [568] has reported that the "Railroad" storm of 1921 and another event in 1909 were both more intense storms than those in 1989 and 2003.

What feature(s) of magnetic storms are the causes for potential damage and disruption? Many possibilities have been discussed in the literature. One of the earliest reports of what is now called "geomagnetically induced currents" (GICs) was presented by Barlow [5] who reported that railroad telegraph magnetic needles were deflected coincident with aurora sightings. More recently Campbell [569] observed electric currents flowing in the Alaskan oil pipeline and deduced that the source of these GICs was the auroral electrojet. The auroral electrojet is a nighttime auroral zone ionospheric current with intensities up to ~$10^6$ A flowing at an altitude of ~100 km above the surface of the Earth. These strong currents can induce substantial currents in ground conductors such as pipelines and power lines (see a review of GICs by Lakhina et al. [570]). Tsurutani and Hajra [191] have studied intense GICs (>10 A) in the Mäntsälä, Finland gas pipeline with 21 years of data [571]. They have found that the most intense events were associated with intense substorms (supersubstorms: [572]) that occurred within magnetic storms, giving credence to the Campbell [569] idea that intense GICs might be caused by the auroral electrojet.

However now the questions become what features of the auroral electrojet cause intense GICs? And are supersubstorms simply more intense "traditional" substorms as first discovered by Akasofu, or are they fundamentally different [190]? The answers to these questions are not known at this time, and research is currently being done by the scientific community on these topics.

**Solar Energetic Particles**

Solar energetic particles have origins at solar flare sites and at shocks in front of interplanetary coronal mass ejections (ICMEs) associated with the solar flares [573-576]. Solar energetic particles and solar flares are discussed more thoroughly in Chapter 16: Solar Energetic Particles, Shocks and the Heliospheric Termination Shock. The energy of the particles can range from MeV to GeV energies and can cause damage to spacecraft solar panels and electronics [577-578]. For particularly intense solar energetic particle events, there can be substantial danger to astronauts in space [579]. At the present time, there is no known mechanism to protect astronauts from such intense radiation.

**Magnetospheric Relativistic Electrons**

The Earth's magnetosphere contains highly variable relativistic (~MeV) electrons [217, 580] whose fluxes can increase and decrease by 3 orders of magnitude during geomagnetic activity. The electron fluxes are highest during the declining phase of the solar cycle [237]. One possible mechanism to explain the observations is that the electrons are accelerated from ~100 keV electrons (injected during substorms and convection events) to higher, relativistic energies through the interaction of electromagnetic whistler mode chorus [221, 581]. Another possible mechanism is that ultra-low frequency (ULF) waves could cause energetic electron energization by radial diffusion [470, 582]. However, Horne et al. [311] have demonstrated that radial diffusion by ULF waves does not work for the Halloween 2003 event, so there are some doubts about the specific mechanism for radiation belt relativistic energization.

Hajra et al. [237] and Tsurutani et al. [74] have shown that relativistic electron acceleration occurs during HILDCAAs [150] which in turn are caused by the southward IMF components of Alfvén waves in the solar wind. The Hajra et al. [237] work showed that HILDCAAs lead to E > 0.6 MeV electrons within one day and E > 4.0 MeV electrons in ~2 days, indicating a "bootstrap" acceleration process. Magnetospheric energetic particles, chorus and substorm ~10 to 100 keV particles are discussed in other Chapters of this paper (see Chapter 6: Radiation Belts/Energetic Magnetospheric Particles, Chapter 8: Plasma Instability Waves and Wave-Particle Interactions and Chapter 5: The Magnetotail and Substorms.

**Low Altitude Orbiting Satellite Drag During Magnetic Storms**

Satellite orbits are selected to be compatible with multi-year lifetimes. For low altitude polar orbiting satellites, the minimum altitude chosen is typically ~400 km or greater. However, during magnetic storms, enhanced upwelling of ionospheric ions and atmospheric atoms and molecules occur associated with energetic particle precipitation and heating caused by friction between the magnetospherically driven plasma and the neutral atmosphere (see Chapter 8: Plasma Instability Waves and Wave-Particle Interactions). This ionospheric/atmospheric heating leads to the increase in ions, atoms and molecules at heights above 400 km [583-584]. Polar orbiting satellites therefore experience additional atmospheric "drag" and lose orbital speed [585]. Because of the additional drag the satellites do not appear at their expected locations at the predicted times and are "lost" to satellite tracking networks until they can be "relocated". At times this can take weeks to "reacquire" the satellites. Since there are thousands of orbiting objects around the Earth, this effect is a major one for satellite monitoring operations. If the satellite drag is particularly severe, the satellite may possibly lose enough altitude to have a shortened mission lifetime.

There is another recently found mechanism for increased satellite drag during magnetic storms. The magnetic storm convection electric field penetrates into the near-equatorial ionosphere and causes **E×B** uplift of the dayside ionosphere and downdraft of the nightside ionosphere [259, 586]. The uplift of parts of the ionosphere to higher altitudes on the dayside brings the ions and electrons to regions of lower recombination rates and the continued photoionization of lower altitude atoms and molecules by solar radiation leads to an increase in the total electron content (TEC) of the dayside near-equatorial ionosphere. Mannucci et al. [262] has shown a ~1200% TEC increase at altitudes above ~400 km (the CHAMP satellite orbit) at ~25° MLAT during the main phase of the October, 30, 2003 "Halloween" magnetic storm. The downdraft of the nightside near-equatorial ionosphere leads to recombination and a reduction in TEC.

Simulations have shown that if a Carrington storm occurred density peaks of oxygen ions would be ~6×10$^6$ cm$^{-3}$ at 700 km altitude, approximately +600% increase over quiet time values [264]. What is presently unknown is the degree to which ion-neutral drag increases neutral oxygen densities at high altitudes, which can lead to additional drag [587]. Satellite drag will be severe for polar orbiting satellites during a Carrington type magnetic storm. Unfortunately, we currently do not know how severe it will be. Deng et al. [588] indicate that for solar wind speeds of 1,000 km s$^{-1}$ and an IMF B$_z$ of -50 nT, at ~400 km altitude, the neutral density will increase by >10 times. We await further development of the computer code so that a full implementation of the Carrington storm can be made.

**Communication and radio blackouts**

The x-ray component of solar flares causes enhanced ionization to the dayside ionospheric D region [251, 589-590], an effect deleterious to radio wave communication and navigation (see a brief review in [591]). These are called sudden ionospheric disturbances or SIDs.

Solar flares also contain radio emissions in the frequency range from 10s of MHz to a few GHz that can interfere with signals from the Global Navigation Satellite Systems (GNSS) such as the Global Positioning System (GPS), which operate in the 1-2 GHz frequency range. One of the strongest flare events on record occurred in 2006 which was approximately ten times stronger than any previously reported event [592]. For tens of minutes, dual-frequency GPS receivers on the sunlit portion of the Earth were degraded such that they could not track enough GPS satellites to compute their locations [593]. See a review of this topic in Yue et al. [594].

Solar energetic particles accelerated at either the flare site or at the ICME shock can cause similar solar flare effects (SFEs). These energetic particles can enter the polar regions of the Earth's magnetosphere and penetrate deep into the polar ionospheres. Since solar flares have durations of tens of minutes with ionospheric D-region recombination time scales being much more rapid, these SFEs are somewhat short-lived. However solar flare particle events can last days (as the ICME shock propagates from the Sun to the Earth and beyond), and therefore the polar SFE can be a greater problem for humanity [595].

**Satellite Navigation and Ionospheric Storms**

Satellite navigation using Global Navigation Satellite Systems (e.g., GPS) may be affected during geomagnetic storms by rapid and large changes in ionospheric densities that occur on global scales. The unpredictable additional signal delay caused by increased ionospheric densities can lead to 10s of meters of positioning error when user receivers acquire only a single GPS frequency, as is currently the case for civil aircraft navigation. For this reason, GPS-based navigation is denied when large ionospheric storms are detected by the system used to augment GPS for aircraft [596]. Although ionospheric storms have been studied for decades [597-598], extreme space weather events that could cause unusually large and rapid ionospheric variations resulting from dayside $\mathbf{E} \times \mathbf{B}$ uplift remain a concern [263].

## 12. Planetary Magnetospheres and Solar/Stellar Wind Interactions with Comets, Moons and Asteroids

As first demonstrated by Carl-Friedrich Gauss [16-17] planetary magnetic fields have two major sources, one internal due to dynamo action in a fluid core, and another external due to electric currents outside the planetary body. While Gauss [16] argued that the external contribution is negligible compared to the internal one, current knowledge tells a different story. The interaction of a planetary body with its plasma environment provides for a physically most interesting new field of plasma and planetary physics.

At the dawn of the space age Thomas Gold [599] coined the name "magnetosphere" for the interaction region of the solar wind and the Earth and its internally generated planetary magnetic field. Gold [599] realized that the circumflow of the dilute ionized solar wind plasma around the Earth is heavily impacted by the Lorentz force $\vec{F} = \vec{u} \times \vec{B}$ due to the presence of the Earth magnetic field. The Lorentz force significantly influences the flow past any magnetized planetary object. Magnetospheric physics opened up an entirely new field of fluid dynamics. The novelty is also due to the extreme scale conditions in the interaction region. The gyro radius of a solar wind proton is of the order of a few thousand kilometres. This is comparable with the scale of the object the flow has to pass, the planetary radius. Such conditions cannot be realized in any terrestrial laboratory. In-situ measurements in space are required to study plasma-planetary interaction regions. The new physical understanding derived in this way is of paramount importance for the understanding of plasma astrophysical processes in general. Furthermore, space is more and more becoming part of the human habitat, which triggers the need to understand the detailed processes in space plasmas.

The past sixty years of space research demonstrated that the Earth's magnetosphere is only a special case of the flow of a magnetized plasma around an obstacle. The term magnetosphere is more and more in use to describe the flow past objects like planets, their moons, asteroids, and comets. The parameter space in which the various types of magnetospheres can be located is at least three dimensional. The Mach number of the flow is another important control parameter. Our eight planets are embedded in the solar wind, a super-sonic flow. The Galilean moons are interacting with the sub-sonic plasma flow of the Jovian magnetosphere. Mercury, Earth, Jupiter, Saturn, Uranus, and Neptune are magnetized objects. They possess *classical* magnetospheres. Asteroids and most moons do not operate planetary dynamos in their interior. Interaction is dominated by the direct interaction of the flow with the surface of these objects. Comets, Venus, and Mars define another regime in the parameter space, the regime of plasma-neutral gas interaction. The various conditions define several different modes of interaction. Three modes are dominant and discussed in further detail: the *classical mode*, the *cometary mode*, and the *lunar mode*.

The *classical mode* magnetosphere is dominated by the interaction of the streaming plasma with the planetary magnetic field. An important property of the solar wind is its almost infinite electrical conductivity, $\sigma \rightarrow \infty$. Basic magnetohydrodynamic considerations show that for such a medium the magnetic flux is frozen into the moving plasma and vice versa. If a fluid parcel is moving around and changing its shape the magnetic field is modified in such a way as to conserve magnetic flux. This is reminiscent to the conservation of vorticity in hydrodynamic rotating flows. The frozen-in flux condition leads to an electric field $\vec{E} = -\vec{u} \times \vec{B}$ in the streaming plasma. And the induction equation is

$$\partial \vec{B} / \partial t = \nabla \ x(\vec{u} \times \overrightarrow{B}) \tag{2}$$

that is any curl of the motional electric field causes a local temporal change in the magnetic field.

The solar wind is a plasma with infinite conductivity. Thus, if a planetary body is immersed in this plasma the planet's magnetic field cannot penetrate the solar wind (see Figure 17 for the solar wind interaction with the Earth' magnetosphere). This induces the build-up of an internal boundary, the magnetopause. Electric currents, magnetopause currents or Chapman-Ferraro currents, flow in this boundary. They ensure that sunward of the magnetopause, the planetary magnetic field is cancelled. On the planetary side the magnetic field is enhanced. Once the dynamic pressure of the flow is balanced by the magnetic pressure on the planetary side the magnetopause reaches its equilibrium position. For a dipolar planetary magnetic field, the magnetopause position $R_{MP}$ is defined by:

$$R_{MP} = \sqrt[6]{B_s^2/p_d}\, R_p \qquad\qquad (3)$$

where $B_S$ denotes the planetary magnetic field at the surface of the obstacle, $p_d = \frac{1}{2}\rho u^2$ the dynamic pressure of the streaming plasma, and $R_p$ the planetary radius. The terrestrial dayside magnetopause is located at about 10 $R_E$. This simple expression for the magnetopause distance has been confirmed many times, at Earth as well as at other planets. It is also in use to estimate the existence and intensity of planetary magnetic field of extra-solar planets [e.g., 600].

The magnetopause is a self-induced internal boundary in the plasma flow. It defines the actual obstacle, the magnetosphere. On its planetary side the magnetic field is strongly compressed, while on the nightside the magnetosphere is stretched out into a long magnetotail with diameter of about four times the magnetopause distance. The magnetotail is divided into northern and southern lopes, separated by an electric current carrying neutral sheet. Would this huge interaction region be visible, it would appear like a comet with its tail.

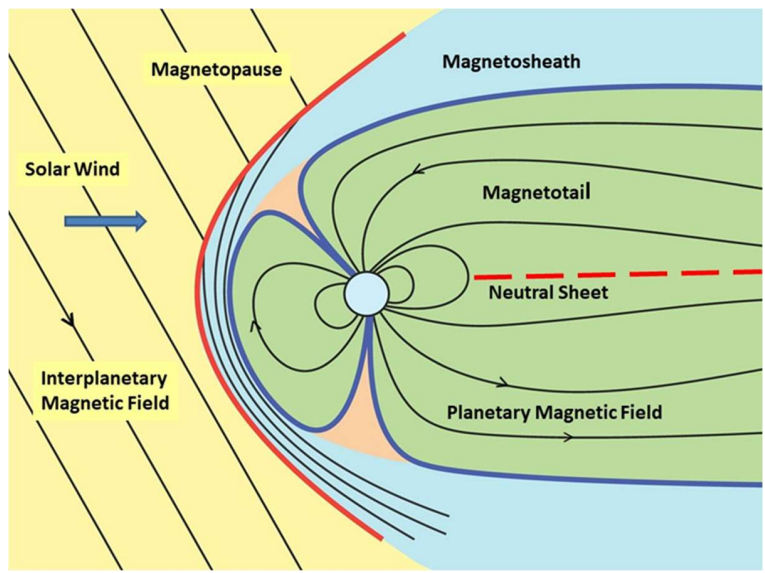

**Figure 17**. The solar wind interaction with the Earth's magnetosphere. The red curved line is the bow shock. The blue region is the magnetosheath where draped magnetic fields (black lines) are convected downstream. The magnetosphere and magnetotail are indicated in green.

Magnetic reconnection [e.g., 601] is the new process identified to explain the tail formation. At the magnetopause magnetic shear $\nabla \times \vec{B}$ causes the flow of strong electric currents in the plasma, of the order of $10^6$ A in total or a density of about $10^{-7}$ A m$^{-2}$. Such an intense current represents a clear deviation from local thermodynamic equilibrium. A multitude of different plasma instabilities may emerge, acting to bring the system back to equilibrium. In essence, these plasma instabilities cause an anomalous electric resistivity in the plasma with local break-down of the frozen-in condition. The resistivity is anomalous as the plasma under consideration is a collisionless plasma. Magnetic flux conservation is no longer possible. This implies a change in magnetic field topology. Magnetic field lines carried towards the magnetopause by the plasma flow are reconnected with magnetic field lines of planetary origin. Strongly bend magnetic field lines occurring on the dayside magnetopause lead to acceleration of the plasma flow towards the nightside. The underlying physics, magnetic reconnection, is a process via which magnetic energy is converted to kinetic energy. It can be viewed at as an anti-dynamo action. Details of this most complex process are still under discussion with electron scale processes playing a major role [e.g., 177]. Reconnection at the dayside is the prime mechanism to generated the magnetotail. The tail has been observed up to distances $\geq 1600$ R$_E$ [e.g., 602-603]. Its two-lobe structure is well observed beyond 200 R$_E$ [e.g., 156].

As the solar wind plasma is a super-sonic flowing plasma stream there must be a region in front of the magnetosphere where the flow is decelerated to sub-sonic speed. This region is the bow shock region. It is a collisionless shock wave structure standing in the plasma flow. Here kinetic energy is converted into thermal energy via plasma wave turbulence. In the bow shock region flow deviation occurs, initiating the flow around the magnetosphere [e.g., 604]. The bow shock is a regime where plasma particles are reflected and accelerated back into the upstream plasma flow. This foreshock region is a playground for a variety of plasma instabilities and intense wave activity [e.g., 394].

The region downstream of the bow shock is termed the magnetosheath. It is the actual regime where the plasma flow streams around the obstacle. The magnetosheath is a very dynamic regime. Towards the magnetopause flow, deceleration occurs. Under stationary conditions the induction equation (2) along the stagnation streamline transforms into:

$$\frac{1}{B}\,\partial_x B = -\frac{1}{u}\,\partial_x u \qquad\qquad (4)$$

During the flow deceleration towards the magnetopause stagnation point, magnetic flux piles up in the magnetosheath. Flow kinetic energy is converted to magnetic energy by dynamo action. The whole process requires a spatial extent. The bow shock typically is a detached bow shock with the stagnation streamline distance $R_{BS}$ related to the magnetopause distance via $R_{BS} \approx 1.3\, R_{MP}$. It should be noted that for the classic magnetospheric mode described here the formation of the magnetopause and magnetosphere proper is the primary process. The bow shock is a secondary process, required to enable a sub-sonic flow around the magnetospheric obstacle.

The classical mode magnetosphere is changing in time depending on solar wind conditions. External events such as solar wind HSSs, CMEs or strong fluctuations in the southward component of IMF are usually driving magnetospheric activity [e.g., 150]. It also exhibits a variety of internal instabilities. Magnetospheric storms and substorms are witnesses of these processes, causing magnetospheric space weather events. Space weather has an increasing impact on technical terrestrial systems [e.g., 605].

The *classical mode* of interaction is also observed at planets Mercury, Jupiter, Saturn, Uranus and Neptune. Of sources, modifications are observed depending on plasma sources, planetary rotation, etc.

The *cometary mode* magnetosphere is very different from this classical mode [e.g., 606]. The obstacle does not have or generate any intrinsic magnetic field. Neutral gas emanates from the cometary nucleus surface as the comet nears the Sun. Due to solar ultra-violet radiation as well as energetic electrons in the developing interaction region, neutral particles, mainly OH- and CO-group molecules are ionized. The region where this happens is a vast region around the cometary nucleus. The physical obstacle to the plasma flow is a spatially much-distributed region.

In cases where the solar wind flow velocity is perpendicular to the interplanetary magnetic field, the newly created ions immediately sense the electric field of the streaming plasma, $\vec{E} = -\vec{u} \times \vec{B}$, and are accelerated to the speed of the stream. Under more general conditions, the new-born ions form beam or ring distributions in the streaming plasma. This constitutes another extreme deviation from local thermodynamic conditions. A plethora of plasma instabilities gives rise to very strong plasma wave activity [e.g., 607-610]. In the low-frequency range, wave amplitudes are largest with $\delta B/B \approx 1$. Non-linear evolution of the waves causes strong turbulence to develop [e.g., 611]. The fluctuating electromagnetic fields act quasi-particles causing strong scattering of the new-born ions and isotropization of the particle distribution. Finally, the new-born ions are picked-up by the solar wind. Momentum and energy are transferred from the streaming plasma to the ions of cometary origin. On the macroscopic scale this pick-up process causes mass loading and deceleration of the streaming plasma. As both momentum and energy must be conserved, mass loading is limited and controlled by the polytropic index γ. Once the mean molecular mass <m> of the pick-up ion enriched plasma exceeds the value $\gamma^2/(\gamma^2-1)m_\infty$, where $m_\infty$ is the molecular mass far away from the comet and γ the polytropic index, a stationary solution for the mass-loaded flow no longer exists. A bow shock forms at some position along the stagnation streamline [612-613]. Thus, a bow shock is the primary internal boundary build-up in the *cometary mode* interaction region. This is different from the *classical mode*, where the magnetopause is the primary boundary, the bow shock being secondary. The *cometary mode* interaction is shown in Figure 18.

Mass loading is strongest along the stagnation stream line. Thus, flow deceleration is most prominent close to this region, deceleration decreasing with increasing distance from the stagnation line. Flow deceleration is accompanied by pile-up of magnetic flux in front of the cometary obstacle. It also causes draping of the frozen-in interplanetary magnetic field around the cometary nucleus. A cometary tail forms as first suggested by Biermann and Cowling [15]. This conjecture was later confirmed by spacecraft observations [e.g., 613-614]. Eventually cometary tail formation is may also be viewed as a large-scale confirmation of the frozen-in theorem.

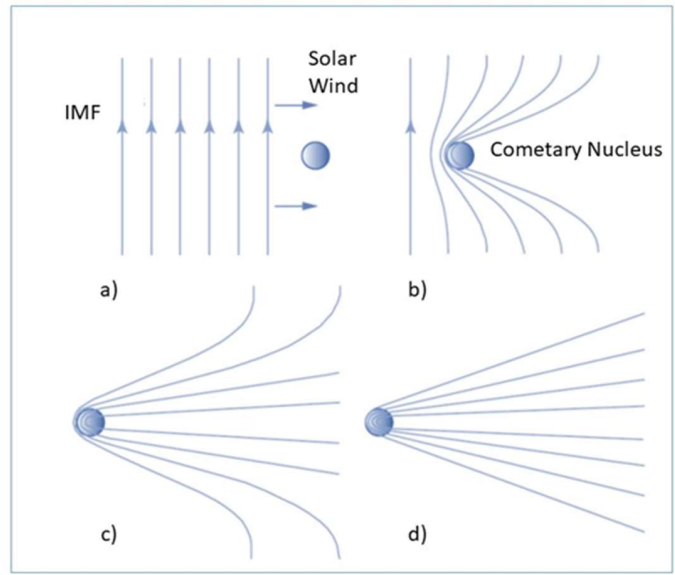

**Figure 18**. A time sequence (from panels a) to d)) of draping of interplanetary magnetic fields around an outgassing cometary nucleus. The figure is taken from Alfvén [615].

Often the *cometary mode* interaction region is called an induced magnetosphere. Planets Venus and Mars exhibit typically induced magnetospheres. Here the interaction with pick-up ions emanating from the planetary atmosphere and direct interaction with the ionosphere determines the physics of the interaction regime [e.g., 616].

A third mode, the *lunar mode* deserves attention. The interaction of the Moon represents a case where the streaming plasma impinges onto the surface of the planetary body, without any planetary magnetic field or emanating neutral gas modifying the plasma. The solar wind particles are absorbed by the obstacle. A wake region forms downstream of the object [e.g., 617-618]. Upstream of the obstacle no bow shock wave exists, different from the other two modes of interaction discussed previously. Due to particle absorption there is no need for the build-up of such a bow shock. This was first observed by magnetic field measurements with the Explorer 35 spacecraft at the Earth's moon [619].

A streaming magnetized plasma may actually be viewed as a kind of two-fluid medium interacting with an obstacle. There is the mechanical part represented by the mass flow. And there is the electromagnetic part, represented by the magnetic field. Although the particles may be absorbed by the surface, this is not necessarily possible for the electromagnetic part. The interaction of the magnetic field transported by the flow towards the obstacle depends very much on the electrical conductive of the obstacle. Non-conducting planetary bodies just allow the magnetic flux to be transported through the body. The obstacle actually does not exist for the magnetic field. If the conductivity is large, however, flux transport through the obstacle induces electric currents in the body interior or any surrounding ionosphere inhibiting the flux flow. Pile-up of the magnetic flux in front of the object and draping around the object may occur. Observed details about the interaction region can thus be used to infer information on the electric properties of the obstacles.

The interaction of the solar wind with asteroids is of the *lunar type.* Asteroids do not exhibit any atmosphere and usually do not have an internally generated magnetic field. Nevertheless, the interaction is special particularly when the scale of the obstacle is comparable to typical plasma scales [620]. Observational evidence of this type of interaction has been presented by Kivelson et al. [621] and Auster et al. [622].

The three modes discussed above provide a kind of coordinate system of a model space where interaction of a streaming plasma with a specific obstacle may be located. A planet with a weak magnetic field like Mercury usually interacts in the *classical mode.* But occasionally, if the dynamic pressure is very strong, the plasma reaches the surface and the *lunar mode* applies. Strong magnetic anomalies have been detected at the Moon. Associated with such anomalies are the mini-magnetospheres and small-scale collisionless shocks upstream of these anomalies [e.g., 623]. Thus locally, the interaction of the solar wind with the Moon occurs in the *classical mode.*

The three modes also differ in another aspect: temporal variations of the obstacle's properties. The Moon or asteroids are rather stable objects with which the streaming plasma is interacting. *Lunar mode* interactions thus do not exhibit any major temporal variations. This is different for the *classical mode.* Planetary magnetic fields change on secular time scales. Mode modifications are possible and led to intensive studies on paleomagnetospheres [624-625]. The *cometary mode* is susceptible to changes of the obstacle properties such as temporal development of the cometary activity [e.g., 626] or sudden outbursts [e.g., 627].

The above classification of interaction regions is based on properties of the planetary bodies, assuming a super-sonic plasma flow. If the flow is subsonic, modifications are necessary. Planetary satellites within magnetospheric plasma environments need to be considered here. Alfvén wings instead of bow shock structures are usually generated in such cases [628-631]. Entirely new aspects of the diversity of the interaction regions emerge [632].

Our solar system is the immediate environment where we can study solar wind-planetary body interactions. Stellar winds are known to exit for most stars. Also, planetary companions are most likely present. One can only hypothesize what type of interaction is taking place. Vernisse et al. [633], for example, presented numerical studies on the lunar type of interaction. Whether exoplanets exhibit planetary magnetic fields is as yet unknown. Scaling laws can be used to estimate global magnetic field strengths [634]. The most promising means to prove the existence and determine the strength of exoplanetary magnetic fields is via the electron maser instability [635-636] driven radio emissions from exoplanets [637]. However, detection sensitivity is still too low to allow any inferences to be made on any exoplanetary radio emissions [638]. Thus, the *classical mode* of interaction is currently a subject of model studies [639].

## 13. Interplanetary Discontinuities, Shocks and Waves

### Discontinuities

In the 1960s and 1970s there were great debates in the literature [640-642] about whether the interplanetary discontinuities were rotational (RD) or tangential (TDs) in nature [643]. The analyses were performed primarily using magnetic field data because the plasma data temporal resolution was slow in comparison. Applying computer codes to identify "directional discontinuities" (DDs: either rotational or tangential), Tsurutani and Smith [289] and Lepping and Behannon [644] indicated that DDs occurred at a rate of one or two per hour in the solar wind. Neugebauer and Giacalone [645] gave a review of the current status of DD discontinuity research.

### Shocks

In 1984 there was an AGU Chapman Conference held in Napa, California on the topic of "Collisionless Shocks". The invited reviews were published in two volumes of AGU monographs and two volumes of JGR special issues. In "*Collisionless Shocks in the Heliosphere: A Tutorial* (1985)" [646], we recommend Kennel et al. [454] and Papadopoulos [647] as two particularly excellent theoretical articles. "*Collisionless Shocks in the Heliosphere: Reviews of Current Research* (1985)" [648] is devoted to writeups of invited reviews of the then recent research. The JGR special issues (January and June, 1985) were devoted to writeups of contributed talks at the conference. A more recent article, dealing with geomagnetic effects of shocks and discontinuities can be found in Tsurutani et al. [649].

**Fast Shocks**

Fast forward shocks have been identified at the regions upstream of fast ICMEs [574]. By "fast" it is meant that the speed of the ICME relative to the upstream solar wind is faster than the upstream magnetosonic speed. By "forward" we mean the direction of propagation is the same direction as the driver, antisunward. The magnetosonic Mach number is typically between ~1 and 3 [574]. Much higher Mach number shocks have been occasionally detected (~28 for the July 23, 2012 event: [650]), but not near the theoretically possible value of ~45 postulated by Tsurutani and Lakhina [265].

For CIRs, at 1 AU there are typically no fast forward shocks bounding the antisolar side of the CIR. For about 20% of cases there are fast reverse shocks on the solar side of the CIR [245]. There are both fast forward and fast reverse shocks bounding the CIRs at distances > 2.5 AU from the Sun [72, 651]. The magnetosonic Mach number of these shocks are also relatively weak.

Planetary "bow" shocks are high Mach number fast reverse shocks. They have been detected at Earth [652-653], Jupiter [272], Saturn [654], Uranus [655], Mars [656-658], Venus [659] and at comets [613, 660]. Since planets are relatively stationary with regard to radial motion relative to the Sun, these Mach numbers are quite high, ranging from ~9 to ~21. The Mach number of bow shocks at comets is low [404, 661] (See more details in the Planetary Magnetospheres chapter).

Charged particles can be accelerated to MeV and even GeV energies at ICME fast forward shocks [662]. See review in Reames [575]. The proposed mechanisms are gradient drift along the shock surface for perpendicular shocks [573, 663] and second order Fermi acceleration for parallel shocks [664-665]. Shocks which are neither perpendicular nor parallel (everything in between) can accelerate charged particles by both mechanisms. One exceptional case of quasi-parallel shock energetic particle acceleration was reported by Kennel et al. [666-667]. See Chapter 16: Solar Energetic Particles, Shocks and the Heliospheric Termination Shock for more detail.

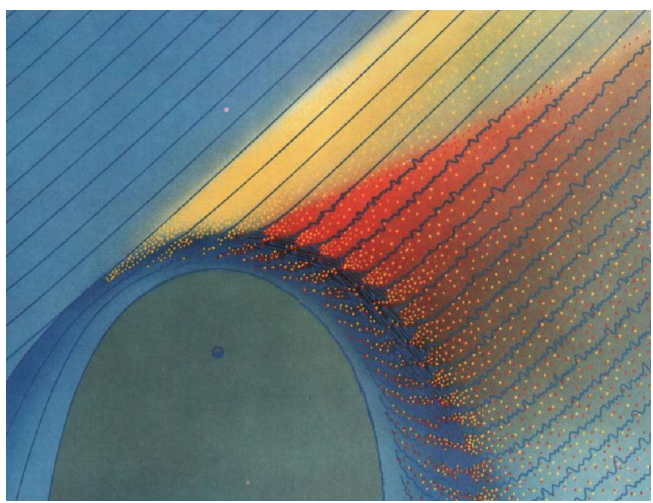

**Figure 19**. The Earth's bow shock and upstream energetic particles and waves (the foreshock region). The Sun is at the top out of view. At 1 AU, the Parker interplanetary magnetic field is nominally at an angle of 45° relative to the Earth-Sun line. Energetic electrons are indicated by yellow dots and energetic ions by red dots. Energetic electrons and ions accelerated at the shock or in the sheath can escape into interplanetary space and form the foreshock region. Upstream energetic electrons and ions can generate plasma waves by electron and ion beam instabilities. The figure is taken from Tsurutani and Rodriguez [394].

Figure 19 shows the Earth's foreshock region. Note that the Earth's bow shock is parallel/quasi-parallel in nature in the morning hours and perpendicular/quasi-perpendicular in the early afternoon hours. ULF waves do not have high enough speeds to propagate into the upstream foreshock region. The waves are generated locally by the escaping energetic ion beams through an anomalous cyclotron resonance instability [668].

Particle acceleration at fast forward and fast reverse shocks also occur at CIR shocks [41, 669]. However, because CIR shocks are low in Mach number, the energy and flux of the accelerated particles is less than at ICME shocks.

The enhanced plasma densities sunward of fast forward quasiperpendicular shocks (or simply interplanetary plasma pressure pulses) can cause compression of dayside magnetospheric preexisting energetic electrons and protons. Compression (conservation of the first adiabatic invariant) causes the perpendicular temperatures to become greater than the parallel temperatures. This can lead to plasma instabilities and both electromagnetic chorus and EMIC wave growth [308; see also Chapter 8: Plasma Instability Waves and Wave-Particle Interactions]. Such shocks cause dayside auroras which propagate from the nose of the magnetosphere tailward, the same antisunward direction as the shock propagation [670-671]. Shock compression of the Earth's magnetosphere/magnetotail can also trigger nightside substorms [188-189].

**Slow Shocks**

Slow shocks have been identified in the Earth's magnetotail located between the plasmasheet and the tail lobes [672-673]. They have also been detected in interplanetary space [674-675]. In one case where an interplanetary pair of forward and reverse slow shocks were detected, energetic particles were noted to have been accelerated at the shocks [676].

**Intermediate Shocks**

To date, there has been only one clear case of an interplanetary intermediate shock observation reported in the literature [677]. If the Tsurutani et al. [678] speculation that interplanetary nonlinear Alfvén waves phase steepen into intermediate shocks is correct, then it is possible that many of the "directional discontinuities" in the solar wind discussed above are intermediate shocks. Although strong plasma heating has been noted to occur, the actual measurements involved in determining the potential shock properties have not been made to date. See also arguments in Lee et al. [679] that intermediate shocks are intrinsically unstable. More will be stated in the topic of interplanetary Alfvén waves below.

**Alfvén Waves, "Switchback" Magnetic Fields, Magnetic Decreases and Magnetic Reconnection**

Alfvén waves [133, 680] is the dominant plasma wave mode present in the solar wind [681]. The wave amplitudes are essentially equal to the ambient magnetic field strength and are highly nonlinear. Because of the nonlinear nature of these waves and the high β of the solar wind, existing kinetic theory [682-683] of smaller wave amplitudes are not applicable for understanding the properties of these waves.

Some Alfvén waves have been shown to be "arc polarized" [684-686]. Tsurutani et al. [687] and Tsurutani and Lakhina [668] showed that the waves were "phase steepened" where the waves were split into two sections, a slowly rotating arc (~180° rotation in phase) and a fast rotation reverse arc (~180° in phase). See Swift and Lee [688] and Vasquez and Hollweg [689] for numerical calculations and simulations illustrating these properties, respectively. The fast rotation, phase-steepened end of the Alfvén wave is a rotational discontinuity. Because Alfvén waves are by nature noncompressive, nonlinear Alfvén wave magnetic perturbations must rotate on the surface of a sphere and are thus spherical waves [681, 690]. Balogh et al. [244] identified events where the radial component of the interplanetary magnetic field reversed sign. Changes in the plasma flows were noted during the magnetic field reversals [691]. All of these features are part of arc-polarized, spherical Alfvén waves called "magnetic switchback events". Some recent works on this topic are: Mozer et al. [692], Larosa et al. [693], Akhavan-Tafti et al. [694] and Neugebauer and Sterling [695].

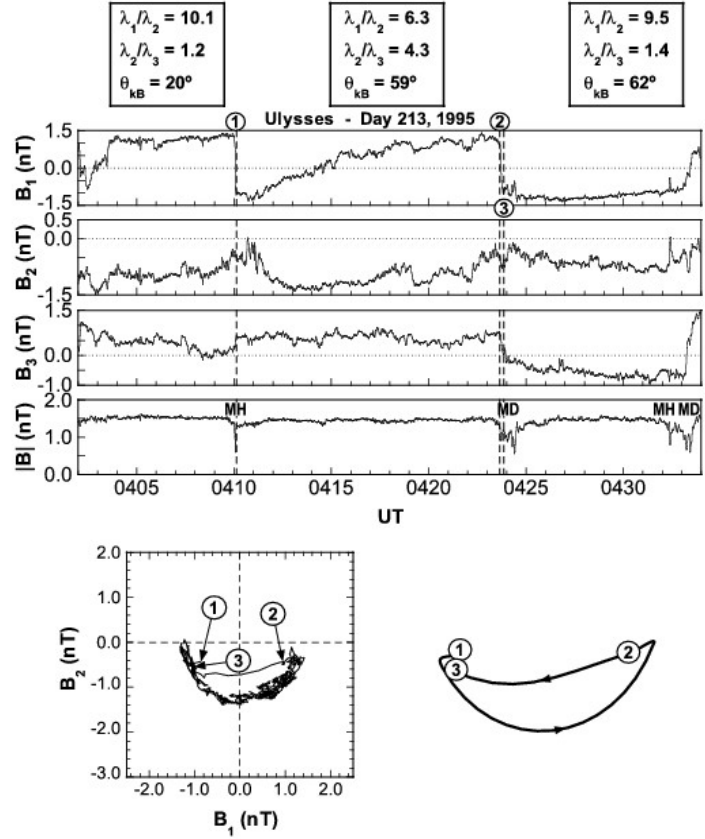

**Figure 20**. A series of 3 interplanetary Alfvén waves detected by *Ulysses*. The waves are phase steepened with MDs/MHs formed at the steepened edges. The dissipation of the steepened waves is believed to be due to the Ponderomotive Force associated with the sharp edges (RDs). The heated plasmas expelling magnetic flux forms the MDs. The lower panel shows that the wave is arc polarized, consistent with a spherical wave. The figure is taken from Tsurutani et al. [696].

Interplanetary magnetic holes (MHs) were first reported by Turner et al. [697]. It was later shown that larger scale magnetic decreases (MDs) were the same thing [696]. We will use the latter name in this paper. MDs are simply defined as short duration decreases in the interplanetary magnetic field magnitude. MDs are pressure balance structures where plasma thermal pressure supplants the decreased magnetic field pressure [698-700].

Fränz et al. [701] and Neugebauer et al. [699] have shown that inside MDs, the proton perpendicular temperature is typically higher than its parallel temperature. Tsurutani et al. [702] argued that the proton perpendicular heating inside MDs was a local process by demonstrating that the temperature anisotropies led to the instabilities where proton cyclotron waves and mirror mode structures were generated. The authors argued that the increased pressure from the locally heated plasma displaced magnetic pressure, leading to the formation of the MDs. Earlier Lin et al. [703-704] showed that electron heating was also occurring inside MDs, indicated by the presence of both electron whistler mode waves and Langmuir waves. What is the source of plasma heating inside MDs? Tsurutani et al. [702] and Dasgupta et al. [705] argue that the Ponderomotive Force associated with the phase-steepened edges of Alfvén waves is doing the heating. It is also possible that slow mode shocks or fast shocks from parametric instabilities are causing the local heating [689, 706]. All of these mechanisms are forms of Alfvén wave energy dissipation which may lead to solar wind acceleration.

Figure 20 shows three cycles of interplanetary Alfvén waves observed by the *Ulysses* magnetometer. The waves are phase-steepened, arc-polarized spherical waves. The steepened edges consist of ~180° of phase rotations and are RDs. The Tsurutani et al. [702] scenario is that the Ponderomotive Force associated with the RDs cause heating of ambient solar wind plasma which then displaces the solar wind magnetic field, creating the MDs.

MDs are typically bounded by sharp edges or tangential discontinuities, TDs [681]. Fränz et al [701] examined 115 thick MDs and found that ~78% of them were bounded by TDs. What are these TDs? Farrugia et al. [707] examined one TD at the boundary of a MD and showed that it was a slow shock.

A recent review [71] has attempted to link all of these interplanetary features together. First, nonlinear Alfvén waves phase-steepen to form rotational discontinuities at their leading edges. These RDs continue to steepen forming intermediate shocks ([678]; however, see Lee et al. [679] who believe intermediate shocks are unstable). The Ponderomotive Force associated with the rotational discontinuities heat the upstream plasma forming the MDs. The heated plasma dissipate energy in the generation of plasma waves. The folded magnetic fields (switchback events) lead to magnetic reconnection [708] of the Petschek [709] type, with further Alfvén wave dissipation.

What are the causes of MDs without TD boundaries, the ~22% noted in Franz et al. [701]? One possibility is that the Alfvén wave amplitude was not sufficient to have folded-back into itself such that opposite field directionalities are adjacent to each other. Another possibility is that reconnection had already taken place and the TD is a "fossil" of a previous slow shock. Research on this topic is currently taking place with *Parker Solar Probe* and *Solar Orbiter* data.

**Formation of interplanetary and interstellar turbulence**

Space plasma researchers [710, and references therein] have noted that the interplanetary magnetic field often take a $f^{5/3}$ power law shape, indicating that Kolmogorov fluid- like processes [711] may be occurring in the solar wind. Matthaeus et al. [712, and references therein] have explained this spectral shape by inverse cascade and quasi-inverse cascade processes. Lee and Lee [713] have examined the local interstellar medium data (*Voyager* 1) and have found that although the magnetic field has a Kolmogorov-like shape, the transverse magnetic field power is higher than the parallel power. They conclude that transverse Alfvén waves or arc-polarizations must be present. Tsurutani et al. [71] have taken a different viewpoint concerning the solar wind turbulence. They have noted that Alfvén waves phase-steepen with the leading edge forming a ~180° rotational discontinuity leaving an elongated trailing portion of the wave consisting of the remainder 180° phase rotation. The former is a form of "wave breaking" and the latter that of a wave "period doubling". This process occurring over many single Alfvén waves can form the magnetic turbulence.

**14. Interplanetary dust**

Signs of interplanetary dust were present well before the space age. A faint white glow is visible in the night sky before sunrise along the zodiac, in the ecliptic plane. This is called the "zodiacal light" or "false dawn". The zodiacal light is due to sunlight scattered by interplanetary dust. There is also a glow visible in the portion of sky directly opposite the direction of the Sun. This is called "gegenschein". It is caused by the backscatter of sunlight by interplanetary dust [e.g., 714]. The zodiacal cloud of dust is a pancake shaped phenomenon in our solar system that straddles the ecliptic plane [e.g., 715]. Hanner et al. [716] using the *Pioneer* 10 spacecraft imaging polarimeter instrument first directly connected the zodiacal light to interplanetary dust in our solar system.

Interplanetary dust particles (IDPs) in the solar system have been studied in situ with dust detectors on-board spacecraft in the space age [717]. It was motivated by the desire to understand the interplanetary dust size distribution. The primary instruments used for in-situ dust detection are impact ionization type of detectors, which measure the plasma electrical pulse created when dust grains impacting the interior walls of the instrument box and then vaporize and ionize into ions and electrons. The zodiacal light, the interplanetary and the in situ interstellar dust have been extensively reviewed in recent papers [715, 718-719] as well as in older, but very useful resources [720-722]. This Chapter introduces the topic of interplanetary dust and describes recent advances in our knowledge of the dust between and around the planets through studies and missions of the last few years. A NASA review/white paper for interplanetary dust is contained in Mann et al. [723] which was written to examine the potential for dust damage to the initial *Solar Probe* mission design, which was intended to reach a close-approach distance of 4 $R_s$ from the Sun. With high spacecraft velocities near the Sun, a sizeable dust impact could torque the spacecraft and the instruments and spacecraft components would then be exposed to intense solar radiation. The instruments and spacecraft would then become seriously damaged.

**The interplanetary dust cloud (IDC)**

Interplanetary dust spans the size range from submicron to up to millimeter sizes and originates mostly from comets [724] along with asteroids, and Edgeworth-Kuiper Belt (EKB) Objects. When asteroids collide, they produce dust and micrometeoroids. Comets sublimate gas when they get close to the Sun. Smaller cometary particles are dispersed and pushed away by the solar radiation pressure force, while larger (micron-sized) particles remain closer to the comet and form a trail along the comet's orbit [725]: these are the meteoroid streams that can produce meteor showers on Earth when the Earth's orbit crosses the cometary tails' orbits. Small dust is dispersed from the meteoroid streams by various mechanisms. Losses processes for interplanetary dust include collisions, sputtering and, especially in the inner solar system, sublimation. The IDC or zodiacal dust cloud dust density distribution generally has an exponential slope of -1.3 with increasing distance to the Sun. The distribution (out of ecliptic plane) is typically described as a 'fan'-like structure [726-727]. Rowan-Robinson and May [728] give a review of the structure of the zodiacal cloud, and fits interplanetary dust models (including an interstellar component) to space-borne infrared observations [728].

Within the solar system, dust grain motion is determined by a combination of gravity, Poynting-Robertson drag [729-730], solar radiation pressure, and electromagnetic interaction with the solar wind. All of these forces act on every dust grain, but the dominant force depends upon each dust grain's mass and surface charge (see [718], and references therein). The motion of dust grains larger than ~ 100 $\mu m$ is determined primarily by gravitational interaction with the Sun (as well as gravitational perturbations from the planets and other solar system bodies). These grains are subject to Poynting-Robertson drag, which causes them to slowly lose angular momentum, reducing the semi-major axis of their orbits, such that they ultimately spiral inward toward the Sun on nearly-circular orbits. The motion of grains with radii between ~1 $\mu m$ and ~100 $\mu m$ is strongly influenced by solar radiation pressure [731]. When radiation pressure dominates gravity, dust grain orbits can become hyperbolic, and grains can be ejected from the solar system as $\beta$-meteoroids [731]. Grains with nanometer radii (< 1 $\mu m$) are often called nanograins. Like all objects immersed in a plasma [732], nanograins experience surface charging. The electromagnetic forces imposed by the flow of the solar wind magnetic field over charged nanograins can dominate gravitation and solar radiation pressure, causing nanograins to exhibit dynamics similar to exceptionally massive pickup ions [733].

Other than interplanetary dust particles (IDP) and meteoroids, more types of dust "in between or around the planets" exist [734]: airless bodies develop dust clouds created by impacting particles [735-736] that thankfully can be used for probing surface compositions using spacecraft with dust mass spectrometers [737-738]. Contemporary interstellar dust moves from interstellar space through the solar system and can also be probed in situ [719, 739-741]. Nanodust from Io's volcanoes or dust originating from the depths of Enceladus' oceans can escape the systems of Jupiter [742] and Saturn [743-744]. Their compositions can be measured by in situ time-of-flight mass spectrometry [745].

**Recent developments**

Over the last decade, significant progress has been made predicting the motions of interplanetary dust grains whose motion is dominated by electromagnetic forces [746]. These 'nanograin' particles have been suggested to play a role in the creation of inner source pickup ions [747-748], cometary evolution [749], and solar wind mass loading [750-751]. Observations of nanograins were reported in association with dust streams from Jupiter [752], between 1 and 5 AU on the Cassini spacecraft [753-754] and at 1 AU by the STEREO spacecraft [755-756]. However, some authors debate whether the STEREO observations were due to nanodust or not [757-758].

Recently, data from the *Parker Solar Probe* mission [759] have enabled new progress understanding the near-Sun evolution of the zodiacal cloud. The *Parker Solar Probe* spacecraft has traveled closer to the Sun than any previous spacecraft, eventually reaching 9.8 $R_S$ over 24 orbits of the Sun. From this unique vantage point, data from the WISPR white light imager [760] was used to confirm a previously unverified prediction from 1929 that a dust-free region exists close to the Sun where dust grains are destroyed by interaction with solar photons [723, 761]. The near-Sun dust density was found to increase with decreasing radial distance until 19 $R_S$. It remains approximately constant until 10 $R_S$, then exponentially decreases closer to the Sun, reaching ~0 density at 3 $R_S$ [762-763]. WISPR data were also used to observe the full longitudinal extension the circumsolar dust ring in the orbit of Venus [763], adding to our current understanding of terrestrial planet circumsolar dust rings [764-768].

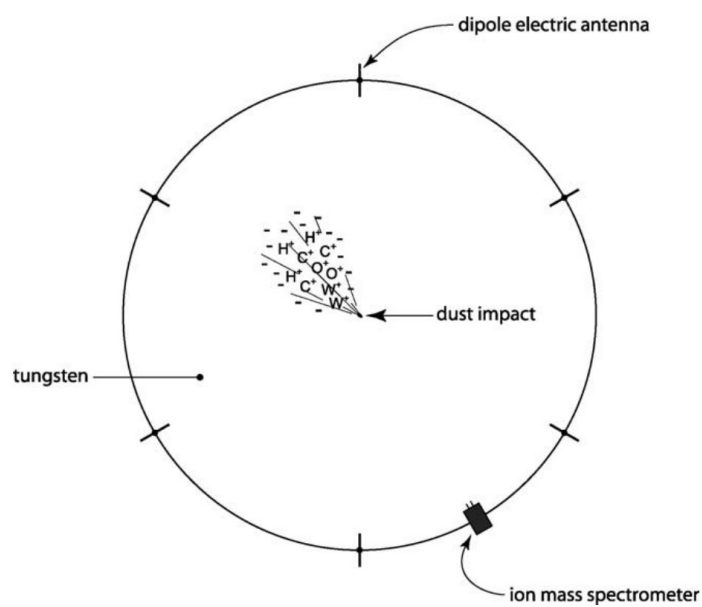

**Figure 21**. A schematic for a future dust detector for space missions. The detector is a flat circular plate with short electric dipole sensors at the edges. The figure was taken from Tsurutani et al. [769].

A new technique for not only detecting dust impacts on spacecraft but also determining their composition was developed using the electric sensor data on the DS1 spacecraft encounter with comet Borrelly [769]. The technique uses the space charge effects of the electron, proton and ion clouds created by the dust impacts. A schematic for a future space dust detector is shown in Figure 21. The advantage is that the detector has a very large cross-sectional area and the plate portion of the detection device can also serve as a spacecraft heat shield. More recently, the FIELDS electric sensors on *Parker Solar Probe* [770] studied dust in-situ via impact ionization [771-773]. Using this data set, it was determined that most of the dust observed in the inner heliosphere is a combination of $\alpha$- and $\beta$-meteoroids (bound versus unbound), where the relative proportion of the two populations observed is determined by variability in the spacecraft orbital velocity [771, 773]. Data-model comparisons [748] were used to constrain the zodiacal mass loss rate (100 - 200 kg s$^{-1}$), the source region for $\beta$-meteoroids (10-20 R$_S$), and the maximum size of dust ($\sim$50 *nm*) that contributes to the inner source pickup ion population [747]. A persistently observed two-peak dust structure (one inbound, one outbound) was observed from *Parker Solar Probe* [774]. Existing zodical cloud models that include only $\alpha$- and $\beta$-meteoroids cannot reproduce this feature [748], indicating a third unknown source of dust in the inner solar system. Data analysis and modeling of this feature suggest that it may be a $\beta$ shower, created by a collisional grinding rate enhancement that occurs when a meteoroid debris stream encounters the dense portion of the zodical cloud [748, 774]. If fully verified, this interaction represents a new process capable of significant modification to both meteoroid debris streams and the near-Sun dust populations.

*New Horizons* was launched in 2006 with the main goal to study the Pluto system during its fly-by. Indeed, it has taken unprecedented images and measurements of the Pluto and Charon surfaces and environments [775]. It is now further on its way out of the solar system, in the Kuiper belt and moving more or less towards the 'nose' of the heliosphere, at an ecliptic longitude [776] of 293° (similar to *Voyager* 2 but more or less in the ecliptic plane. At the time of writing, the spacecraft was at ca. 50 au from the Sun [776]. New Horizons carries the Student Dust Counter instrument, designed to measure interplanetary dust particles of sizes above about 0.6 $\mu$m, or interstellar dust particles of about 0.3 $\mu$m and above. The instrument consists of 14 permanently polarized polyvinylidene fluoride (PVDF) impact sensors of which two are not exposed to space but serve as a reference to help distinguishing noise from dust impact events. The total surface of exposed panels is 1.292 m$^2$. Dust flux observations to date seem to follow the currently available models for the EKB dust. The interstellar dust component will be increasingly important as the distance from the Sun increases [776].

The Juno spacecraft, launched in 2011 for investigations in the Jovian system, carries four-star cameras for attitude determination of the magnetometer system. One of these has been used to search for asteroids. This camera also detected impact ejecta from dust particles hitting the Juno solar panels [777]. The panels are large (60 m$^2$) and the spallation products are estimated to be in the order of 1-100 $\mu$m, making this method interesting for the monitoring of bigger interplanetary dust particles impacting at speeds of 5-15 km $s^{-1}$. A new model of the IDP cloud origin and structure was proposed, based on these data [778], but was refuted in Pokorný et al. [779]. Juno also carried a plasma wave instrument that detected dust impacts on the spacecraft [780], sensitive to slightly smaller particles than the camera.

## 15. Space Dusty Plasmas

Dusty plasmas are plasmas consisting of electrons, ions and charged dust grains. Dusty plasmas are observed in astrophysical and space environments, for example nebulas, interstellar clouds, cometary environments, planetary rings, planetary Moons, ionospheres and magnetospheres [781-788]. Dusty plasmas encountered in the laboratories are also called "complex plasmas". The electostatic energy between the charged dust grains is very high as the dust grains can become heavily charged, e.g., a micron-size particle can have ~1000s of elemental charges. This leads to strong electostatic coupling in dusty plasmas as compared to usual electron-ion plasmas. Therefore, it is possible to observe transitions in dusty plasmas from a disordered gaseous-like phase to a liquid-like phase. The formation of ordered structures of dust particles forming plasma crystals is also possible. An excellent review of experimental dusty plasmas is given by Fortov et al. [789].

The dust grains in the dusty plasmas are generally highly charged due to the liberation and capture of additional electrons and ions from the ambient plasma, photoemission, secondary emission, and field emission, etc. [790-794]. The presence of these massive, highly charged dust particles may significantly influence various physical processes in dusty plasmas. Usually, the acquired charges on a dust grain fluctuates and it does not remain constant. This property makes dusty plasmas different from the conventional electron-ion plasmas. Collisions and charge fluctuations in dusty plasmas can lead to momentum loss of electrons, protons and dust grains. Further, the charge to mass ratio of dust particles may significantly vary due to the accumulation or loss of electrons or protons from the dust grains. As the sizes of the dust grains are not the same, the dusty plasmas commonly have a dust mass distribution. Since the charge-to-mass ratio of a dust particle is much smaller than that of a singly charged ion or electron, the dynamics of massively charged dust occurs on much longer time scales than those associated with ion or electron dynamics. Therefore, a dusty plasma can support plasma wave modes with much smaller frequencies than the usual electron-ion plasma, the most notable are the dust-acoustic (DA) wave [795] and dust ion-acoustic (DIA) wave [796]. It is interesting to note that the DA wave is supported by the pressures of both electrons and ions providing the restoring force, whereas the inertia is provided by the charged dust grains.

Electrostatic waves in dusty plasmas having constant dust charge have been investigated by several workers [795, 797-808]. Arshad et al. [809] haves studied the dust-acoustic mode instability driven by solar wind streaming through a dusty cometary plasma. The effect of dust charge variations on the dust-acoustic wave has been studied by several authors [810-816]. It is found that dust charge fluctuations lead to damping of the dust-acoustic wave because of phase differences between the electostatic wave potential and the dust charge.

It has been shown that dust grain charge fluctuations as well as collisions can give rise to the dissipative term which can lead to the formation of DIA shock waves in a dusty plasma [817-818]. Popel and Gisko [787] have given an excellent review of the shock phenomena in dusty plasmas encountered in our solar system. Recently, there has been great interest in the study of dust and dusty plasma at the Moon [819-820]. Popel et al. [788] have reviewed the results of theoretical investigations on lunar dusty plasmas performed by Russian scientists in preparation for future Russian moon missions.

Study of dust-acoustic solitary waves and double layers has been an active area of research for the past few decades. Mamun et al. [821] showed that in a plasma system having negatively charged dust grains and non-thermal ions, dust-acoustic solitary waves of negative and positive potential can coexistence. In an unmagnetized dusty plasma consisting of negatively charged warm dust grains, non-thermal ions and Boltzmann electrons, positive potential double layers were found to limit the existence of the domain of positive dust-acoustic solitary structures from the high-Mach-number region [822-824]. The existence domain of dust-acoustic solitary waves in a two-dust system consisting of positively and negatively charged dust grains, non-thermal electrons and non-thermal ions have been considered by several workers [825-827]. The charge-to-mass ratios of the positive and negative dust grains are found to control the possible existence domains. Maharaj et al. [828] have shown the existence of positive potential dust-acoustic solitons and supersolitons (having Mach numbers greater than that of double layer) in a plasma system consisting of cold negative dust, adiabatic positive dust, Boltzmann electrons, and non-thermal ions.

It is interesting to note that, because of a large mass range of the charged dust grains, dusty plasmas allow situations where gravitational and electromagnetic forces can become comparable. Under such situations, Jeans and Buneman instabilities can arise which can have profound effect on the formation of spokes in the planetary rings, stars and galaxies [829-832].

Dusty plasmas have been found to support nonlinear dust ion-acoustic freak or rogue waves which are the rational solution of the nonlinear Schrödinger equation (NLSE) [833]. Singh and Saini [834] have shown that modulational instability of dust-acoustic waves can give rise to different kinds of dust acoustic (DA) breathers and rogue waves in a dusty plasma system comprising of negatively charged dust, Maxwellian electrons, and nonthermal ions. On the other hand, the presence of dust is shown to have a damping effect on the stellar wind driven instability [835].

In the presence of charged dust grains, the Alfvén speed is substantially reduced leading to important changes in the propagation characteristics of electromagnetic modes in a dusty plasma [836-837]. This fact has generated a lot of interest in the study of linear and nonlinear evolution of various low-frequency electromagnetic modes, e.g., namely dust Alfvén waves, dust magnetosonic waves, mixed modes, etc., in dusty plasmas encountered in space and astrophysical plasmas [838-843].

To summarize, charged dust grains are ubiquitous in most space, astrophysical and laboratory plasmas. The dusty plasmas give rise to a variety of new low-frequency dust plasma waves and instabilities. The presence of these new modes can lead to the occurrence of new phenomena in interplanetary space, the interstellar medium, interstellar or molecular clouds, comets, planetary rings, and the Earth's environment.

## 16. Solar Energetic Particles, Shocks and the Heliospheric Termination Shock

The properties of space plasma and energetic particles and the interactions between both depend on where in the heliosphere they are detected. Energetic particles observed at Earth orbit are mostly messengers from remote locations and bear the imprints of these interactions. Within this mix of particles different sources and prevalent interaction mechanisms can be identified in them. Scientists are well aware that acceleration and propagation mechanisms are not exclusive but can, albeit with different contributions, coexist.

High fluxes of ~1 MeV to ~5 GeV energetic particles are created during solar flares [844-851]. Energetic protons, helium, carbon, nitrogen, oxygen, and high Z ions, and relativistic electrons are created by several different processes. Ion acceleration at the flare/CME release magnetic reconnection site has been proposed by Freier and Webber [852] and Kallenrode [853]. Various mechanisms for relativistic electron acceleration have been proposed, including stochastic acceleration by cascading magnetosonic waves ([854]; see review by Vilmer [855]) and reconnection-related stochastic acceleration [856-860].

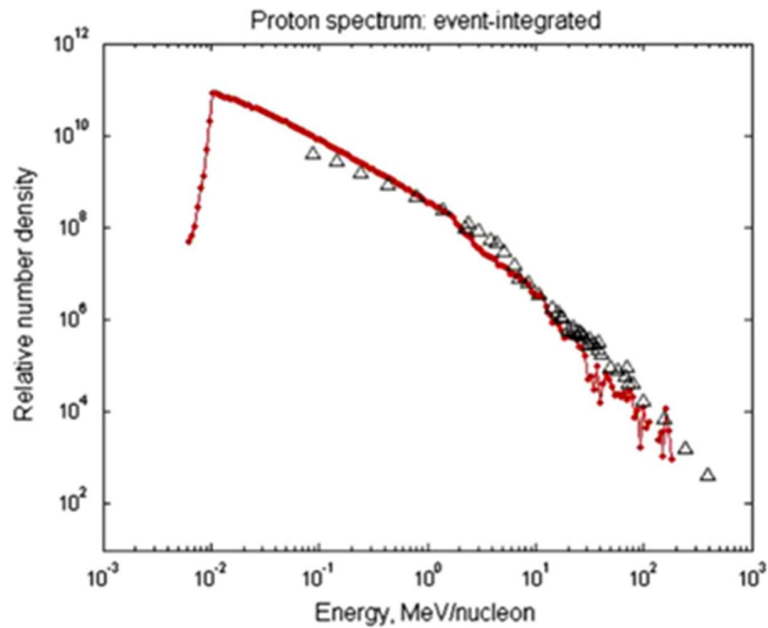

**Figure 22**. Event-integrated proton spectrum for the 13 December 2006 SEP event. Here the PATH (Particle Acceleration Throughout the Heliosphere: [861]) code results are shown by the solid line. Fluences obtained from ACE, STEREO, GOES-11 and SAMPEX observations are shown by triangles. The figure is taken from Verkhoglyadova et al. [862].

There are two different intensity-time profiles of the energetic ion fluxes. Those with relatively sharp rises, called "prompt" or "impulsive" events [e.g., 863-864]. Those energetic ion events with slower rises are called "gradual" events. Prompt events typically correspond to ions accelerated near the Sun at the flare site, whereas gradual events are larger, exhibit different compositional properties than impulsive events, and are associated with interplanetary coronal mass ejection events [e.g., 851]. An example of an observed spectrum of solar energetic protons is illustrated in Figure 22 showing both the observed and theoretically predicted event-integrated spectrum for the 13 December 2006 SEP event. The distinction between the two classes is not always clear-cut [865] and shape alone is inconclusive: a series of four impulsive events at 0.3 AU appears as one gradual event at 1 AU [866]. These are often described as "mixed events" and exhibit characteristics of both impulsive and gradual events [867-868]. The upcoming observations by *Parker Solar Probe* and *Solar Orbiter* certainly will add important new aspects to the distinctions.

Solar energetic particles (SEPs) stream through interplanetary space along interplanetary magnetic field lines and can enter the polar regions of the Earth's atmosphere [869-872]. Energetic protons lose their kinetic energy by ionization of upper atmospheric atoms and molecules at heights of ~80 km down to ~50 km above the surface of the Earth. The atmospheric ionization absorbs radio waves and thus transpolar ionospheric radio communications are blocked. These communications outages are called polar cap absorption events, or PCAs. In addition, the ionization affects atmospheric chemistry, in particular the depletion of ozone [873].

When the flare protons reach energies > 100 MeV, through a nuclear cascade process, neutrons and other charged daughter particles can reach ground level creating "ground level events" or GLEs [844, 874-877] with even larger consequences for atmospheric ionization [878].

*Pioneers* 10 and 11 were the first spacecraft to fly to the outer heliosphere. McDonald et al. [879] noted that the ~0.1 to 10 MeV/nucleon particle fluxes were increasing by an order of magnitude as the spacecraft traveled from 1 to 4 au from the Sun. See also Van Hollebeke et al. [880], Barnes and Simpson [881], Tsurutani et al. [882], and Pesses et al. [573, 883]. These particles must have been accelerated locally in interplanetary space! The obvious idea was that interplanetary shocks at the edges of CIRs were somehow involved with the acceleration process. Smith and Wolfe [72] showed that CIR fast forward and fast reverse shocks could form from ~1.5 to ~2.5 AU from the Sun and then persist to 4 AU and beyond. These shocks would continuously accelerate energetic particles as the CIRs and their attached shocks slowly propagated towards the outer heliosphere. Sometimes the particle peaks were located at the shocks and sometimes they were not. Tsurutani et al. [41] developed a statistical test and applied it to ~0.5 to 1.8 MeV ion fluxes at 123 CIRs detected between 1 and 6 AU, finding that indeed particle peaks were statistically correlated to the shocks. *Ulysses'* observations showed that CIRs and particles accelerated at CIRs extend to even higher solar latitudes [e.g., 884].

International Sun Earth Explorer 3 (ISEE-3) experimenters (E.J. Smith, R.D. Zwickl, J.T. Gosling and B.T. Tsurutani) were tasked by the NASA/ESA ISEE Project to try to determine the properties of shocks (Mach numbers and shock normal angles) upstream of fast ICMEs from in situ interplanetary magnetic field and plasma measurements (shock determination is discussed in Chapter 13 of this paper). This two-year list of shocks was used by the full ISEE science team for their data analyses. Tsurutani and Lin [574] used the list to identify shock acceleration of > 2 keV electrons and > 47 keV ions. Kennel et al. [666-667] presented a comprehensive examination of shock acceleration of 1-6 keV protons and electrons and > 30 keV/Q ions at a supercritical quasi-parallel interplanetary shock.

Reames [575, 851] argued convincingly that the preponderance of solar energetic particles observed at 1 au are due to diffusive shock acceleration associated with fast CMEs/ICMEs and the subsequent upstream escape into the interplanetary medium of energetic ions. Shocks first form at 2 to 5 $R_s$ from the Sun upstream of fast CMEs [101] and continue to accelerate energetic particles as the shock/ICME propagates to 1 AU and beyond. The accepted mechanism for energetic ion acceleration at parallel shocks is first order Fermi acceleration [664-665, 885-888]. For perpendicular shocks, particle gradient drift along the shock surface can energize ions [662, 889-890]. For quasi-perpendicular and quasi-parallel shocks (> 95% of all cases), both gradient drift and diffusive shock acceleration [891-893] have been identified as acceleration mechanisms. Models examining shock acceleration of energetic ions contain both mechanisms.

Modeling energetic ion acceleration is an extremely difficult task. The shock Mach number, shock normal angle, and upstream (seed) environment varies with distance from the Sun. Additionally the upstream and downstream waves vary as a consequence of the evolving shock and seed properties. Progress on this difficult task have been given in Zank et al. [861], Rice et al. [894] and Li et al. [895]. An excellent textbook summarizing interplanetary energetic particles can be found in Kallenrode [853].

**The Heliospheric Termination Shock**

The heliospheric termination shock (HTS) is the largest shock wave in the heliosphere, decelerating abruptly the unimpeded expanding supermagnetosonic solar wind to a subsonic flow that is heated and compressed, which forms the heliosheath. *Voyager* 1 crossed the HTS in December 2004 in the northern hemisphere when the spacecraft was at 94 AU from the Sun [896-899]. *Voyager* 1 did not directly observe the HTS since the crossing occurred during a data gap.

*Voyager* 2, following a southern trajectory, crossed and observed the termination shock in August 2007 at a distance of 84 AU [900-904], providing the first magnetic and plasma measurements of the HTS itself, and the first plasma measurements of the heliosheath. The observed HTS was observed to be comparatively weak (compression ratio ~1.7 in one case) and highly perpendicular with as many as 5 crossings observed. The multiple crossings of the HTS by *Voyager* 2 suggest that the HTS was in continual motion, possibly moving back and forth across *Voyager* 2. Alternatively, or as well as, the HTS may have ripples sweeping along the shock front. As illustrated in Figure 23, at least one crossing resembled the structure of a proto-typical perpendicular shock with a well-defined foot, ramp, and overshoot, the difference being that the reflected ions were almost exclusively reflected pickup ions, as was predicted already in 1996 [905], and the downstream thermal plasma was only ~180,000 K [903]. Most of the shock dissipation energy went into the heating of pickup ions, making the HTS quite different compared to inner heliospheric shocks. Both the *Voyager* 1 and 2 HTS crossing revealed the presence of an extended foreshock and instabilities, due to both backstreaming MeV protons and energized pickup ions. This situation is quite different than the easily identified sudden transitions at ICME or CIR shocks at 1 AU. The HTS crossing and related results are summarized extensively in Zank [906].

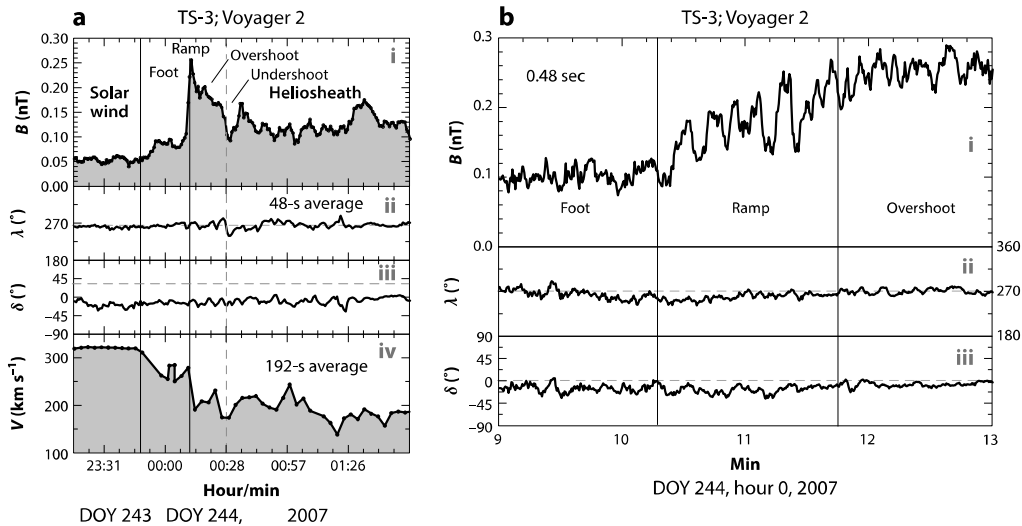

**Figure 23**. Structure of the heliospheric termination shock as observed by *Voyager* 2. (*a*) (*i*) Magnetic field strength *B* measured using 48-s averages, (*ii*) its direction λ, (*iii*) elevation angle δ, and (*iv*) 192-s averages of the solar wind speed *V* across the third of the heliospheric termination

shock crossings TS-3 by *Voyager* 2. Clearly visible are an extended foot, a ramp, the magnetic field overshoot, and trailing oscillations. (*b*) (*i*) The internal structure of the ramp of TS-3, based on observations of the magnetic field strength *B*; (*ii*) azimuthal angle λ; and (*iii*) elevation angle δ at 0.48-second intervals. The figure is taken from Burlaga et al. [900].

Finally, the different distances at which *Voyagers* 1 and 2 crossed the HTS suggest an asymmetric three-dimensional termination shock and heliosphere, and indeed the inferred orientation of the local interstellar magnetic field is consistent with such an inference. The three-dimensional configuration of the shock and heliosphere is believed to have a blunt (nonspherical) front-side shape [907-908]. Energetic particle observations are consistent with this picture. A recent review of the termination shock can be found in Jokipii [909].

## 16. Final Comments

Each chapter of this review were written by 2 and sometimes 3 or more different experts so that a balance point-of-view could be obtained. We hope we have been able to present the readership with a reasonable objective and accurate accounting of Space Plasma Physics/Space Physics/space weather.

## Acknowledgements

Portions of the work were performed at the Jet Propulsion Laboratory, California Institute of Technology under contract with NASA. G.P. Zank acknowledges the partial support of a Parker Solar Probe contract SV4-84017 and an NSF EPSCoR RII-Track-1 cooperative agreement OIA-1655280 and OIA-2148653. V.J. Sterken received funding from the European Union's Horizon 2020 research and innovation programme under grant agreement N 851544. GSL thanks the Indian National Science Academy for the support under the INSA-Honorary Scientist Scheme. Richard B. Horne was supported by NERC Highlight Topic Grant NE/PO1738X/1 (Rad-Sat) and NERC National Public Good activity grant NE/R0164451. The work of R.H. is funded by the Science and Engineering Research Board (SERB, grant No. SB/S2/RJN-080/2018), a statutory body of the Department of Science and Technology (DST), Government of India through the Ramanujan

Fellowship. P.F. Chen was supported by the National Key Research & Development Program of China (2020YFC2201200),

American Geophysical Union (AGU) and the Solar Physics Division (SPD) of the American Astronomical Society. https://connect.agu.org/tess2018/home

To appear in IEEE Transactions on Plasma Science

**Biosketches and photos of all authors**

**Bruce T. Tsurutani (ORCID: 0000-0001-7299-9835)**

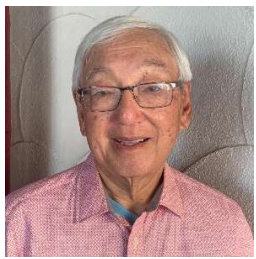

Bruce T. Tsurutani, PhD in physics, University of California at Berkeley (1972). Scientist at the Jet Propulsion Laboratory (JPL), California Institute of Technology, Pasadena, California from 1972 to 2019. At JPL he did scientific research, design of space instrumentation and design of spacecraft. Currently Emeritus at JPL. Dr. Tsurutani has been a Visiting Professor, Visiting Associate, or Adjunct Professor at the University of Cologne, Germany, Caltech, Pasadena, California, the Technical University of Braunschweig, Braunschweig, Germany, Kyoto University, Kyoto, Japan, the University of Southern California, Los Angeles, and the University of Alaska, Fairbanks, Alaska.

Dr. Tsurutani has published 700+ scientific articles, edited or coedited 8 books, organized or co-organized 5 American Geophysical Union (AGU) Chapman Conferences, 8 magnetic storm workshops and 10 Nonlinear Wave and Chaos workshops. He has published 10 NASA technical briefs. Tsurutani has an H-index of 96 an i10 index of 439 and 36,326 citations. Tsurutani has served as President and President-Elect of the Space Physics and Aeronomy Section of the AGU (1988-1992). Under his presidential term he established the James A. Van Allen, Marcel Nicolet and Eugene N. Parker AGU lecture series and the Fredrick L. Scarf PhD. thesis prize.

Among Bruce Tsurutani's honors are the International Kristian Birkeland medal for Space Weather and Space Climate (2019), the COSPAR Space Science Medal (2018), Asteroid 6313 named Tsurutani (2018), the John A Fleming (AGU) medal (2009), AGU Fellow (2009), the Latin American Geophysical Society inaugural Gold Medal (2001), NASA Exceptional Service medals (2001, 1985), a Von Humboldt Research Fellowship (1993-1995), the Brazilian National Space (Werner Von Braun) medal (1992), 10 NASA/ESA spacecraft instrument achievement awards and 11 "excellence in refereeing" awards (JGR, Wiley, EOS, Solar Physics, JASTP, and GRL). Dr. Tsurutani acknowledges the much appreciate mentoring that he had from scientific giants in the field: S.-I. Akasofu, H. Alfvén, K.A. Anderson, J.W. Dungey, S. E. Forbush, A. Hasegawa, C.F. Kennel, T. Obayashi, J.A. Simpson, E.J. Smith and J.A. Van Allen. Email: bruce.tsurutani@gmail.com

**Gary P. Zank (ORCID: 0000-0002-4642-6192)**

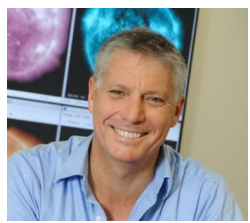

Gary P. Zank (B.Sc (Hons.), PhD, University of Natal, South Africa) is a space plasma physicist who works on the physics of the solar wind, especially its interaction with the local interstellar medium, the acceleration and transport of energetic particles, turbulence, and shock waves. He is currently the first University of Alabama Board of Trustees Trustee Professor, and holds the Aerojet/Rocketdyne Chair in Space Science. He is also an Eminent Scholar and Distinguished Professor in the Department of Space Science at the University of Alabama at Huntsville (UAH), of which he is Chair and founder, and Director of the Center of Space Physics and Aeronomic Research at UAH. He was a Max-Planck Post-Doctoral Fellow in Germany and Bartol Research Institute Post-Doctoral Fellow, before joining the faculty of the Bartol Research Institute of the University of Delaware. Prior to his joining UAH in 2008, Zank was the Chancellor's Professor of Physics and Astronomy at the University of California, Riverside. He was the System-wide Director of the Institute of Geophysics & Planetary Physics at the University of California and the campus Director of the Institute of Geophysics & Planetary Physics at the University of California, Riverside.

Dr. Zank has published in excess of 650 publications in space and solar physics, plasma physics, and astrophysics, has edited 23 books, and published two textbooks, 7 National Research Council publications. His current h-index is 84 (Google Scholar), his i10-index is 343, and his approximate number of citations is 24,434. Dr. Zank is the recipient of numerous honours including being a Fellow of the American Physical Society, the American Association for the Advancement of Science, the American Geophysical Union, an Honorary Member of the Asia, Oceania Geosciences Society (AOGS), the Zeldovich Medal (COSPAR), the Axford Medal (AOGS), NASA Silver Achievement Medal, and the 2018 AAS Neil Armstrong Space Flight Achievement Award. He was elected a Member of the US National Academy of Sciences in 2016. Email: garyp.zank@gmail.com

**Veerle J. Sterken (ORCID: 0000-0003-0843-7912)**

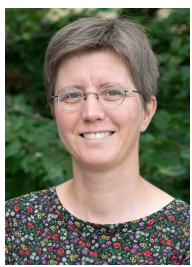

Photo credits: ETH Zürich/D-PHYS/Heidi Hostettler

Veerle J. Sterken (Belgian), M.Sc in Aerospace Engineering (2005) at Delft University of Technology, the Netherlands, and PhD in geophysics (2012) at the Technical University (TU) of Braunschweig, Germany. Scientist and group leader at the Eidgenössische Technische Hochschule Zürich (ETHZ), Switzerland since 2020. Principal Investigator of the European Research Council Starting Grant ASTRODUST N° 851544.

Dr. Sterken worked for the European Space Agency (ESA) on space systems engineering, and for TU Delft, ESA, and Thales Alenia Space (Cannes) on the Darwin space interferometry mission. During and after her PhD, she focused on the dynamics of Interstellar Dust in the solar system, while staying at the Max Planck Institute for Nuclear Physics in Heidelberg (Germany) and at the International Space Science Institute in Bern (Switzerland). She also performed instrument calibration experiments, co-authored several ESA mission proposals, organized workshops and was co-editor of two scientific books. After having worked for the University of Bern in the field of precise satellite orbit determination, and subsequently in science administration (Swiss National Science Foundation), she returned to academia with an ERC Starting Grant for research on Interstellar Dust and heliosphere sciences. She received a few prizes, among which the Amelia Earhart award (2007) and the Christophe Plantin prize (2021). Email: vsterken@ethz.ch

**Kazunari Shibata (ORCID: 0000-0003-1206-7889)**

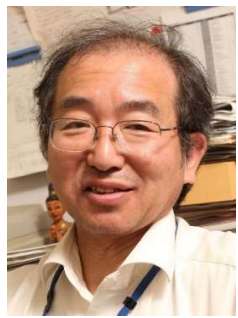

Department of Environmental Systems Science, School of Science and Engineering, Doshisha University, Kyoto 610-0394, Japan. Kazunari Shibata is an emeritus professor at Kyoto University, Japan. He got PhD in astrophysics from Kyoto University in 1983. His research covers magnetic reconnection, solar flares, jets, active galactic nucleus, accretion disk dynamics, etc. He has published more than 360 papers in refereed journals, with more than 20000 citations (h-index of 76). He served as the director of Kwasan & Hida Observatories, Kyoto University from 2004 to 2018. He received the 2009 NISTEP Award, 2001 Chushiro Hayashi prize, 2019 S. Chandrasekhar Prize of Plasma Physics, and 2020 George Ellery Hale Prize, and the 2021 Kristian Birkeland Medal. Email: shibata@kwasan.kyoto-u.ac.jp

**Tsugunobu Nagai (ORCID: 0000-0002-3238-864X)**

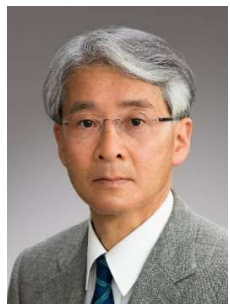

Tsugunobu Nagai (Tokyo, Japan) received the B.S. and M.S. degrees in astronomy from the University of Tokyo. He worked at Kakioka Magnetic Observatory from 1977 to 1981. He received the Ph.D. degree in geophysics from the University of Tokyo in 1981.

From 1981 to 1983, he was an NRC research associate at NASA/Marshall Space Flight Center (under Dr. C. R. Chappell). At Meteorological Research Institute (Japan Meteorological Agency) from 1983 to 1993, he developed the space weather forecast (prediction of MeV electron intensity at geosynchronous orbit, 1988) and the atmosphere-ocean coupled model for El Niño (1992). He focused on space physics at Tokyo Institute of Technology (1993–2018) and at Institute of Space and Astronautical Science (2018–2022). He was editor of Geophysical Research Letters (AGU) (1995–1997) and PI of magnetic field experiment (MGF) on the Geotail spacecraft from 1999.

Dr. Nagai is a Fellow of American Geophysical Union (2008) and Japan Geoscience Union (2019). Email: nagai@stp.isas.jaxa.jp

**Anthony J. Mannucci (ORCID: 0000-0003-2391-8490)**

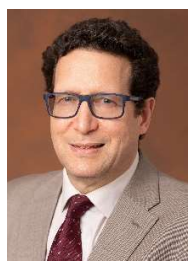

Dr. Anthony J. Mannucci received the B.A. degree in physics (with honors) from Oberlin College in 1979 and the Ph.D. degree in physics from the University of California, Berkeley in 1989.

From 1999 to 2018, he served as Group Supervisor of the Ionospheric and Atmospheric Remote Sensing group at NASA's Jet Propulsion Laboratory. He is currently Deputy Manager of the Tracking System and Applications Section (335) at JPL. He is a Principal Member of the Technical Staff, and a Senior Research Scientist. Dr. Mannucci's research interests include Earth and space science remote sensing applications with signals of opportunity such as GPS. Dr. Mannucci co-developed the widely used Global Ionospheric Mapping technique, and is co-inventor of the rate-of-TEC-index (ROTI) used to monitor ionospheric irregularities using GPS. His scientific focus areas include ionospheric behaviour during large geomagnetic storms and during high-speed solar wind streams. Dr. Mannucci played a major role in developing the Federal Aviation Administration's GPS navigation system for aircraft (called WAAS) and he is a charter member of the WAAS Integrity Performance Panel. He is lead author of book chapters on GNSS radio occultation (2021), the ionosphere and thermosphere responses to extreme geomagnetic storms (2017), and a chapter in the 1999 URSI Reviews of Radio Science on the use of GPS receivers for ionospheric measurements. Dr. Mannucci was lead organizer of the Chapman Conference "Scientific Challenges Pertaining to Space Weather Forecasting Including Extremes" held in February, 2019. He co-organized the 2007 Chapman Conference on "Mid-Latitude Ionospheric Dynamics and Disturbances" and is co-editor of the associated American Geophysical Union Monograph Volume 181. He has published more than 144 peer-reviewed journal articles with an h-index of 44 (Web Of Science).

Dr. Mannucci is a member of the American Geophysical Union and the IEEE. He serves as Associate Editor of the IEEE journal Transactions on Geoscience and Remote Sensing. He is the recipient of several NASA Space Act awards (2004-2013) and holds patents pertaining to the design of differential GPS systems and using GPS data for remote sensing purposes. Email: Anthony.j.mannucci@jpl.nasa.gov

**David M. Malaspina (ORCID: 0000-0003-1191-1558)**

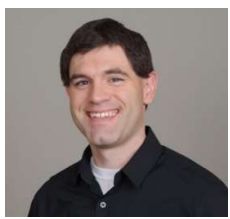

David M. Malaspina, Astrophysical and Planetary Sciences (APS) department, University of Colorado, Boulder, USA. David M. Malaspina is an assistant professor in the Astrophysical and Planetary Sciences (APS) department at the University of Colorado, Boulder (CU), and a research scientist at the Laboratory for Atmospheric and Space Physics (LASP). He received his PhD in Physics from the University of Colorado, Boulder (2010). His research focuses on data analysis-based study of plasma wave-particle interactions and plasma-spacecraft interactions in the solar wind and planetary magnetospheres. He has published more than 200 papers in peer-reviewed journals.

In pursuit of his research goals, Dr. Malaspina designs, builds, and operates scientific instruments for spacecraft, focusing on the measurement of electric and magnetic fields. He has carried out critical aspects of the design, development, and operation of spaceflight instruments, including the Van Allen Probes / Electric Fields and Waves (EFW), Magnetospheric Multiscale mission (MMS) / FIELDS, Parker Solar Probe (PSP) / FIELDS, Geospace Dynamics Constellation (GDC) / AETHER, the Lunar Surface Electromagnetics Experiment (LuSEE), the CANVAS cubesat electric field instrument, and the Rapid Active Plasma Sounder (RAPS) instrument on the COUSIN sounding rocket. He led the recent Plasma Imaging LOcal and Tomographic experiment (PILOT) mission concept study. Email: david.malaspina@lasp.colorado.edu

**Gurbax S. Lakhina (ORCID: 0000-0002-8956-486X)**

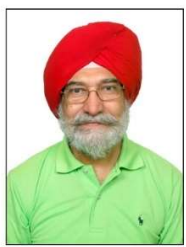

Gurbax S. Lakhina graduated from Panjab University, Chandigarh, and got his Master and Ph.D. degrees in Physics from Indian Institute of Technology, Delhi. He is a Former Director of Indian Institute of Geomagnetism (IIG), Mumbai, India (1998 – April 2004). After his retirement, he became CSIR Emeritus Scientist, then INSA Senior Scientist, then NASI-Senior Scientist Platinum Jubilee Fellow, and currently an INSA-Honorary Scientist at the IIG, Navi Mumbai, India. He has been a Visiting Scientist at Instituto Nacional de Pesquisas Espaciais (INPE) (September – October 2012), NASA Goddard Space Flight Center and Astronomy Dept. University of Maryland (August – September 2008), Visiting Professor at RISH, Kyoto University, Japan (February – July 2005), STEL, Nagoya University, Japan (September – December 2004). He was a Senior NRC Associate at Jet Propulsion Laboratory, Pasadena, USA (September 1996 – June 1998). Before that Prof. Lakhina was a Research Scientist at Courant Institute of Mathematical Sciences, New York University (1988 – 1989) and a Humboldt Fellow at the Ruhr Universität Bochum, Bochum, Germany (1981 – 1982, 1985).

Prof. Lakhina has served as Chairman of *Indian Joint National Committee for COSPAR-URSI-SCOSTEP(*2016-2019), Member AGU Ambassador Award Committee (2016 – 2017), Co-Chairman of *COSPAR Scientific Commission D (*2012-2021),  Associate Editor of *Advances in*

*Space Research*, Elsevier (2008 – present), Associate Editor of *JGR-Space Physics*, AGU( 2006-2007), Chairman of COSPAR Commission D-3 (2004-2012), Member of the Executive Council of the International Association of Geomagnetism and Aeronomy (1999 – 2007).

Prof. Lakhina's main areas of scientific interest are: linear and nonlinear waves in space and astrophysical plasmas, solar wind interaction with magnetospheres, magnetic storms and Space Weather. He has written 353 research papers; 292 papers are in peer reviewed journals and 61 are in Books/Proceedings. Prof. Lakhina has h-index of 49, i10-index of 182 and 8031 citations (https://scholar.google.com/citations?hl=en-US&user=Qq1Y2yEAAAAJ).

Among Prof. Lakhina's honours are the 2015 AGU Space Weather and Nonlinear Waves and Processes Prize, 2014 COSPAR Vikram Sarabhai Medal, the Kalpathi Ramakrishna Ramanathan Medal (2005) of Indian National Science Academy (INSA), New Delhi, the Decennial Award-Gold Medal 2000, Indian Geophysical Union, Hyderabad, India, Senior Associate, International Centre for Theoretical Physics Trieste, Italy (1997 – 2002), NASA Senior Resident Research Associateship, National Research Council, USA (1996 – 1998), and Alexander von Humboldt Research Fellowship, Germany (1981 – 1982 and in 1985). Email: gslakhina@gmail.com

**Shrikanth G. Kanekal (ORCID: 0000-0001-8157-4281)**

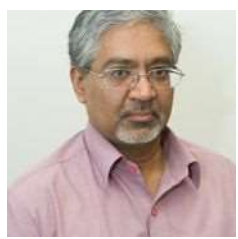

Dr. Kanekal's research interests include energization and loss processes of relativistic electrons in the Earth's magnetosphere, solar energetic particles, Jovian electrons, magnetospheric energetic particle boundary dynamics, space Instrumentation, and space weather. He was involved in several NASA missions including the HERMES/Lunar Gateway, Van Allen Probes as well as two CubeSats, CeREs, and CusP. He is currently building the MERiT (Miniaturized Electron Proton Telescope), for the Lunar Gateway.

On Van Allen Probes, he has been involved in the design and calibration of the REPT (Relativistic

Electron Proton Telescope) instrument and made significant contributions to refining science objectives, instrument requirements, and REPT hardware. He was fully responsible for the GEANT4 simulation of REPT and developed software to simulate the detector geometry with high fidelity. He has also used GEANT4 simulations for optimization of event trigger types, estimation of backgrounds due to inner belt protons, GCRs, ultra-relativistic electrons and impact of magnetometer boom within field of regard. He has directed and participated in the REPT beam test calibration at UC Davis, Indiana cyclotron facility, Lawrence Berkeley Laboratories for protons and at the Idaho Electron accelerator facility for electrons.He has done extensive work in cross calibrating REPT with MagEIS (Magnetic Electron Ion Spectrometer) sensor.

Dr. Kanekal's involvement in scientific analyses of energetic particle data obtained from SAMPEX since its launch in 1992 has been extensive. He has been fully responsible for writing the browse and level-2 data software and participated in calibrating the PET sensor onboard SAMPEX. He continues to carry on science analyses using SAMPEX data. He has been the PI of a NASA grant to make the SAMPEX data (2004 onwards) available to the public by expanding the SAMPEX data server. On the Polar mission, he has been extensively involved with the CEPPAD suite of instruments. He has used energetic electron and proton data from sensors onboard SAMPEX and Polar to study various physical phenomena in the Earth's radiation belts, such as electron energization and loss, trapping of solar protons, access of low energy cosmic rays and space weather.

Dr. Kanekal's Ph.D. research involved precision measurements of the hadronic branching fractions $\tau \to K^* \upsilon_\tau$ and $\tau \to \rho\upsilon_\tau$, which resulted in the first measurements of a fundamental constant of nature, the Cabbibo angle using $\tau$ lepton decays. His research included decay modes of charmed mesons, the $\tau$ lepton, the $\Upsilon$ resonances, as well as detector simulation software. Dr. Kanekal has extensive experience in different types of particle detectors and associated detection techniques. As a postdoctoral fellow, he maintained the detector simulation package of the CUSB detector at CESR, the Cornell electron positron storage ring; at Fermilab, he was a member of the D0 team, where he carried out Geant modeling of D0, development of silicon vertex detector software, and physics analysis of hadronic decay modes of the top quark. Email: shrikanth.g.kanekal@nasa.gov

**Keisuke Hosokawa (ORCID: 0000-0001-7389-7128)**

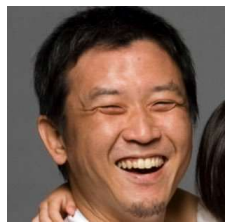

Graduate School of Communicating Engineering and Informatics, University of Electro-Communications, Tokyo, Japan. Dr. Keisuke Hosokawa completed a doctorate in solar-terrestrial physics at Department of Geophysics, Graduate School of Science, Kyoto University. He is now a professor in Department of Communication Engineering and Informatics, The University of Electro-Communications, Tokyo. His research interest covers various phenomena in the planetary upper atmosphere including aurorae, ionospheric waves and plasma bubbles. His team is currently operating a number of radio and optical instruments in the world, for example in high Arctic region. He has published more than 170 papers in refereed journals, with more than 2600 citations (h-index of 28). He received the Young Career Award of the Society of Geomagnetism and Earth, Planetary and Space Sciences in 2009. He has been serving as a Topical editor for Annales Geophysicae since 2012. Email: <u>keisuke.hosokawa@uec.ac.jp</u>

**Richard B. Horne (ORCID: 0000-0002-0412-6407)**

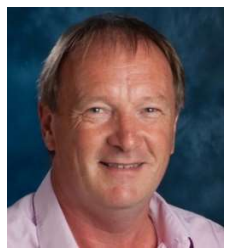

Professor Richard B. Horne, MA, ScD, FRS is Head of Space Weather at the British Antarctic Survey and Honorary Professor at the University of Sheffield. He has published over 200 research papers on wave-particle interactions, wave propagation and space weather. He led the EU SPACECAST project (2011-2014) to develop a space weather forecasting system for satellites, and the EU SPACESTORM project (2014-2017) which showed that the risk to satellites from space weather is much higher than previously thought. Richard's work led to revised hazard assessments for the UK National Risk Register of Civil Emergencies in 2017 and 2020. He is Chair of the Space Environment Impacts Expert Group which provides advice on space weather to the UK government. He has also served as Vice President of the Royal Astronomical Society (1997-1999), and Chair of Commission H for the International Union of Radio Science (2005-2008).

Richard was elected Fellow of the Royal Society in 2021. He is also a Fellow of the American Geophysical Union (2011), the International Union of Radio Science (2017), The Royal Astronomical Society (1981) and St Edmund's College Cambridge (2013). Richard was awarded the Gold Medal from the Royal Astronomical Society (2022), the International Kristian Birkeland Medal for space weather and space climate (2020), the URSI Appleton Prize (2020), and Doctor of Science from the University of Cambridge (2020) for distinguished research. Email: rh@bas.ac.uk

**Rajkumar Hajra (ORCID: 0000-0003-0447-1531)**

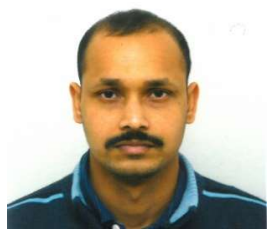

Rajkumar Hajra received his Master of Science in physics, and PhD in space science from Calcutta University, India (2012). He has been awarded postdoctoral fellowships by *Fundação de Amparo à Pesquisa do Estado de São Paulo* (FAPESP), Brazil to work at Instituto Nacional de Pesquisas Espaciais (INPE), Brazil (August 2012 –January 2015, March 2015 – September 2015) and Arecibo Observatory, Puerto Rico (December 2014 – March 2015), by *Agence Nationale de la Recherche* (ANR), France and *Centre National d'Etudes Spatials* (CNES), France to work at Laboratoire de Physique et Chimie de l'Environnement et de l'Espace (LPC2E), Centre National de la Recherche Scientifique (CNRS), France (March 2016 – October 2018). Presently he is a *Ramanujan fellow* of *Science & Engineering Research Board (SERB), Department of Science & Technology (DST), Government of India*, India, working at Indian Institute of Technology Indore (IITI), India. Rajkumar's scientific interest is interplanetary space weather involving near-Earth plasma, comets, and planets. He has published 66 scientific papers in refereed journals, 1 NASA Technical Briefs, 2 book chapters. Dr. Hajra has an h-index of 20, an i10 index of 41, and 1244 citations. Email: rajkumarhajra@yahoo.co.in

**Karl-Heinz Glassmeier (ORCID: 0000-0003-4327-5576)**

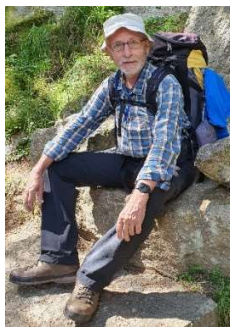

Karl-Heinz Glassmeier studied geophysics, physics, philosophy, and economy at the University of Muenster, Germany. He received his Diploma in physics 1979 and his PhD in geophysics in 1985. From 1985 until 1991 he worked as an Ass. Prof. (equiv.) at the University of Cologne, where he finished his Habilitation in geophysics in 1989. From 1991 until 2020 he was Professor of Geophysics and Chair at the Technische Universitaet Braunschweig. He was Principal and Co-Investigator of space missions like Giotto, Cluster, Cassini, Themis/Artemis, VenusExpress, Rosetta, BepiColombo, and Juice. His main research areas are planetary magnetic fields and magnetospheres, waves and turbulence in space plasmas, cometary physics, data analysis methods, and history of science. In 1990 he was awarded the Yakov B. Zeldovich Medal of COSPAR, in 2010 the Julius-Bartels Medal of the European Geoscience Union, and in 2014 the Basic Science Award of the International Academy of Astronautics. In 2018 he became a Fellow of the American Geophysical Union. In 2021 asteroid *2000 GQ141* was renamed into *27506 Glassmeier*. Karl-Heinz Glassmeier served as Editor and Co-Editor of Annales Geophysicae, Geophysical Research Letters, and Reviews of Geophysics. From 2011-2020 he was member of the Board of Reviewing Editors *of Science*. He served as a member of ESA's *Space Science Advisory Committee* and Vice-President of COSPAR. He is author and co-author of more than 500 referred scientific papers. He edited or co-edited six scientific books. His bibliometric information reads: h-index 79, i10-index 417, 27200 citations. Email: kh.glassmeier@tu-bs.de

**C. Trevor Gaunt (ORCID: 0000-0002-4531-1430)**

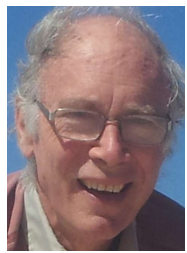

C. Trevor Gaunt B.Sc Eng. (Elec.) Natal, MBL South Africa, PhD Cape Town is an Emeritus Professor and Senior Scholar in the Department of Electrical Engineering at the University of Cape Town. He is the principal investigator on research funded by the Open Philanthropy Project to investigate the mitigation of effects of geomagnetically induced currents that can introduce extreme distortion into power systems. Email: ct.gaunt@uct.ac.za

**Peng-Fei Chen (ORCID: 0000-0002-7289-642X)**

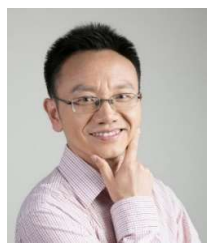

Peng-Fei Chen, School of Astronomy and Space Science, Nanjing University, China. Peng-Fei Chen is a professor at Nanjing University, China. He got PhD in astrophysics from Nanjing University in 1999. His research covers magnetic reconnection, solar flares, coronal mass ejections, EUV waves, and solar filaments, etc. He has published more than 150 papers in refereed journals, with more than 4300 citations (h-index of 34). He received the Young Career Award of the Asia-Pacific Solar Physics in 2017, and served as a member of the Steering Committee of the Solar and Heliosphere Division of IAU (2013-2015), the vice chair of the E2 Subcommission of COSPAR (2018-) and a science committee member of the International Space Science Institute (2019-2022). Currently he is an associate editor of the journals *Reviews of Modern Plasma Physics*, and *Acta Astronomica Sinica*, a scientific editor of *Science China Physics Mechanics Astronomy*, and *Universe*. Email: chenpf@nju.edu.cn

**Syun-Ichi Akasofu (ORCID: 0000-0003-4579-1558)**

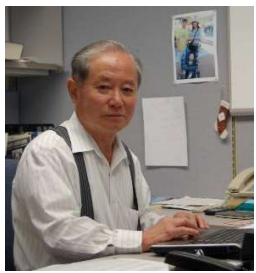

Dr. Syun-Ichi Akasofu is a professor of physics and director emeritus of the University of Alaska Fairbanks. He was the director of the International Arctic Research Center at the University of Alaska Fairbanks since its establishment in 1999 until his retirement in 2007. Prior to that, he was director of the UAF's Geophysical Institute for 13 years from 1986 to 1999. He helped establish the institutes as a key research center in the Arctic, and played a critical role in the genesis of the Alaska Volcano Observatory and the modernization of the Poker Flat Research Range.

Akasofu came from Japan to the University of Alaska Fairbanks in 1958 as a graduate student to study the aurora under the guidance of Sydney Chapman, receiving his PhD in 1961. He has been a professor of geophysics since 1964. Akasofu has written more than 550 professional journal articles and authored or co-authored 11 books. Akasofu is an expert on the aurora borealis and solar-terrestrial physics. His paper on the auroral substorm in 1964 is still often cited.

In 1976, the Royal Astronomy Society of London presented Akasofu with its Chapman Medal. In 1980, UAF named him a Distinguished Alumnus. In 1981, he earned mention as one of the "1000 Most Cited Scientists" in all fields of science, and in 2002, one of the most cited authors in space physics. In 1985, Dr. Akasofu became the first recipient of the Chapman Chair Professorship at the University of Alaska Fairbanks, and in 1987, the National Association of State Universities and Land Grant Colleges named him as one of its "Centennial Alumni." He has also been honored with the Japan Academy Prize (1977), the John Adams Fleming Award of the American Geophysical Union (1979), and the Hannes Alfvén Medal from the European Geosciences Union (2011). In 2003, the Emperor of Japan bestowed on him the Order of the Sacred Treasure, Gold and Silver Star.

In addition, he has received award of appreciation for his efforts in support of international science activities from the Ministry of Foreign Affairs of Japan in 1993 and from the Ministry of Posts and Telecommunications of Japan in 1996. He was also the recipient of the University of Alaska Edith R. Bullock Prize for Excellence in 1997, and was named a Fellow of the American Geophysical Union in 1977, and of the American Association for the Advancement of Science in 2001. He received the 1999 Alaskan of the Year Denali Award, and the 2003 Aurora Award from the Fairbanks Convention and Visitors' Bureau.

Upon his retirement in 2007, the University of Alaska Board Regents officially named the building that houses the International Arctic Research Center the "Syun-Ichi Akasofu Building" in recognition of "his tireless vision and dedicated service to the university, the state, and country in advancing arctic science." Email: sakasofu@alaska.edu